\pdfoutput=1
\documentclass{elsart}
\usepackage{graphicx}
\usepackage{amssymb}
\usepackage{wasysym}
\usepackage{color}
\usepackage{lineno}
\usepackage{pgf}
\RequirePackage{xspace}

\def\TReg{\textsuperscript{\textregistered}}

\newcommand{\ee}{e$^+$e$^-$\kern -0.2em\xspace}
\newcommand{\off}{\ensuremath{o\kern -0.1em f\kern -0.2em f\kern
    -0.1em set}\xspace}

\newcommand{\ev}{\ensuremath{\mathrm{\,e\kern -0.1em V}}\xspace}
\newcommand{\mev}{\ensuremath{\mathrm{\,Me\kern -0.1em V}}\xspace}
\newcommand{\mevc}{\ensuremath{{\mathrm{\,Me\kern -0.1emV\!/}c}}\xspace}
\newcommand{\mevcc}{\ensuremath{{\mathrm{\,Me\kern -0.1em V\!/}c^{2}}}\xspace}
\newcommand{\gev}{\ensuremath{\mathrm{\,Ge\kern -0.1em V}}\xspace}
\newcommand{\gevc}{\ensuremath{{\mathrm{\,Ge\kern -0.1emV\!/}c}}\xspace}
\newcommand{\gevcc}{\ensuremath{{\mathrm{\,Ge\kern -0.1em V\!/}c^{2}}}\xspace}

\newcommand{\dedx}{\ensuremath{dE/dx}\xspace}
\newcommand{\PH}{\ensuremath{P\kern -0.1em H}\xspace}
\newcommand{\ppm}{\ensuremath{\mathrm{\,ppm}}\xspace}

\newcommand{\perc}{\ensuremath{\%}\xspace}
\newcommand{\dia}{\ensuremath{\diameter\,}\xspace}
\newcommand{\ft}{\,$--$\,}

\newcommand{\degree}{\ensuremath{\,\kern -0.1em ^\circ}\xspace}
\newcommand{\degreeC}{\ensuremath{\,\kern -0.1em ^\circ \mathrm{C}}\xspace}
\newcommand{\mrad}{\ensuremath{\,\mathrm{mrad}}\xspace}
\newcommand{\rad}{\ensuremath{\,\mathrm{rad}}\xspace}

\newcommand{\nm}{\ensuremath{\,\mathrm{nm}}\xspace}
\newcommand{\um}{\ensuremath{\,\mu\mathrm{m}}\xspace}
\newcommand{\mm}{\ensuremath{\,\mathrm{mm}}\xspace}
\newcommand{\cm}{\ensuremath{\,\mathrm{cm}}\xspace}
\newcommand{\m}{\ensuremath{\,\mathrm{m}}\xspace}
\newcommand{\cmperus}{\ensuremath{\,\mathrm{cm}/\kern -0.15em \mathrm{\mu s}}\xspace}

\newcommand{\uJ}{\ensuremath{\mathrm{\,\mu J}}\xspace}

\newcommand{\mV}{\ensuremath{\mathrm{\,m\kern -0.08em V}}\xspace}
\newcommand{\V}{\ensuremath{\mathrm{\,V}}\xspace}
\newcommand{\Vpermm}{\ensuremath{\mathrm{\,V\kern -0.18em /mm}}\xspace}
\newcommand{\Vpercm}{\ensuremath{\mathrm{\,V\kern -0.18em /cm}}\xspace}
\newcommand{\kVpercm}{\ensuremath{\mathrm{\,k\kern -0.06em V\kern -0.18em/ cm}}\xspace}
\newcommand{\Vperm}{\ensuremath{\mathrm{\,V\kern -0.18em /m}}\xspace}
\newcommand{\kV}{\ensuremath{\mathrm{\,k\kern -0.1em V}}\xspace}
\newcommand{\A}{\ensuremath{\,\mathrm{A}}\xspace}
\newcommand{\kA}{\ensuremath{\,\mathrm{kA}}\xspace}
\newcommand{\uA}{\ensuremath{\,\mu\mathrm{A}}\xspace}
\newcommand{\T}{\ensuremath{\,\mathrm{T}}\xspace}
\newcommand{\Tm}{\ensuremath{\,\mathrm{Tm}}\xspace}
\newcommand{\Hz}{\ensuremath{\mathrm{\,Hz}}\xspace}
\newcommand{\kHz}{\ensuremath{\mathrm{\,kHz}}\xspace}
\newcommand{\MHz}{\ensuremath{\mathrm{\,MHz}}\xspace}
\newcommand{\MOhm}{\ensuremath{\mathrm{\,M\Omega}}\xspace}
\newcommand{\kOhm}{\ensuremath{\mathrm{\,k\Omega}}\xspace}
\newcommand{\Ohm}{\ensuremath{\mathrm{\,\Omega}}\xspace}
\newcommand{\nF}{\ensuremath{\mathrm{\,nF}}\xspace}
\newcommand{\pF}{\ensuremath{\mathrm{\,pF}}\xspace}
\newcommand{\fC}{\ensuremath{\mathrm{\,f\kern 0.1emC}}\xspace}
\newcommand{\mVperfC}{\ensuremath{\mathrm{\,m\kern -0.08em V\kern -0.18em /f\kern 0.1emC}}\xspace}

\newcommand{\tb}{\ensuremath{\,\mathrm{time\kern 0.4em bin}}\xspace}
\newcommand{\ns}{\ensuremath{\,\mathrm{ns}}\xspace}
\newcommand{\us}{\ensuremath{\,\mathrm{\mu s}}\xspace}
\newcommand{\ms}{\ensuremath{\,\mathrm{ms}}\xspace}
\newcommand{\s}{\ensuremath{\,\mathrm{s}}\xspace}
\newcommand{\perh}{\ensuremath{\kern -0.1em \mathrm{/h}}\xspace}

\newcommand{\mbar}{\ensuremath{\mathrm{\,mbar}}\xspace}

\newcommand{\mosfetf}{\ensuremath{M_{\kern -0.06em f}}}
\newcommand{\capf}{\ensuremath{C_{\kern -0.06em f}}}
\newcommand{\Vgs}{\ensuremath{V_{\kern -0.1em gs}}}
\newcommand{\enc}{\ensuremath{\,e^{-}}\xspace}
\newcommand{\ADCcounts}{\ensuremath{\,\mathrm{ADC\kern 0.4em counts}}\xspace}
\newcommand{\pasa}{\ensuremath{P \kern -0.115em AS \kern -0.09em A}\xspace}
\newcommand{\kB}{\ensuremath{\,\mathrm{kB}}\xspace}

\begin{document}

\begin{frontmatter}

\title{The CERES/NA45 Radial Drift Time Projection Chamber}

\author[6]{D.~Adamov\'{a}},
\author[1]{G.~Agakichiev},
\author[1]{D.~Anto\'{n}czyk},
\author[11]{H.~Appelsh{\"a}user},
\author[3]{V.~Belaga},
\author[9]{J.~Biel\v{c}\'{\i}kov\'a},
\author[1]{P.~Braun-Munzinger},
\author[7,2]{R.~Campagnolo},
\author[4]{A.~Cherlin},
\author[2]{S.~Damjanovi\'{c}},
\author[2]{T.~Dietel},
\author[2]{L.~Dietrich},
\author[5]{A.~Drees},
\author[2]{W.~Dubitzky},
\author[2]{S.I.~Esumi},
\author[2]{K.~Filimonov},
\author[4]{Z.~Fraenkel},
\author[1]{C.~Garabatos},
\author[2]{P.~Gl{\"a}ssel},
\author[1]{G.~Hering},
\author[1]{J.~Holeczek}
\author[6]{V.~Kushpil},
\author[1]{A.~Mar\'{\i}n},
\author[2]{J.~Milo\v{s}evi\'{c}},
\author[4]{A.~Milov},
\author[1]{D.~Mi\'{s}kowiec},
\author[7]{L.~Musa},
\author[3]{Y.~Panebrattsev},
\author[3]{O.~Pechenova},
\author[2]{V.~Petr\'{a}\v{c}ek},
\author[7]{A.~Pfeiffer},
\author[9]{J.~Rak},
\author[4]{I.~Ravinovich},
\author[2]{M.~Richter},
\author[1]{H.~Sako},
\author[2]{E.~Sch{\"a}fer},
\author[2]{W.~Schmitz},
\author[7]{J.~Schukraft},
\author[2]{W.~Seipp},
\author[7]{A.~Sharma},
\author[3]{S.~Shimansky},
\author[2]{J.~Stachel},
\author[6]{M.~\v{S}umbera},
\author[2]{H.~Tilsner}, 
\author[4]{I.~Tserruya},
\author[10]{J.~P.~Wessels},
\author[2]{T.~Wienold},
\author[2]{B.~Windelband},
\author[9]{J.~P.~Wurm},
\author[4]{W.~Xie},
\author[2]{S.~Yurevich},
\author[3]{V.~Yurevich}

\address[6]{NPI/ASCR, \v{R}e\v{z}, Czech Republic}
\address[1]{Gesellschaft f{\"u}r Schwerionenforschung, Darmstadt, Germany}
\address[11]{University Frankfurt, Germany}
\address[3]{Joint Institute for Nuclear Research, Dubna, Russia}
\address[9]{Max-Planck-Institut f{\"u}r Kernphysik, Heidelberg, Germany}
\address[7]{CERN, Geneva, Switzerland}
\address[4]{Weizmann Institute, Rehovot, Israel}
\address[2]{University Heidelberg, Germany}
\address[5]{Department of Physics and Astronomy, SUNY Stony Brook, USA}
\address[10]{University M{\"u}nster, Germany}

\newpage
\begin{abstract}
The design, calibration, and performance of the first radial drift
Time Projection Chamber (TPC) are presented. The TPC was built and
installed at the CERES/NA45 experiment at the CERN SPS in the late
nineties, with the objective to improve the momentum resolution
of the spectrometer. The upgraded experiment took data
twice, in 1999 and in 2000. After a detailed study of residual
distortions a spatial resolution of $340\um$ in the azimuthal and
$640\um$ in the radial direction was achieved, corresponding to 
a momentum resolution of $\Delta p/p = 
\sqrt{\left(1\% \cdot p/{\rm GeV}\right)^2 + \left(2\%\right)^2}$. 
\end{abstract}

\end{frontmatter}

  \section{Introduction}

   Heavy-ion collisions at ultra-relativistic energies offer the
   possibility to study the behavior of nuclear matter at high density
   and/or temperature where one expects the formation of the Quark
   Gluon Plasma (QGP). A valuable tool to explore the early stage of
   heavy-ion collisions are electromagnetic probes. They are not
   subject to the strong interaction and can freely escape the
   surrounding hadronic medium.

   CERES/NA45 (Cherenkov Ring Electron Spectrometer) is the only
   experiment at the CERN Super Proton Synchrotron (SPS) dedicated to
   the study of \ee-pairs produced in \mbox{p-A} and \mbox{A-A}
   collisions in the low mass range of
   $0.1\gevcc<m_{ee}<1.2\gevcc$. It was installed in 1990 at the H8
   beam line of the SPS North Area and started its operation in
   1991. Systematic studies have been done using \mbox{S-Au}
   interactions in 1992 and proton-induced reactions \mbox{p-Be} and
   \mbox{p-Au} in 1993. An energy scan of \mbox{Pb-Au} interactions
   has been performed in beam times from 1995 to 2000. One of the main
   achievements of the CERES experiment is the measurement of an
   enhanced dilepton yield in \mbox{S-Au} and \mbox{Pb-Au} collisions
   as compared to expected contributions from vacuum decays of
   hadrons~\cite{Aga95,Aga98,Aga98b,Ada03c,Aga05}.

   In 1998 the experiment was upgraded with a radial drift Time
   Projection Chamber (TPC)~\cite{App98,Mar99} in order to
   significantly improve the mass resolution in the range of the
   $\phi$-meson from ${\Delta m/m = 7\perc}$ \cite{Aga05} to
   $3.8\perc$~\cite{Mar04,Mis05}. The additional information from the
   differential energy loss $\dedx$ in the TPC further improved the
   pion/electron separation capability. At an electron efficiency of
   $0.68$ the pion misidentification rate dropped from $5\cdot10^{-4}$
   to $2.5\cdot10^{-5}$ at a particle momentum of $1\gevc$
   \cite{Yur05,Bus06}. The dilepton invariant mass spectra measured
   after the experimental upgrade with the TPC allowed for the first
   time to discriminate between different theoretical
   approaches~\cite{Yur05,Ada06}. Finally, the TPC also opened the
   possibility to study hadronic channels. Since then many interesting
   topics have been addressed like particle multiplicities, elliptic
   flow, event-by-event fluctuations, particle correlations, and
   strangeness analyses (see for example~\cite{Mis05}).

\begin{figure}[ht]
    \centering
    \pgfimage[width=1.\textwidth]{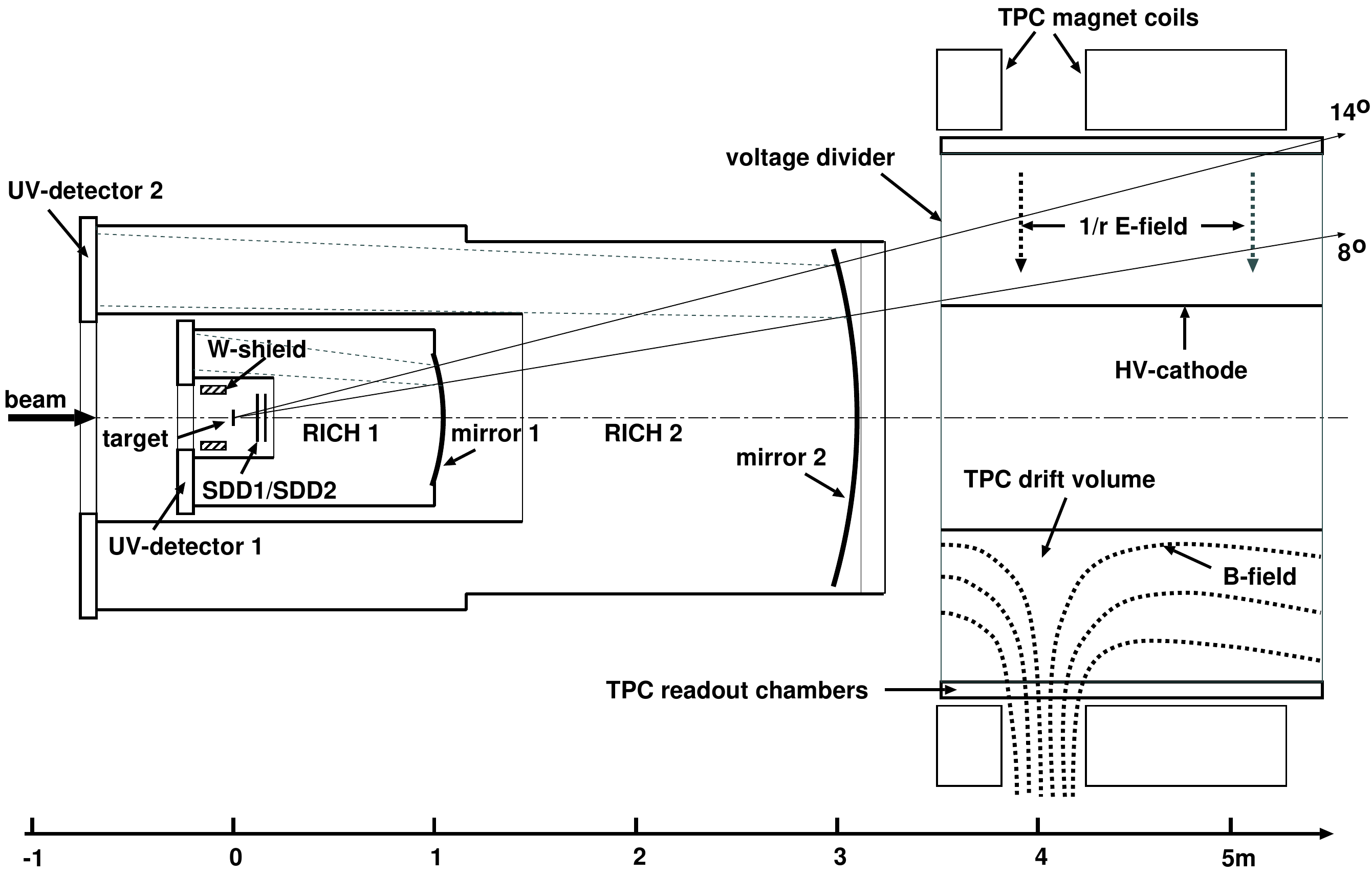}
    \caption{\label{fig:ceres} {\small Schematic view of the
	CERES/NA45 spectrometer. The TPC is operated inside an
	inhomogeneous magnetic field generated by two opposite
	polarity coils. The electric field inside the TPC is radial,
	pointing from the readout chambers to the
	high voltage cylinder.}}
  \end{figure}

   The upgraded CERES spectrometer, as it was used during the
   heavy-ion runs in 1999 and 2000, is shown in
   Fig.~\ref{fig:ceres}. The main components of the experiment are two
   Silicon Drift Detectors (SDD1, SDD2) for vertex reconstruction,
   two Ring Imaging Cherenkov Counters (RICH1, RICH2) for electron
   identification, and a radial drift Time Projection Chamber (TPC) for
   the measurement of the particle momentum and particle
   identification. All detector components of the spectrometer cover
   full azimuth at polar angles $8\degree < \theta < 14\degree$,
   corresponding to a pseudorapidity range of $2.10 < \eta < 2.65$.

  \section{TPC design considerations}

  The aim of an improved mass resolution required the addition of a
  charged particle spectrometer to the existing CERES experiment,
  consisting of a magnet system and a large volume electron tracking
  device \cite{Cer96a,Cer96b}. The TPC technology has been
  successfully applied for this task in heavy-ion experiments,
  combining high resolution tracking with large acceptance, low
  material budget, and reasonable cost.

  The new spectrometer system had to preserve the polar angle
  acceptance and the azimuthal symmetry of the existing experiment.
  This led to a radial electric field configuration resulting in the
  first radial drift TPC ever to be operated. In addition, a magnetic
  field configuration with two new magnet coils around the TPC has
  been adopted providing a strong radial component for momentum
  determination (cf. Fig.~\ref{fig:ceres} and
  Sect.~\ref{sec:mfield}). With a momentum dependent curvature in the
  $r$-$\phi$~plane the ionization electrons drift almost radially
  outwards to the readout chambers which are installed on the outer
  circumference of the TPC.

  For significant parts of the drift volume this design implies
  crossing electric and magnetic field lines leading to a finite
  variable Lorentz angle of the drifting electrons. The drift gas
  mixture Ne/CO$_2$ (80\perc/20\perc) has been found as an optimum
  with regard to small diffusion coefficients and Lorentz angle,
  sufficient primary ionization, long radiation length, and detector
  stability.  In addition, the maximum drift time should be kept low
  in order not to compromise the trigger rate without requiring
  excessively high drift fields.
  
  \section{Mechanical layout}
  \label{sec:Layout}

\begin{figure}[htb]
    \centering
    \pgfimage[width=0.5\textwidth]{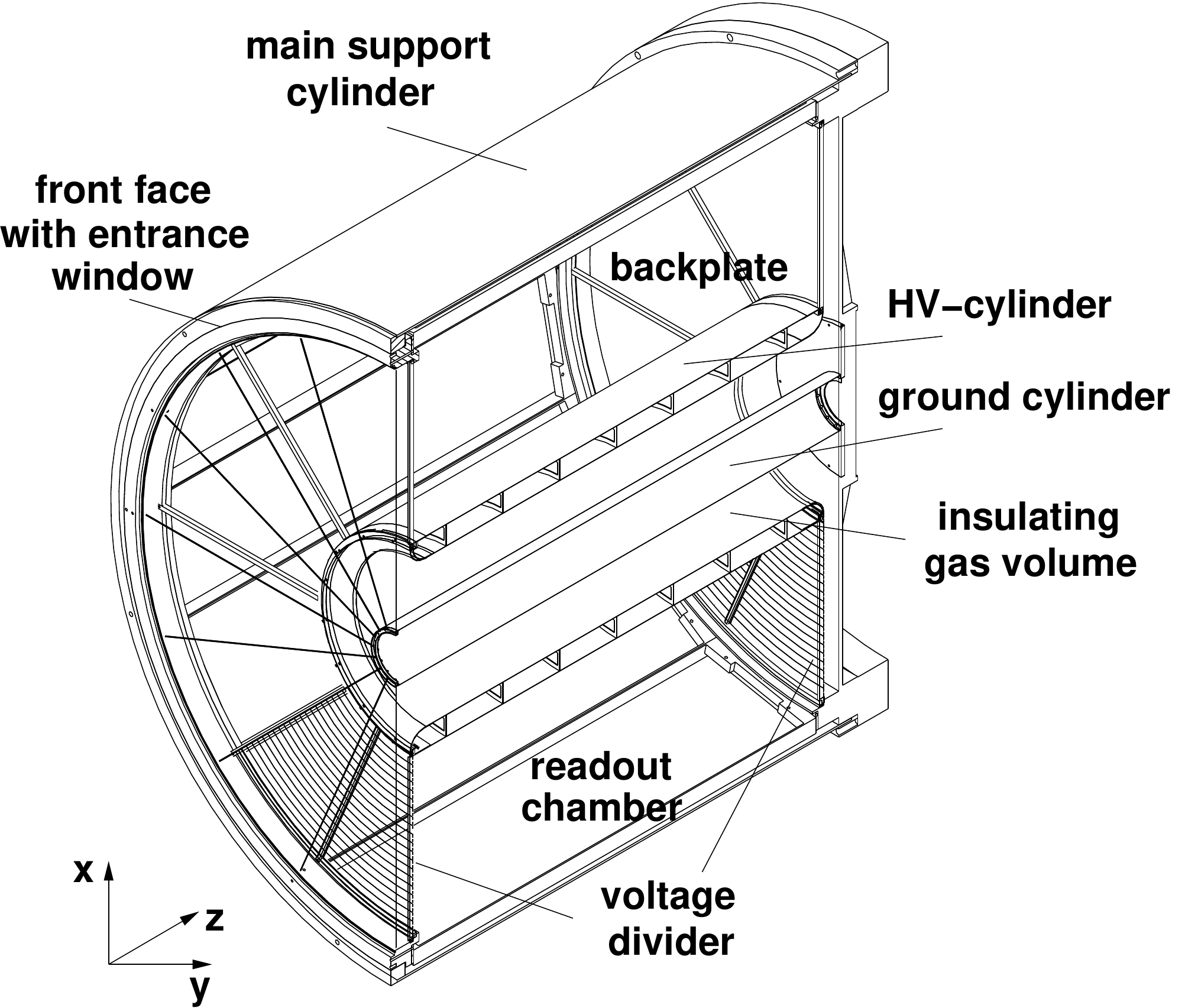}
    \caption{\label{fig:fieldcage} {\small Cross-section of the CERES
      TPC. The front and end faces are closed with Kapton\TReg foils which,
      together with the inner high voltage cylinder and the readout
      chambers, form a field cage and enclose the drift volume filled
      with Ne/CO$_{2}$. An insulating gap between the two inner
      cylinders of $8\cm$ width is flushed with CO$_{2}$.}}
  \end{figure}

  The TPC was installed downstream of the existing spectrometer, at a
  distance of $3.8\m$ from the target system.  A cross-section of the
  TPC is shown in Fig.~\ref{fig:fieldcage}. The mechanical stability
  of the TPC is provided by a massive backplate and an outer cylinder,
  both in aluminum.  The active volume of about $9\m^3$ is filled with
  a gas mixture of Ne/CO$_{2}$ (80\perc/20\perc). It is enclosed by
  the inner high voltage cylinder at $r=486\mm$, the 16 readout
  chambers at $r\approx1.3\m$, and two $50\um$
  Kapton\TReg\footnote{Kapton\TReg is a registered trademark of E. I. du
  Pont de Nemours and Company} foils with thin Cu voltage
  dividers at the front and the end faces. The active length of the
  TPC is $2\m$.  The ionization electrons created along charged
  particle tracks drift outwards and are detected in the readout
  chambers, Multi Wire Proportional Chambers (MWPC) with cathode
  pad-readout.  The TPC is operated in a magnetic field which
  predominantly deflects charged particles in azimuth, and provides
  the reconstruction of up to 20 space points along the particle
  track.

  \section{Electric field}
  \label{sec:efield}

\begin{figure}[htb]
    \centering
    \pgfimage[width=0.5\textwidth]{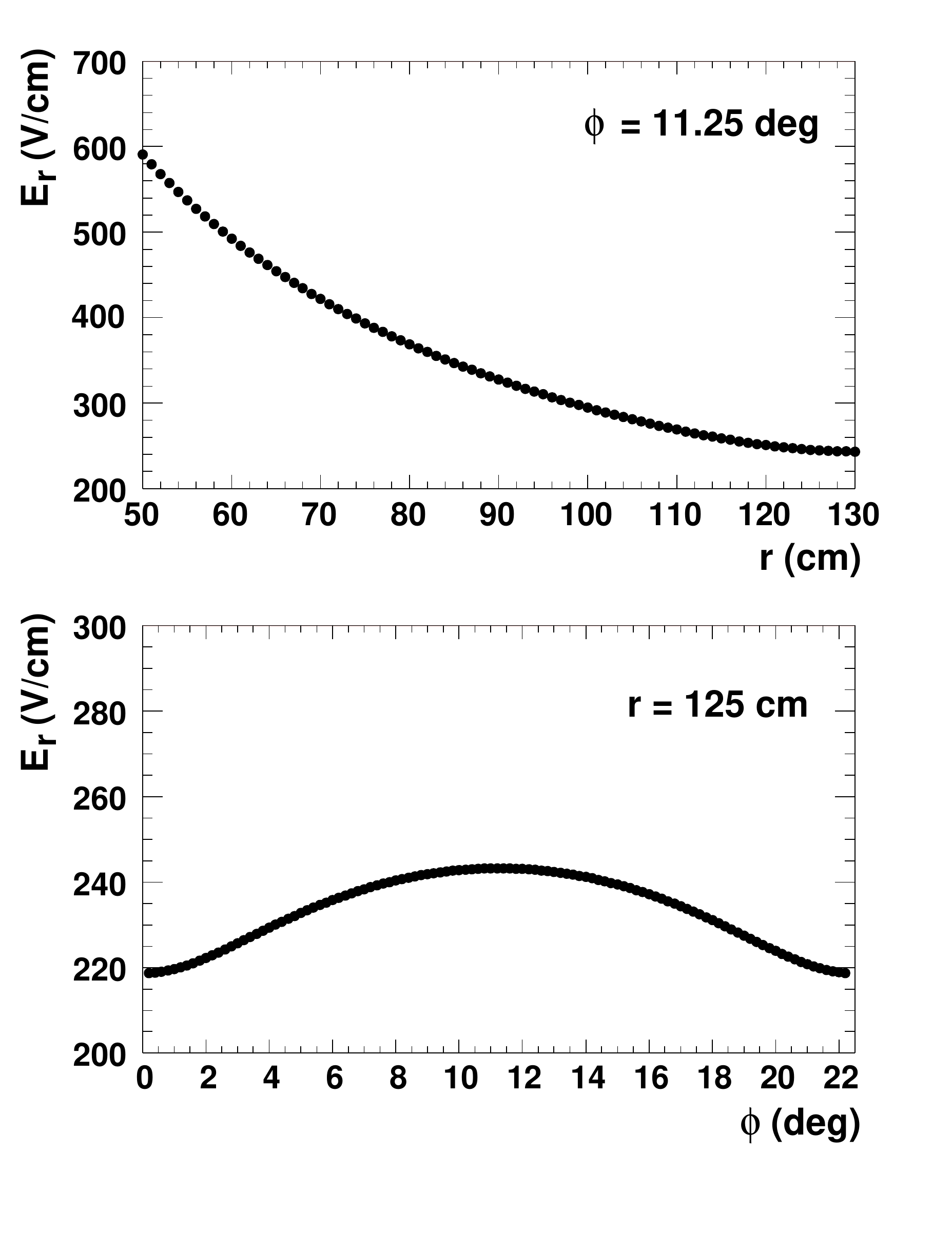}
    \caption{\label{fig:efield} {\small Radial component of the
      electric field vector as a function of the radius $r$ and the
      azimuthal coordinate~$\phi$ ($1$ readout chamber $\equiv
      22.5\degree$).}}
  \end{figure}

  The electric drift field follows a $1/r$-dependence varying between
  $600$ and $200\Vpercm$ (Fig.~\ref{fig:efield}). In Ne/CO$_{2}$
  (80\perc/20\perc) this corresponds to non-uniform drift velocities
  between $0.7$ and $2.4\cmperus$ and a maximum drift time of
  $71\us$. The electric field has a small azimuthal component due to
  the polygonal outer shape of the TPC and a small longitudinal
  component at the end caps.

  \subsection{Field cage}
  \label{sec:fieldcage}
 
  The field cage foils of the TPC are supposed to match the radial
  drift field at the ends of the drift volume and to provide gas
  tightness. This must be achieved at a minimum of material budget and
  with a mechanically and electrostatically stable structure.

  The foils are made of $50\um$ thick Kapton\TReg coated on both sides with
  42 copper concentric circular strips. The copper strips have a
  thickness of $200\um$. They are $15\mm$ wide and $5.46\mm$ apart from
  each other. The foils are composed of four pieces covering 90
  degrees and are supported by 8 stesalit bars of $1\cm$ width and
  thickness. The bars are wrapped in Kapton\TReg foil with the same copper
  pattern. In this case the copper layer is $4\um$ thick. The bars are
  connected to an inner and an outer Al ring which are attached to the
  HV-cylinder and the outer structure ('hamster cage'), respectively.
  The metal of the two sides of each foil is connected at the edges
  via small folded copper sheets of $0.5\mm$ thickness soldered to
  each strip. The copper sheets are used to solder the foils to the
  supporting bars.  The intermediate bars are connected to the field
  cage with silver conducting epoxy.  For mechanical stability and gas
  tightness, epoxy (Araldit 106) is applied along all the bars.

\begin{figure}[htb]
    \centering
    \pgfimage[width=0.5\textwidth]{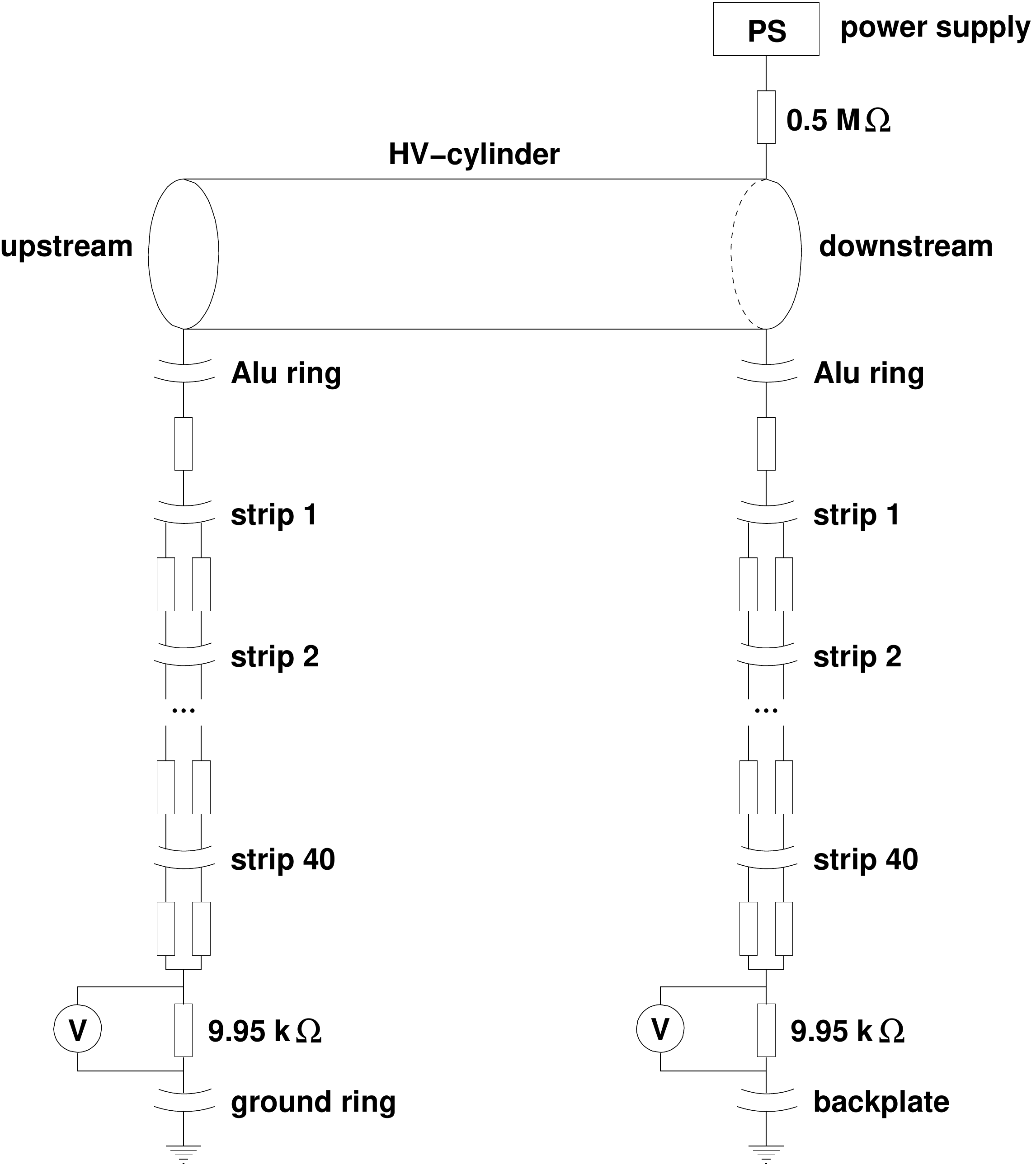}
    \caption{\label{fig:electrical}{\small Electric scheme of the field
      cage.  A high voltage of $29.2\kV$ is applied to the inner
      cylinder. Two resistor chains supply potentials to the upstream
      and downstream rings.  The two field cage currents are monitored
      via the voltage drop on the $9.95\kOhm$ test resistors.}}
  \end{figure}

  The electric diagram of the voltage divider is shown in
  Fig.~\ref{fig:electrical}.  A voltage of $-29.2\kV$ is applied to
  the inner cylinder via a $0.5\MOhm$ resistor. Two resistor chains
  are connected to the cylinder on one side and to the ground on the
  other. The resistors are regular carbon film resistors. They are
  soldered onto the strips of one bar. Because the matching field is
  radial and the strip pitch is constant, the resistor values must
  follow a 1/r-dependence.  The resistor values range from $16\MOhm$
  to $4\MOhm$.  The total current in each chain is $80\uA$.  The
  currents are measured via the voltage drop on the external
  $9.95\kOhm$ test resistors in series with the chains. The bars and
  the resistors are inside the drift volume.
  
  The resistors in the voltage dividers were chosen such that the
  strips should ensure a longitudinally homogeneous electric field
  inside the drift volume. However, it turned out that the resistance
  of some of the resistors used in the voltage dividers changed with
  the applied voltage. This resulted in a distortion of the drift field
  near the ends of the TPC which produced deviations of up to
  $4\cm$, as observed with laser tracks. A three dimensional
  calculation of the electric field with detailed geometry and
  appropriately adjusted resistor values accounted for this
  effect and reduced the track distortions by a factor of 6.

  \subsection{Field calculation}
  \label{sec:fieldcalculation}
 
\begin{figure}[htb]
    \centering
    \pgfimage[width=0.5\textwidth]{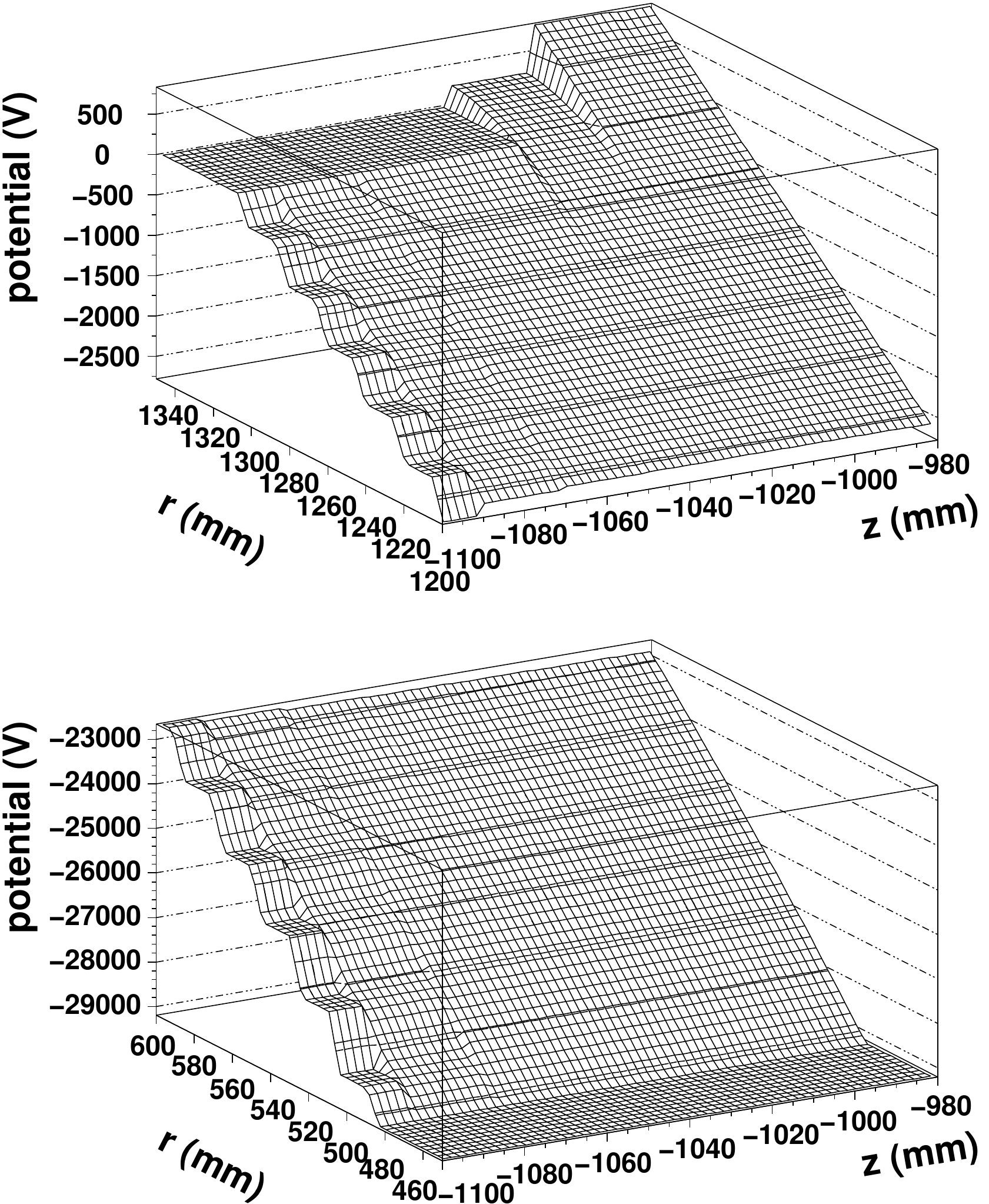}
    \caption{\small Electric potential $V(r,z)$ in the vicinity of the
      upstream end of the TPC close to the outer cylinder (top) and
      close to the inner cylinder (bottom). Both calculations were
      performed at an azimuthal angle corresponding to the center of a
      chamber. The steps are caused by the field cage rings. The
      calculation grid is $(dr,d\phi,dz)=(1\mm, 0.25\degree,
      2\cm)$. The sensitive volume is within $-1000\mm<z<1000\mm$.}
    \label{fig:mymax-a}
  \end{figure}

  The electric field calculation was performed using a custom program
  based on the relaxation method, including the proper knowledge of
  the ring potentials (Sect.~\ref{sec:fieldcage}) and the leakage of
  the amplification field through the cathode wire plane of the
  readout chambers (Sect.~\ref{sec:fieldconf}). An example of the
  potential calculated in the vicinity of the upstream field cage
  rings is shown in Fig.~\ref{fig:mymax-a}. The calculation was
  performed for half of a chamber (corresponding to $11.25\degree$)
  assuming chamber symmetry. The code heavily exploited the symmetries
  of the detector and thus allowed to calculate a potential map within
  several hours running time on a personal computer. This would be
  impossible using a standard field computation tool like
  \cite{maxwell}. Indeed, the drift field obtained by running this
  package for 6 days was still exhibiting non-physical inhomogeneities
  on the level of several percent.

  \subsection{Corrections at the ends of the TPC}
  \label{sec:corefieldTPCend}

\begin{figure}[ht]
    \centering
    \pgfimage[width=0.5\textwidth]{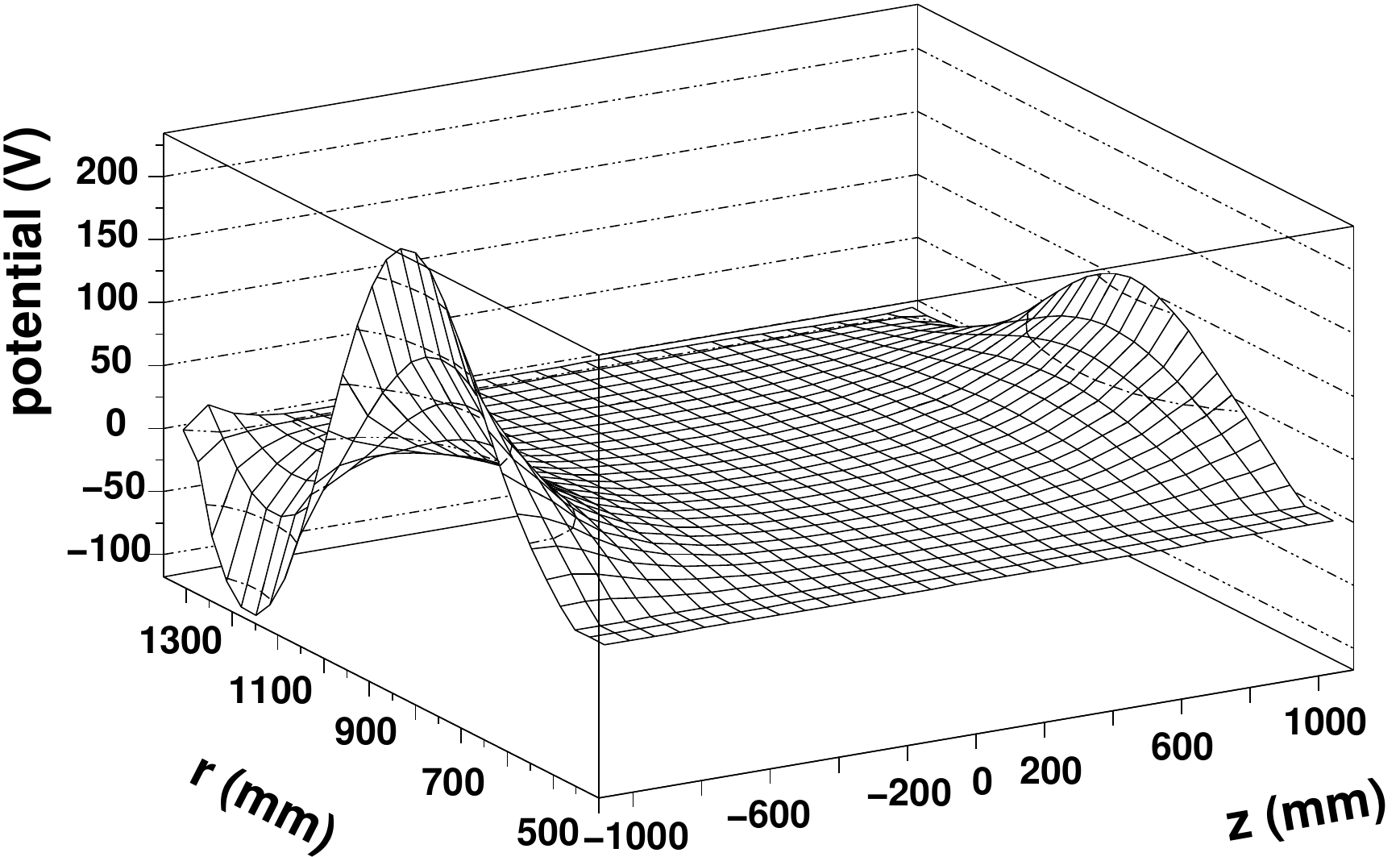}
    \caption{\small The residual correction potential map needed to
      straighten the reconstructed laser tracks. The correction is
      larger at the upstream end of the TPC.}
    \label{fig:mymax-c}
  \end{figure}

  After accounting for the variation of the field cage resistors with
  the voltage the deviations of the reconstructed laser tracks from a
  straight line went down from $4\cm$ to $5\ft7\mm$.  The final tuning
  was done by adding a phenomenological correction potential
  calculated requiring $V=0$ at the inner cylinder and the readout
  chambers, with the shape optimized to straighten the laser tracks
  (Fig.~\ref{fig:mymax-c}).

  \subsection{Correction for misalignment}
  \label{sec:corefieldmisalignment}

\begin{figure}[ht]
    \centering
    \pgfimage[width=0.5\textwidth]{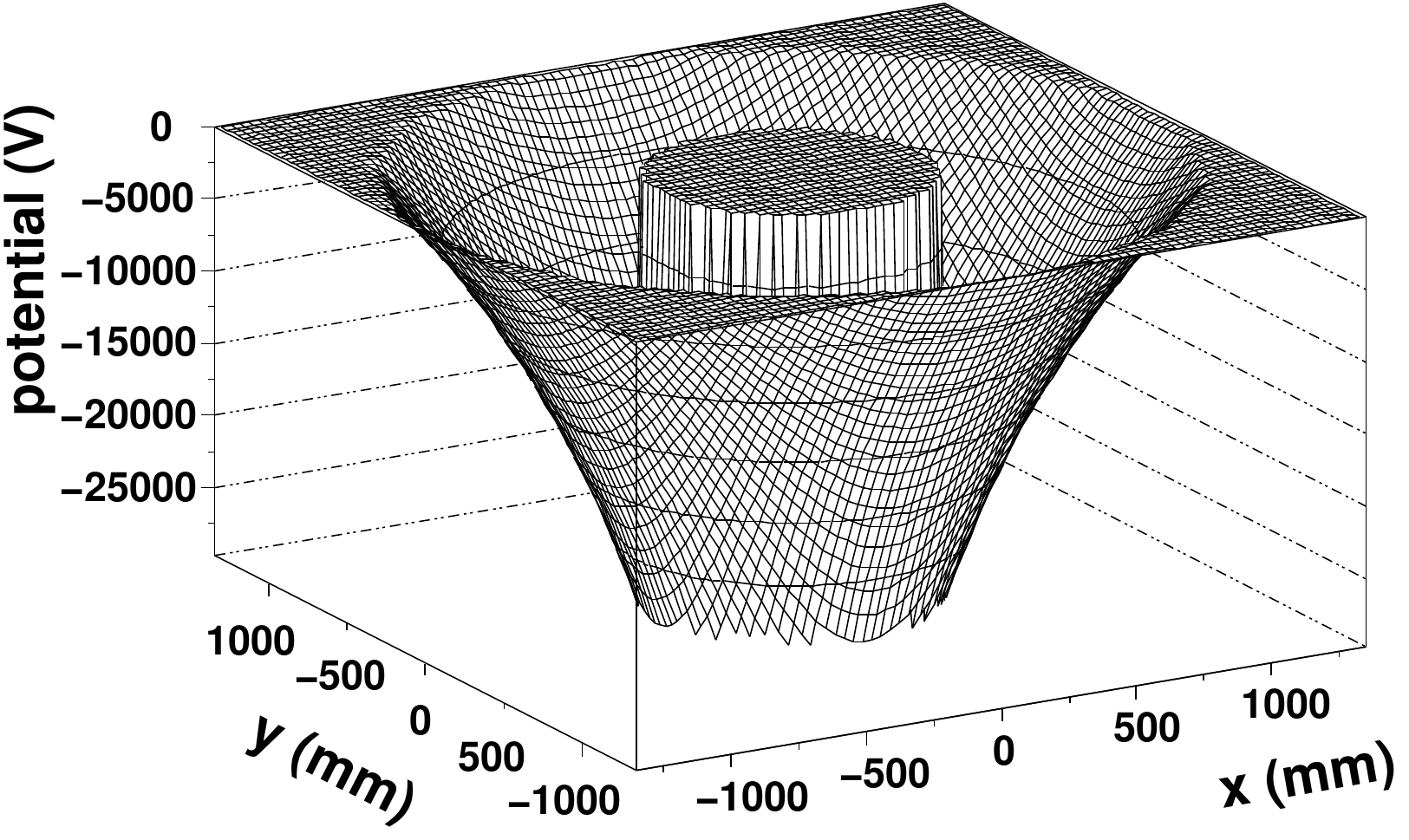}
    \caption{\small Two-dimensional electric potential $V(r,\phi)$ calculated
      with all chambers at their nominal positions.}
    \label{fig:mymax2d-a}
  \end{figure}

\begin{figure}[ht]
    \centering
    \pgfimage[width=0.5\textwidth]{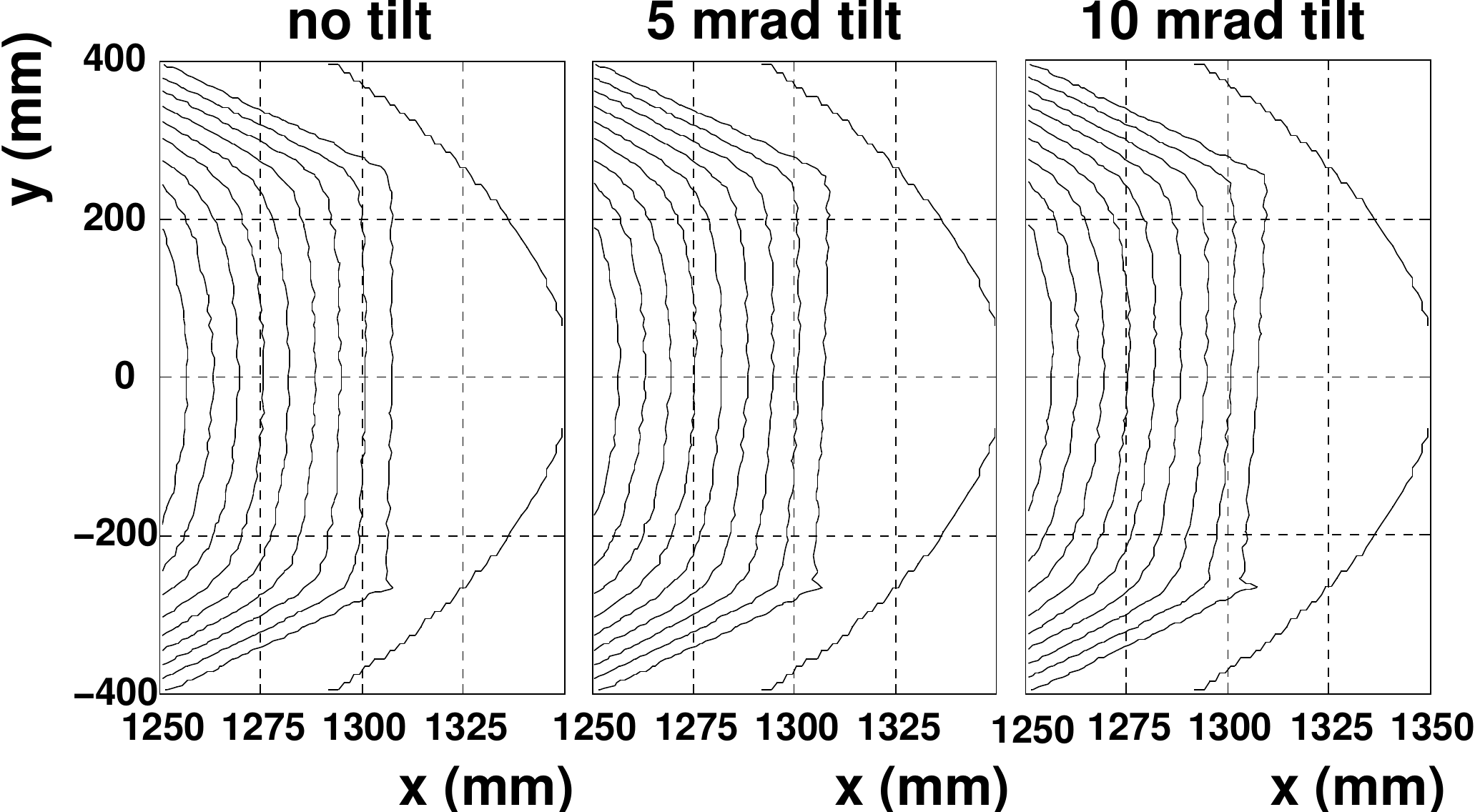}
    \caption{\small Equipotential lines calculated with the chamber
      at its nominal position (left) and with two values of tilt (middle and
      right). Correction maps, defined as the difference between a displaced
      map and a nominal one, were added to the three-dimensional potential
      calculated with the ideal geometry.}
    \label{fig:mymax2d-b}
  \end{figure}

  The three-dimensional field calculation, needed to account at the
  same time for the polygonal readout chamber configuration ($r(\phi)\neq const$)
  and for the track curvature in $r(z)$, was done for performance
  reasons only for a 11.25$^{\rm o}$ slice in $\phi$ of the TPC,
  i.e. for half a readout chamber.  The azimuthal symmetry, however,
  is broken by slight misalignments of the 16 chambers.  To account for this,
  a number of two-dimensional potential calculations $V(r,\phi)$ was
  performed for a complete set of 16 chambers
  (Fig.~\ref{fig:mymax2d-a}), and then repeated with one of the
  chambers being slightly displaced in $r$ or tilted in $r$ vs. $\phi$
  (Fig.~\ref{fig:mymax2d-b}).
  The correction potentials, defined as the difference between a two-dimensional
  calculation for a displaced (or tilted) and a perfectly aligned chamber
  \begin{equation}
    V_{cor}(r,\phi) = V_{misaligned}(r,\phi) - V_{ideal}(r,\phi)
  \end{equation}
  were then added to the three-dimensional calculations, separately
  for each chamber and each of the 20 $z$-planes. The displacements in
  $r$ and the tilts in $r(\phi)$ were chosen such as to move the
  reconstructed position of the high voltage cylinder as close as
  possible to the nominal $r=486\mm$ (maximum measured deviation of
  $0.3\mm$). The high voltage cylinder is visible in the data either
  via the electrons knocked out from aluminum by the laser light or as
  the edge in the radial distribution of hits in physics events. One
  should note that displacing a chamber affects the reconstructed
  cylinder position in two ways: via the modified field and via the
  change in the path length.  Both mechanisms were studied separately
  and in combination.  Displacing a chamber by $1\mm$ without changing
  the field shifts the reconstructed cylinder by $3\mm$ (this is
  because the drift velocity close to the cylinder is three times
  higher than at the chamber).  Using a field calculated with a
  chamber displaced by $1\mm$ but keeping the path length unchanged
  also gives a $3\mm$ shift. The combination of the two (as expected
  in a real case) gives a $4\ft5\mm$ apparent shift.
  
\begin{figure}[t]
    \centering
    \pgfimage[width=0.9\textwidth]{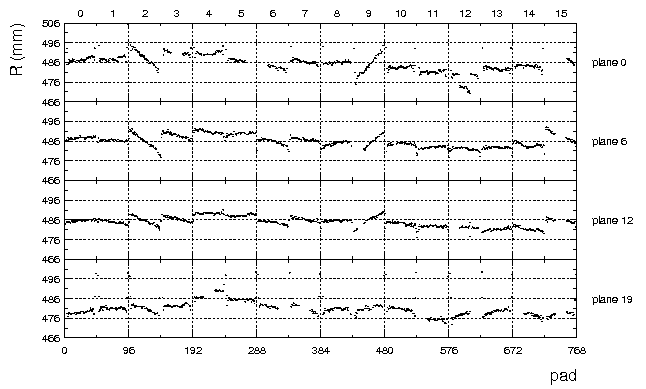}
    \caption{\small Reconstructed edge of the radial distribution of
      hits before correcting for chamber positions.  The
      data have been corrected for the pad-to-pad trace length
      variations.}
    \label{fig:cylinder_before_b}
  \end{figure}

\begin{figure}[t]
    \centering
    \pgfimage[width=0.9\textwidth]{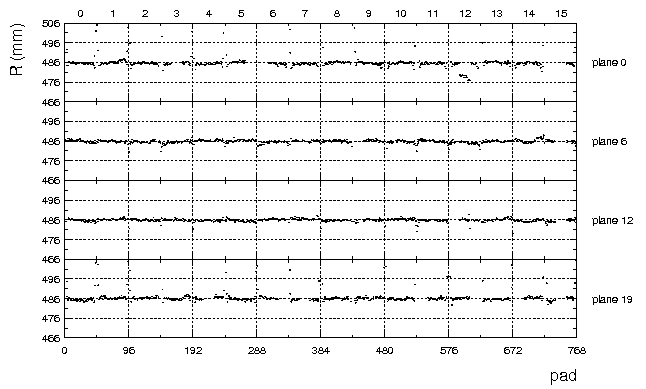}
    \caption{\small Reconstructed edge of the radial distribution of
      hits after correcting for chamber positions.  The
      data have been corrected for the pad-to-pad trace length
      variations.}
    \label{fig:cylinder_after_b}
  \end{figure}

  Displacements and tilts by up to $1.2\mm$ and $6\mrad$,
  respectively, are needed to get the reconstructed cylinder to its
  nominal position and shape. Figs.~\ref{fig:cylinder_before_b} and
  \ref{fig:cylinder_after_b} show the HV-cylinder position
  reconstructed before and after applying this correction,
  respectively.  The data in the figures have been corrected for the
  pad-to-pad trace length variations (Sect.~\ref{sec:padtopad}).

  \section{Magnetic field}
  \label{sec:mfield}
  
\begin{figure}[htb]
    \centering
    \pgfimage[width=0.5\textwidth]{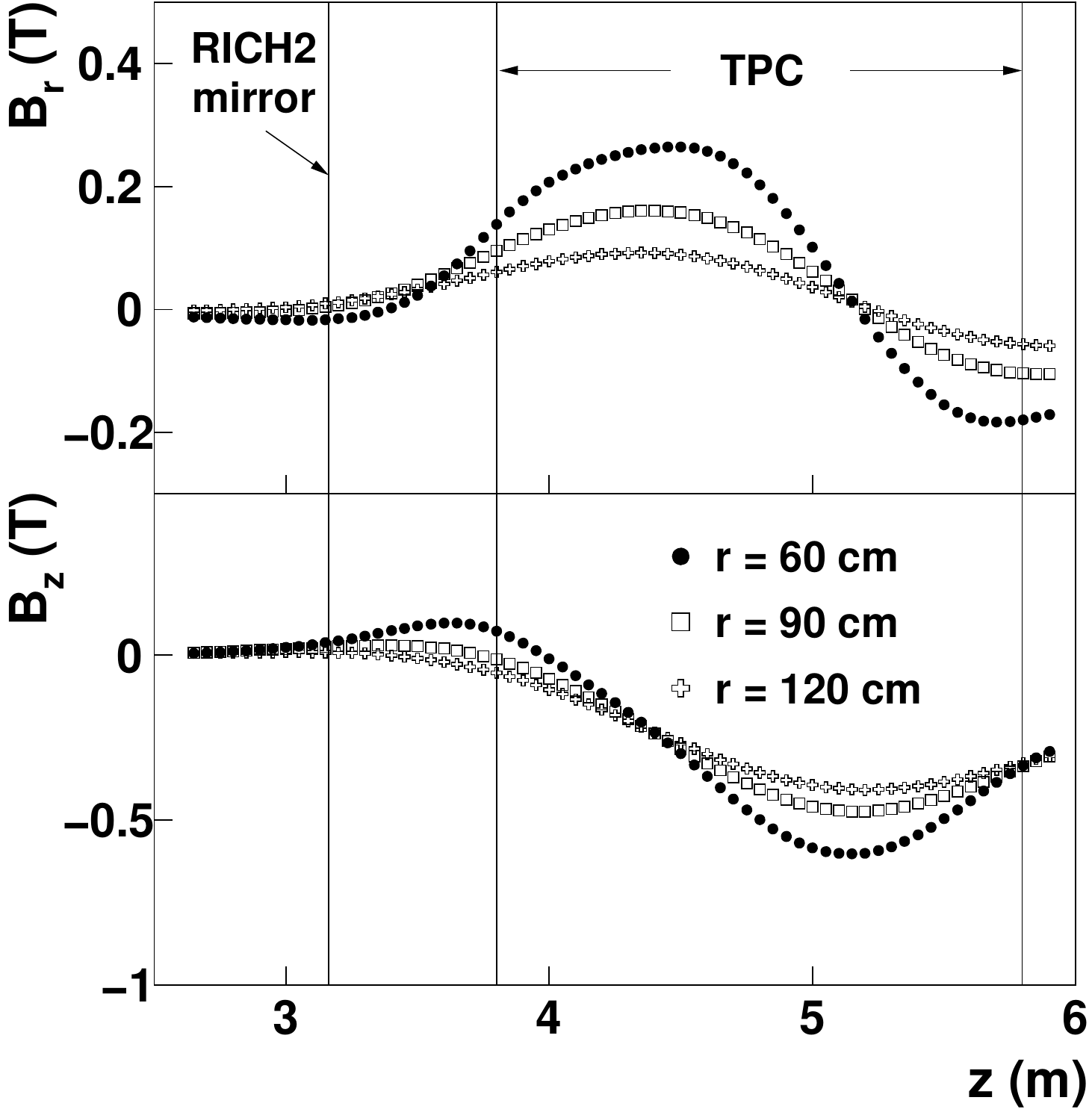}
    \caption{\label{fig:bfield} {\small Radial and longitudinal
      component of the magnetic field as a function of the
      longitudinal coordinate $z$ for different radii $r$. The active
      region of the TPC and the position of the mirror of the RICH2
      detector are indicated by vertical lines.}}
  \end{figure}

  The TPC is operated inside an inhomogeneous magnetic field generated
  by two solenoidal coils with opposite sense currents of up to
  $4\kA$. Fig.~\ref{fig:bfield} shows a calculation of the magnetic
  field using the CERN POISSON package \cite{POI06}. In the region
  between the two coils a field strength of $0.5\T$ is
  reached. Particles are deflected primarily in azimuthal
  direction. The field integral is $0.18\Tm$ at $\theta=8\degree$ and
  $0.38\Tm$ at $\theta=14\degree$. The magnetic fringe field outside
  the TPC drops off rapidly and is negligible at the position
  of the mirror of the RICH2 detector of the CERES spectrometer. The
  magnetic field was mapped (in presence of all TPC support structures)
  prior to installation of the TPC in the experimental area. The
  measurements correspond to the calculations with deviations from
  azimuthal symmetry at the percent level. These deviations are
  included in the analysis as corrections to the nominal field map
  \cite{Yur05}.

  \section{Counting gas}
  \label{sec:gas}

  The commonly used Ar/CH$_4$ ($90\perc$/$10\perc$) mixture was
  excluded as counting gas for its large Lorentz
  angles ($60\degree\ft70\degree$ in the relevant field range),
  diffusion coefficients, and multiple scattering angle. In good
  approximation, the width of the angular distribution caused by
  multiple scattering is given by \cite{Hig75,Lyn91}
  \begin{equation}
    \theta_{ms} = \frac{13.6\mev}{\beta c p} \, Z \, \sqrt{\frac{x}{X_{0}}}
    \left(1+0.038 \ln{\frac{x}{X_{0}}}\right),
  \end{equation}
  where $p$ is the momentum in $\mevc$, $\beta c$ is the velocity, $Z$
  is the charge number of the traversing particle, and $x/X_{0}$ is
  the thickness of the medium in units of its radiation length. Thus,
  a counting gas with large radiation length minimizes the influence
  of multiple scattering. In terms of multiple scattering, Ne-based
  gas mixtures are preferable to Ar-based mixtures.
  \subsection{Drift velocity}
  
  The drift velocity of the ionization electrons in the gas as derived 
  from the Langevin equation can be written as a function of the electric 
  field $\vec{E}$ and the magnetic field $\vec{B}$:
  \begin{equation}
    \label{eqDriftVel}
    \vec{v}_{d} = \frac{\mu}{1+(\omega
      \tau)^2} \left(\vec{E} + \omega \tau
    \frac{\vec{E} \times \vec{B}}{B} + (\omega \tau)^2
    \frac{(\vec{E} \cdot \vec{B}) \vec{B} }{B^2}\right) .
  \end{equation}
  In this equation $\tau$ is the mean time between two
  collisions. The cyclotron frequency $\omega$ is given by:
  \begin{equation}
    \omega \tau = \frac{e}{m} B \tau= B \mu
  \end{equation}
  where $\mu$ is the electron mobility. The
  mobility is a function of the electric field, gas composition,
  pressure, and temperature.

  \subsection{Composition}
  \label{sec:composition}

\begin{figure}[ht]
    \centering
    \pgfimage[width=0.7\textwidth]{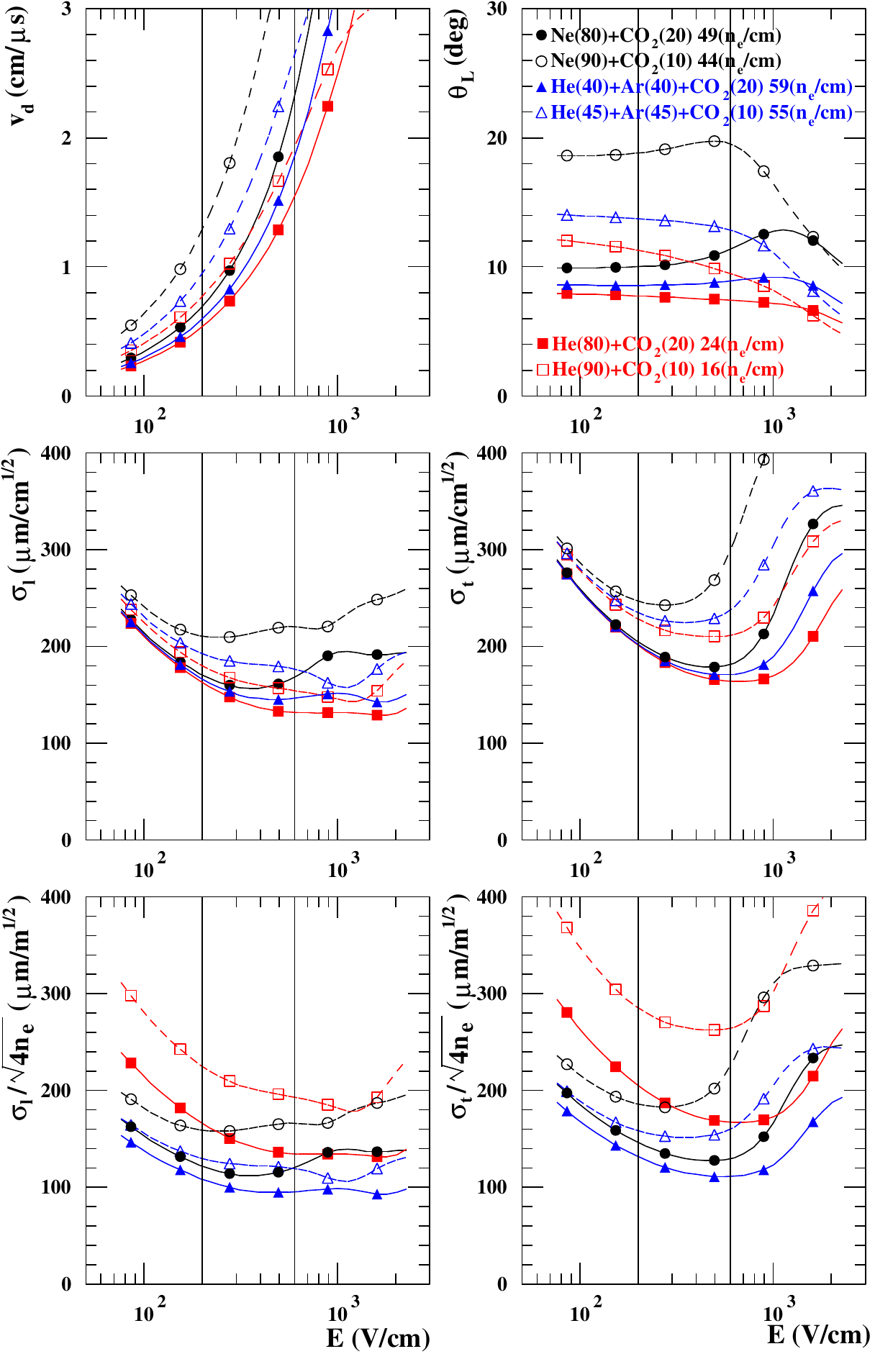}
    \caption{\label{fig:gasOpt} {\small MAGBOLTZ simulations
    for a magnetic field of $0.5\T$ perpendicular to the electric
    field. The drift velocity (top left), Lorentz angle (top right),
    longitudinal (middle left), and transverse (middle right)
    diffusion coefficients and the diffusion coefficients normalized
    to the square root of the number of primary electrons (bottom)
    were calculated for various gas mixtures. The vertical lines
    indicate the relevant range of the electric field in the CERES
    TPC.}}
  \end{figure}

  The outlined considerations were the baseline for detailed simulations
  with the package MAGBOLTZ \cite{Bia99}. For a large
  number of gas mixtures the drift velocity, Lorentz angle,
  longitudinal, and transverse diffusion coefficients were calculated
  as a function of the electric field for a given magnetic field at a
  fixed angle. The calculations cover magnetic fields from $0\T$ to
  $0.5\T$ at angles between $0\degree$ and $90\degree$ with respect to
  the electric field. Fig.~\ref{fig:gasOpt} shows results for various
  gas mixtures for the most disadvantageous case of a magnetic field
  of $0.5\T$ perpendicular to the electric field \cite{Cer96b}. The
  gas mixtures under considerations are Ne, He and Ar/He with
  admixtures of $10\perc$ to $20\perc$ CO$_2$. Also shown at the
  bottom of Fig.~\ref{fig:gasOpt} are the diffusion coefficients
  normalized to the square root of the number of primary electrons
  $n_e$, since this ratio is the relevant quantity for the position
  resolution at the readout chamber.

  Another limiting factor is the rate of clean interaction at a beam
  rate of $5\cdot 10^5$/s ($2\cdot 10^6$/burst). This means that no
  further interaction should occur within a time window of twice the
  maximum drift time. A maximum drift time of $34\us$ makes Ne/CO$_2$
  ($90\perc$/$10\perc$) the preferred solution. He/Ar/CO$_2$
  ($45\perc$/$45\perc$/$10\perc$) has the advantage of lower cost and
  yields, in terms of resolution, very similar performance but has a
  maximum drift time of $70\us$. The final choice of $80\perc$ Ne and
  $20\perc$ CO$_2$ is a compromise between an optimum resolution and
  an acceptable loss of primary electrons, both of which are
  increasing with the fraction of CO$_2$. The latter is caused by the
  attachment of free electrons to oxygen impurities in the counting
  gas, with subsequent deexcitation through collisions with CO$_2$
  molecules (Sect.~\ref{sec:eAttach}). In addition, a higher
  concentration of quencher allows for a safer operation against glow
  discharges of Ne-based gas mixtures at relatively high gains.
  On the other hand, CO$_2$-based mixtures are very sensitive to
  temperature and pressure fluctuations.

  \subsection{Gas system}

\begin{figure}[t]
    \centering
    \pgfimage[width=1.\textwidth]{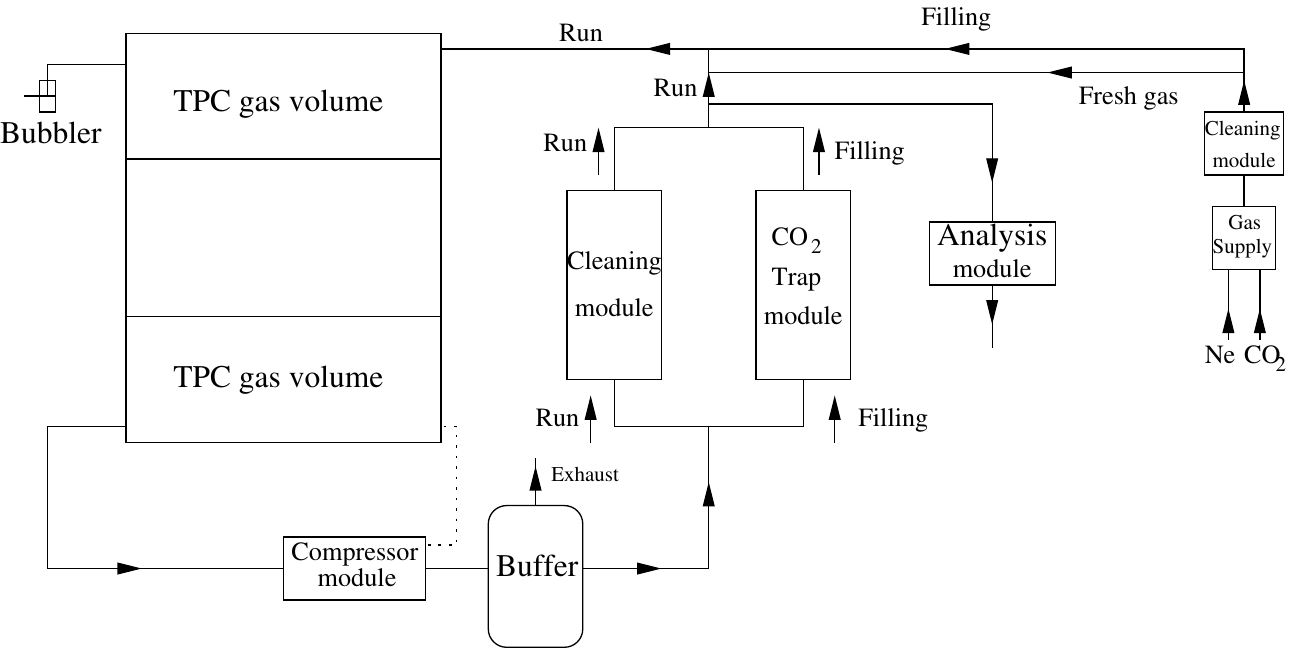}
    \caption{\label{fig:gasSystem} {\small Layout of the gas supply
    system of the CERES TPC. The copper catalyzers (cleaning module)
    remove oxygen and water vapor from the gas. The molecular sieves
    (trap module) trap CO$_2$ to bring the TPC to the correct
    gas mixture of Ne/CO$_2$ ($80\perc$/$20\perc$).}}
  \end{figure}

  In order to minimize the overall Ne consumption the gas system for
  the CERES TPC was designed as a closed loop circuit. The schematic
  is shown in Fig.~\ref{fig:gasSystem}. The majority of the gas
  mixture recirculates through the gas system cleaning part and the
  TPC. During a normal run small amounts of fresh gas
  ($0.12\m^3\perh$) are replaced. The gas circulation rate is about
  $1.4\m^3\perh$, corresponding to about $15\perc$ volume exchange per
  hour.

  The gas system contains two compressors with a capacity of
  $3.7\m^3\perh$ each, located at the exit of the TPC. The
  recirculating gas flows through a buffer vessel and subsequently
  through one of two cartridges, filled with a copper catalyzer, to
  remove oxygen and water vapor. The purity and composition of the gas
  is monitored using oxygen, water, and CO$_2$ analyzers. Ambient
  pressure fluctuations are compensated for by filling or emptying gas
  from the buffer vessel, thus keeping a constant overpressure in the
  TPC.  The pressure in the TPC is regulated at about $1\mbar$ above
  the atmospheric pressure. The operation of the system is controlled by a
  Programmable Logic Controller (PLC). A safety bubbler is installed
  at the detector to prevent any under- or overpressure at the TPC
  higher than $2\mbar$, in particular during power failures.

  The gas system can be operated either in an open system
  configuration for purging with CO$_2$ or in a special mode foreseen
  to bring the TPC to the correct mixture of Ne/CO$_2$ filling. The latter is
  achieved with two parallel cartridges containing 2.7 kg of molecular
  sieve 13X each to absorb CO$_2$. The cartridges are equipped with heating
  and cooling devices. A pump is connected to the output of the
  cartridges to vent the absorbed CO$_2$. During the filling
  process the gas of the TPC is passed through one of the cartridges
  and the CO$_2$ is trapped. At the same rate Ne gas is injected into
  the system. Since Ne is not trapped the amount of Ne in the mixture
  increases with time. Once the cartridge is saturated with CO$_2$,
  the system changes automatically to the second cartridge and the first
  one is regenerated. Regeneration is achieved by heating the
  cartridge to 220$^\circ$C while purging it with Ar and
  pumping the liberated CO$_2$ out of the system. After 1 hour the
  cooling line brings the cartridge to ambient temperature and the
  system switches over to the other cartridge.
  In this manner, no Ne is waisted while filling the detector. A total
  of 24 passes is necessary to trap $7.2\m^3$ of
  CO$_2$. The filling process takes about 50 hours.

  \subsection{Cooling system}
  
  The TPC requires a stable operating temperature which was set to
  $24\degreeC$.  The sources of temperature variation are the
  proximity to the RICH detectors which are kept at $50\degreeC$, the
  magnet, the TPC readout electronics, and the hall environment.
  Three types of cooling circuits were installed: an insulating screen
  between the magnet and the TPC, cold screens against the heat from
  electronic cards, and CO$_2$ circuits.
  \begin{itemize}
  \item {\bf Insulating screen}. The screen consists of two circuits
    of copper pipes with water circulating through them. The pipes 
    are attached to the TPC's
    external aluminum cylinder by heat-transfer cement. Each circuit
    is connected to a temperature adjustable assembly to guarantee a
    constant temperature outside the TPC.
  \item{\bf Cold screens.}  Copper plates with a welded-on pipe are
    attached to each electronics card.  The plates are
    water-cooled. The temperature is maintained above the dew point
    and it is the same on all the plates.
  \item{\bf CO$\mathbf{_2}$ insulating volumes.} Two closed CO$_2$
    circuits, one in the TPC entrance window and one in the backplate,
    each $10\cm$ thick, are used to stabilize the temperature inside
    the TPC from variations in the hall, and/or RICH.
  \end{itemize}
  The water circuits are supplied via a Leakless Cooling System
  Version 2 (LCS2) \cite{COO00}, where the pressure is kept below
  atmospheric pressure. All the circuits are remotely controlled by a
  PLC. Individual PID controllers adjust the temperatures.
  
  With this cooling system the short-term temperature variations could
  be stabilized to below $0.3\degreeC$ during data taking periods.

  \subsection{Slow control}

\begin{figure}[ht]
    \centering
    \pgfimage[width=0.5\textwidth]{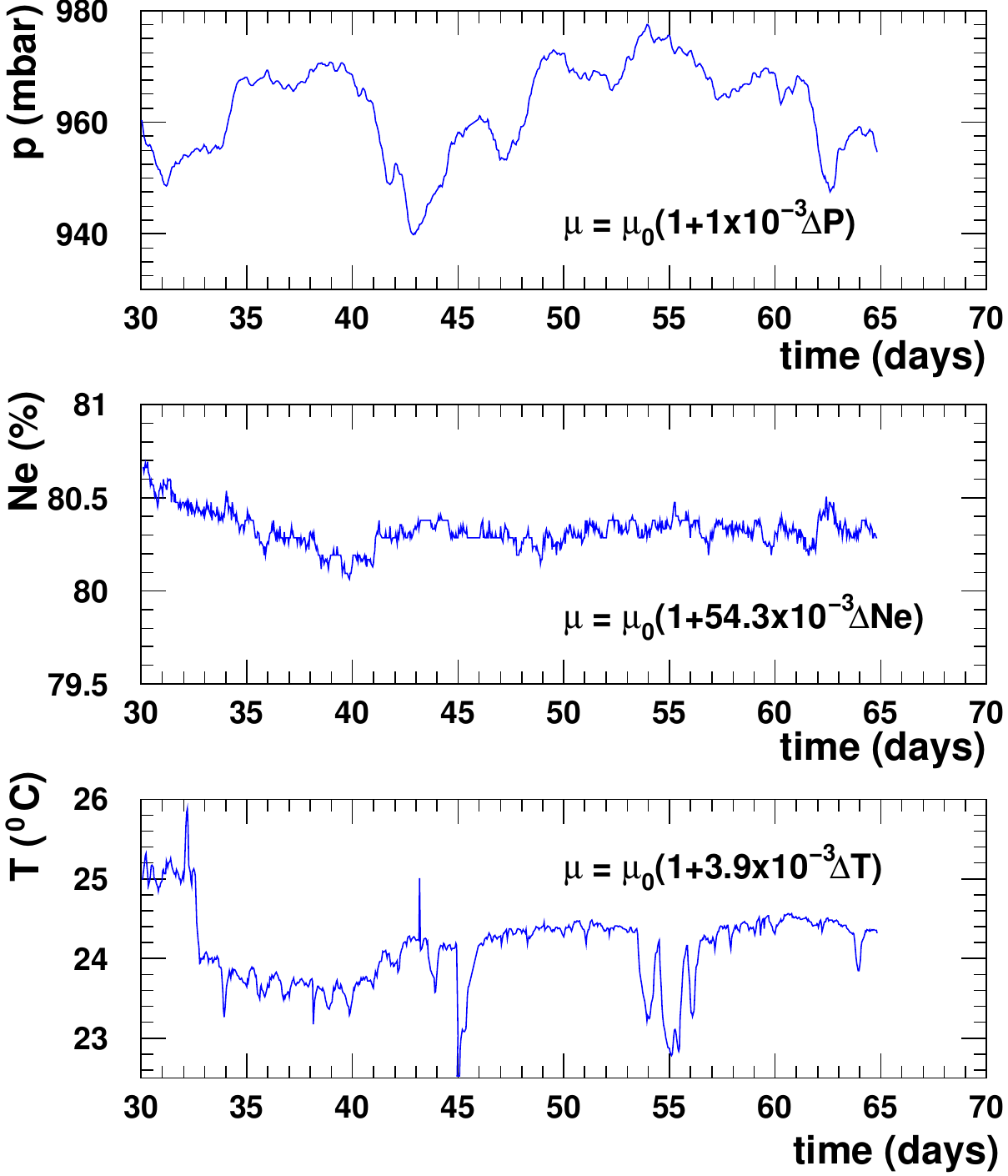}
    \caption{\label{fig:slowctrl} {\small The monitoring system of the
      CERES experiment surveyed all parameters that could influence
      the detector response, e.g. the pressure (top), gas composition
      (middle), and temperature (bottom). The changes of the nominal
      mobility $\mu$ due to these fluctuations are in the percent
      range.}}
  \end{figure}

  A computer driven system monitors all the parameters that could have
  an influence on the response of the TPC during the beam time
  (Fig.~\ref{fig:slowctrl}). These parameters include pressure,
  temperature, composition, oxygen and water content of the drift gas,
  drift velocity of electrons in the TPC, and the stability of the
  electric drift field. The gas monitoring system consists of a
  personal computer (PC) running LINUX, a 16 channel $330\kHz$ 12-bit
  Analog-to-Digital-Converter (ADC) with programmable gain installed
  in the PC, a 192-bit Digital-Input-Output board (DIO), and a 48
  channel Form C, $6\A$ relay.

  The devices measuring the pressure, gas mixture and water content
  provide a voltage level which is read directly by the ADC. The TPC
  is equipped with three 16-bit PT100 temperature sensors. They are
  supplied with a constant current from an ISO-5B32 current input
  module. The reading is done sequentially using the relay which is
  driven by the DIO module. The oxygen level is recorded via a
  Orbisphere sensor with voltage level output. The drift velocity
  measurement is provided by a drift velocity monitor, known as GOOFIE
  \cite{Mar95}, which is operated with the TPC gas mixture. The drift
  velocity monitor is read out with a LeCroy CAMAC Waveform Recorder
  6810 and sent to a dedicated PC running WINDOWS. Via SAMBA the data
  are redirected to the LINUX PC, where they are stored and
  analyzed. The electric field stability is monitored by recording the
  high voltage, the total current through the voltage divider, and the
  voltage drop through two external resistors of the field cage.

  \subsection{Calibration of the drift velocity}
  \label{sec:drift}

  Since the drift velocity in first order scales with the electric
  field, it is convenient to consider the ratio of the two, called
  electron mobility, in the absence of a magnetic field.  The
  knowledge of the mobility for different values of the electric field
  is essential for understanding drift in a radial field.  Seven
  approximately parallel laser rays in one sector of the TPC, together
  with the signal from the HV-cylinder (photoelectric effect in the Al), were
  used to determine the electron mobility function.  This was achieved
  by minimizing radial distances of the reconstructed TPC hits from
  the expected track positions.  The expected track positions were
  calculated from the laser beam position before entering the TPC,
  measured with position sensitive diodes, and the known mirror
  positions and angles (cf. Sect.~\ref{sec:laser}). In the case of the
  cylinder signal the reconstructed radius was compared to the nominal
  $486\mm$.  The 12 fit parameters included three factors for electric
  field corrections at each TPC end (Sect.~\ref{sec:corefieldTPCend}),
  five parameters of the electron mobility curve, and the electronics
  time offset.
  
\begin{figure}[ht]
    \centering
    \begin{tabular}{cc}
    \pgfimage[width=0.5\textwidth]{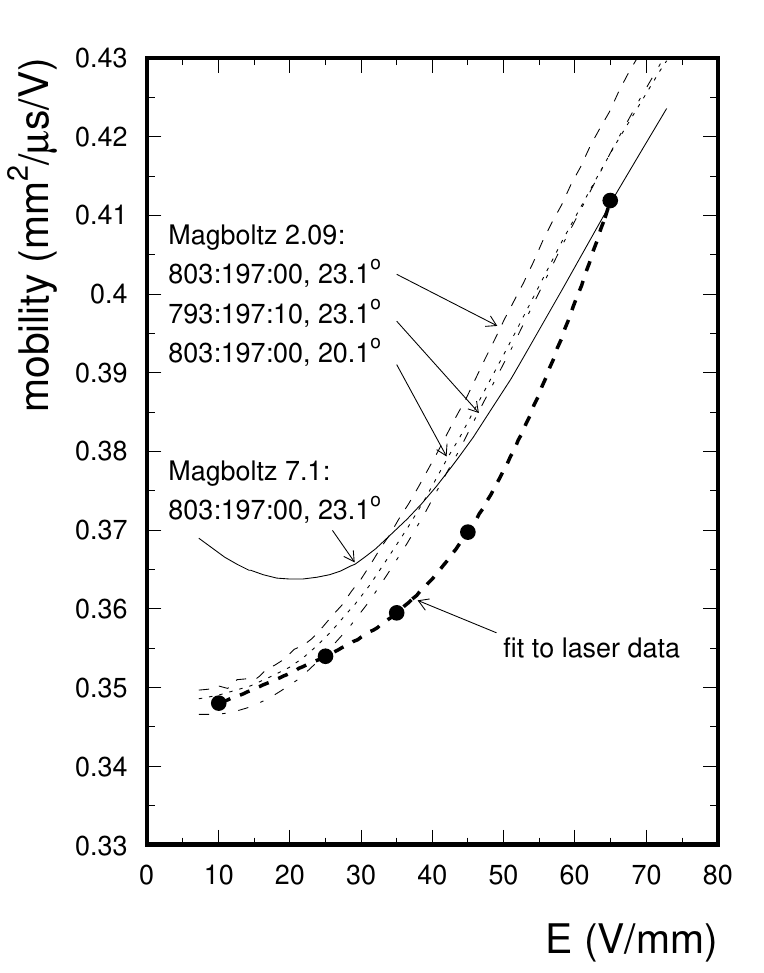}&
    \pgfimage[width=0.5\textwidth]{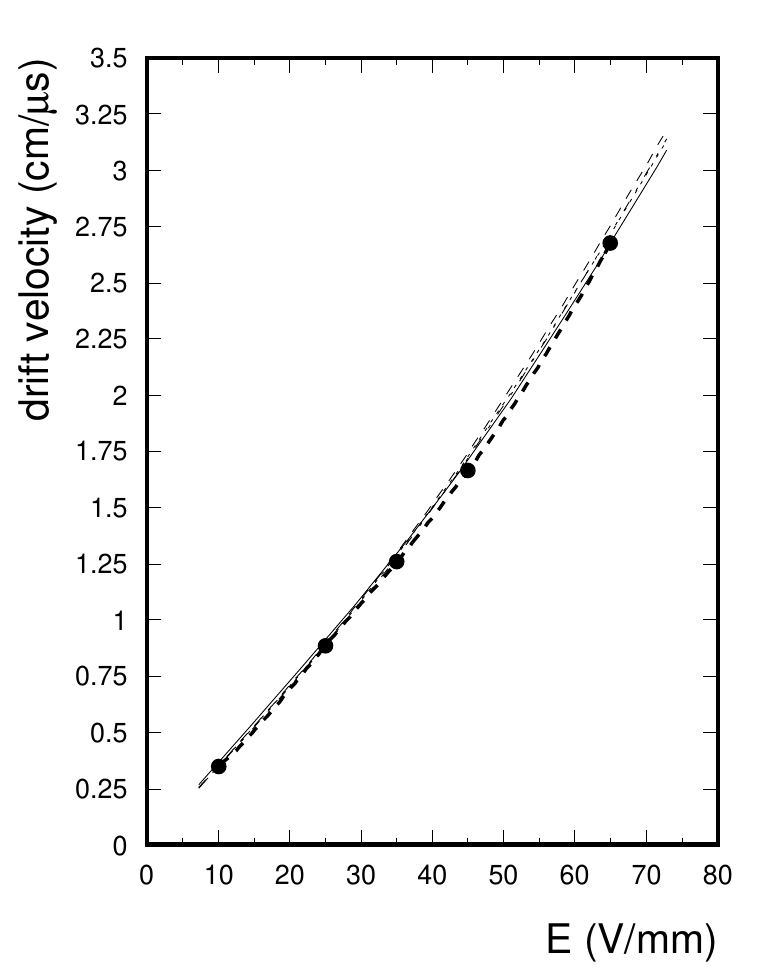}\\
    \end{tabular}
    \caption{\small Left: Mobility obtained from the fit to the laser
      data (dashed line with full dots). For comparison, the dashed,
      the dotted, and the dash-dotted lines show mobility calculations
      with MAGBOLTZ 2.09 for three gas compositions (Ne:CO$_2$:N$_2$)
      and temperature cases at a pressure of $971.5\mbar$. The solid
      line shows an analogous calculation done with MAGBOLTZ 7.1.  The
      difference between different versions of MAGBOLTZ originate from
      updates of its cross section values. Right: The curves become
      nearly indistinguishable when the drift velocity is plotted
      rather than the mobility.}
    \label{fig:mobility-fit87b}
  \end{figure}

  The five-parameter mobility curve obtained from the fit is shown as
  a dashed line with full dots in Fig.~\ref{fig:mobility-fit87b}. The
  five parameters are the mobility values at $E = 10$, 25, 35, 45, and
  $65\Vpermm$, covering the range of electric field values at
  different positions in the TPC (cf. Fig.~\ref{fig:efield}). The rest
  of the curve is an interpolation \cite{divdif} between these points.
  This technique couples the parameters to specific drift regions and
  thus helps the fit to converge quickly. The measured mobility is
  compared to various versions of the simulation package MAGBOLTZ
  \cite{Bia99} in which the mobility is calculated microscopically
  based on the energy dependent cross sections for collisions of
  electroncs on atoms and molecules. The deviation is significant and
  cannot be removed by even significant changes in the input gas
  composition or temperature (solid, dashed, dotted and dash-dotted
  lines).

  \subsection{Electron attachment}
  \label{sec:eAttach}

  The noble gas neon in the TPC has an admixture of $20\%$ carbon
  dioxide as quencher. The CO$_{2}$ improves the drift properties
  (cold gas) and prevents multiple discharges.  It absorbs the photons
  emitted by excited atoms or deexcites the atoms directly through
  collisions. The energy mainly goes into rotational and vibrational
  excited molecular states and into ionization of the quencher.

  However, the CO$_{2}$ molecules also interact with gas impurities
  like oxygen. Oxygen has the undesired property of attaching free
  electrons from primary ionization processes:
  \begin{equation}
    e^{-} + O_{2} \rightarrow O_{2}^{-*}.
  \end{equation}
  In dilute media the oxygen loses its energy by the reemission of the
  electron or by radiation. Unfortunately, at atmospheric pressure, as
  present in the TPC, the dominating process is the interaction with
  another molecule M:
  \begin{eqnarray}
    & & O_{2} + M + e^{-} \\ \label{proc1}
    O_{2}^{-*} + M  \begin{array}{c}
      \; \nearrow \\ \; \searrow 
    \end{array}  & &\nonumber \\
    & &  O_{2}^{-} + M^{*}. \label{proc2}
  \end{eqnarray}
  In the second case the electron is lost. The abundant excitation modes of
  CO$_{2}$ enhance the process of Eq.~(\ref{proc2}). Therefore, high
  requirements for the oxygen purity of the gas mixture in
  the TPC are indispensable.
  
  The electron attachment can be parametrized as a function of the
  drift length and therefore drift time by
  \begin{equation}
    \label{eq:elecAttach}
    N(t) = N_{0} \cdot e^{-p(M) p(O_{2})  K t},
  \end{equation}
  where $p(M)$ is the operation pressure of the counting gas,
  $p(O_{2})$ is the partial pressure of the oxygen impurity, and $K$
  is the electron attachment coefficient. The average value of
  $p(O_{2})$ during the beam time in the year 2000 was $0.011\mbar$,
  i.e. $11\ppm$, while the operation pressure $p(M)$ in the TPC was
  about $1\mbar$ above the atmospheric pressure. It is clear from
  Eq.~(\ref{eq:elecAttach}) that the effect of the electron loss due
  to attachment increases with the drift length.
  
  The decrease of the induced signal with the drift time was
  parametrized with an exponential description
  (Eq.~\ref{eq:elecAttach}) and corrected individually for each of the
  $20$ planes in the TPC in units of one hour data taking. The
  correction accounted for different particle composition varying as a
  function of the polar \mbox{angle $\theta$}.

  \section{Readout chambers}
  \label{sec:roc}

\begin{figure}[b]
    \centering
    \pgfimage[width=0.5\textwidth]{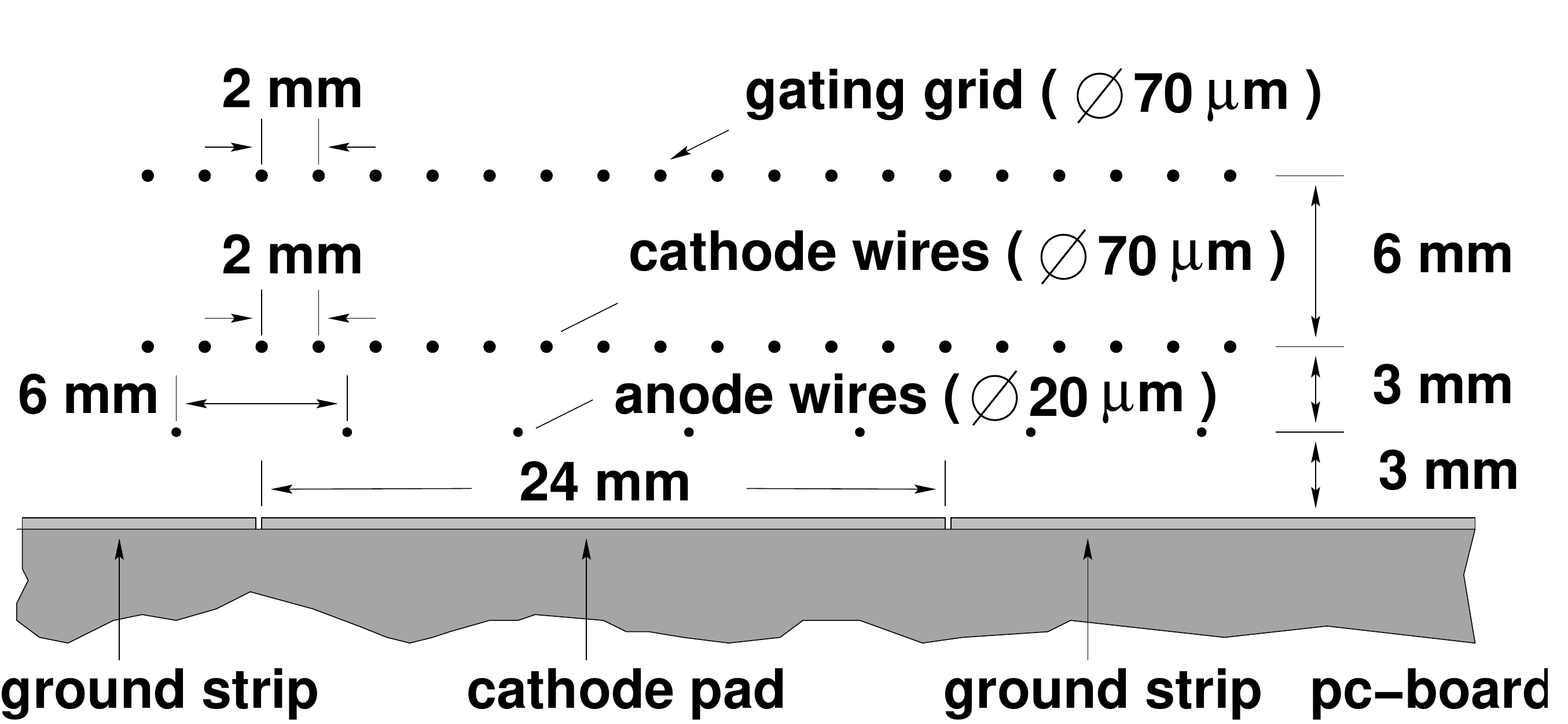}
    \caption{\label{fig:readoutchamber} {\small Cross-section of a
      readout chamber. The wires are stretched in azimuthal direction
      above the pad plane. The anodes are operated at about $1.3\kV$,
      the cathodes are grounded. The potential of the gating grid
      controls the passage of electrons and ions.}}
  \end{figure}

  The readout chambers are conventional Multi Wire Proportional
  Chambers. The electrode configuration is shown in
  Fig.~\ref{fig:readoutchamber}.
  Thin parallel
  and equally spaced anode wires are sandwiched between a cathode wire
  plane and the pad plane. The anodes are operated at about
  $1.3\kV$. The cathode wires are held at ground potential.
  With the given gas mixture
  this corresponds to an electron amplification of about $8 \cdot 10^{3}$.
  The potential of the gating grid, located $6\mm$
  above the cathode wire plane, controls the passage
  of electrons and ions. When the gate is closed electrons from the
  drift volume cannot reach the anode wires. It also prevents the
  backdrift of ions generated in the avalanche process from
  reaching the drift volume. The
  gating grid is operated at an offset potential of \mbox{$U_{\off} =
  -140\V$} to ensure full transparency to drifting electrons.
  In the closed mode, an additional bias potential of
  \mbox{$U_{bias} = \pm 70\V$} is applied between neighboring wires of
  the gating grid. This ensures full opacity to drifting electrons and
  limits the ion leakage fraction to about $10^{-4}$.
  Only at a trigger signal
  the gating grid is switched to the transparent mode at \mbox{$U_{bias}
  = 0\V$} for the duration of the maximum drift time of electrons,
  using fast gating grid pulsers based on MOSFET technology \cite{Vra00}.
 
  \subsection{Field configuration}
  \label{sec:fieldconf}

\begin{figure}[b]
    \centering
    \pgfimage[width=0.7\textwidth]{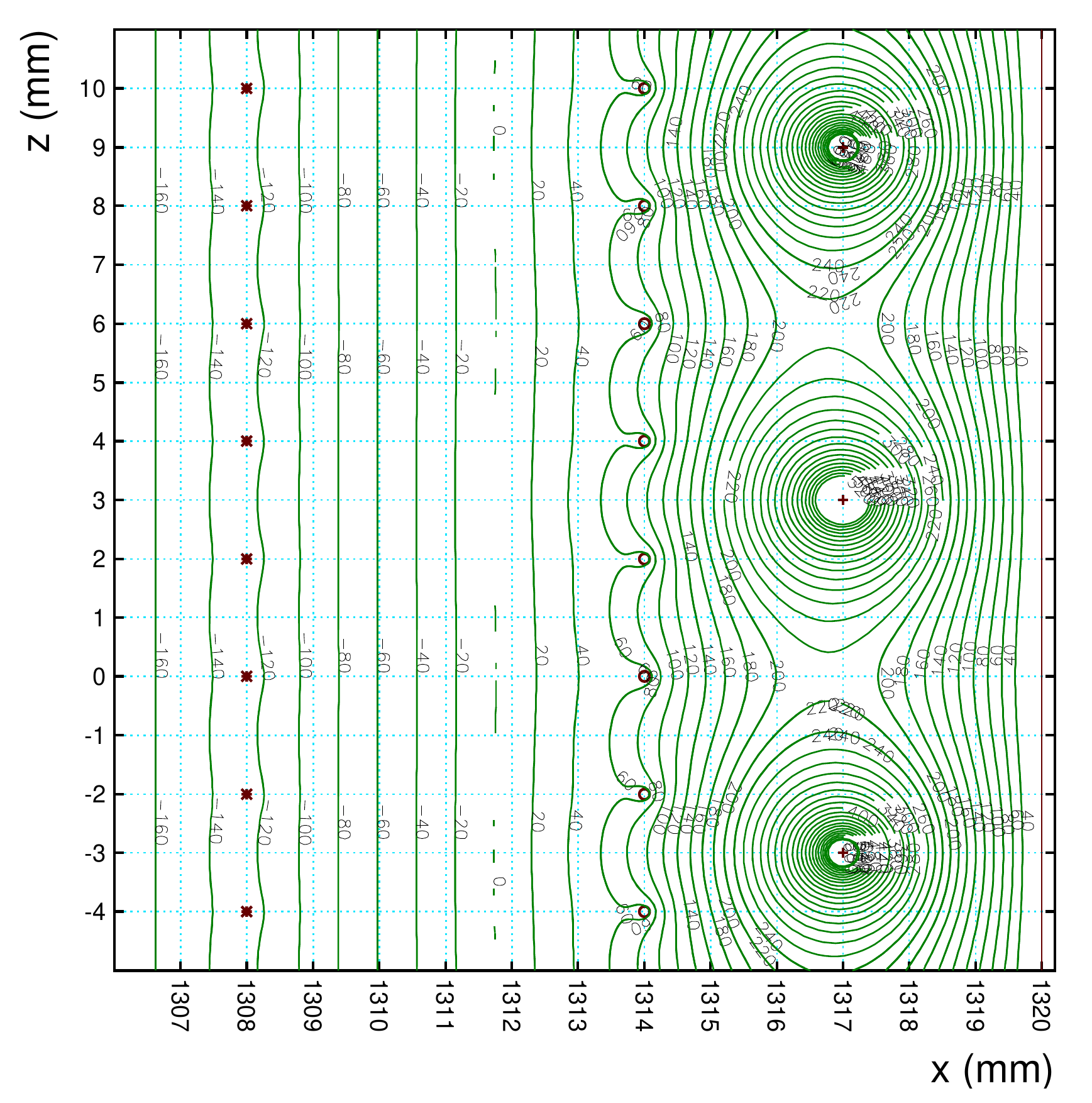}
    \caption{\small Equipotential lines inside the readout chamber
      calculated with the GARFIELD package.  The gating grid wires at
      $x=1308\mm$ and the cathode plane wires at $x=1314\mm$ are kept
      at $-140\V$ and $0\V$, respectively.  However, because of the
      leakage of the field, the effective $-140\V$ and $0\V$ planes are
      at $x=1307.5\mm$ and $1311.7\mm$, respectively. }
    \label{fig:garfield247}
  \end{figure}

  The electric potential map of a readout chamber with the gating grid
  open, calculated with the simulation package GARFIELD \cite{Vee98},
  is shown in Fig.~\ref{fig:garfield247}.
  A gating grid potential of $-140\V$ would match exactly a drift
  field defined by the $-30\kV$ cylinder potential on one side and the
  grounded cathode plane at $x=1314\mm$ on the other. However, the
  leakage of the anode field through the cathode plane, which
  effectively shifts the $0\V$ plane by about $2\mm$, and the fact
  that the actual cylinder potential is $-29.2\kV$, led to a slight
  mismatch, visible at the gating grid plane at $x=1308\mm$. The
  distortion is more significant in $\phi$ close to the edge of the
  chamber, where the distance from the cylinder is
  larger. The resulting potential at the gating grid plane at
  $r=1308\mm$, averaged in $z$ over the distance between two grid
  wires, acquires values between $-110\V$ and $-126\V$, depending on
  $z$ and $\phi$. The potential map $V_{gg}(z,\phi)$ was used as the
  boundary condition when calculating the three-dimensional electric
  potential map (Sect.~\ref{sec:efield}). It has been checked that a
  variation of the anode potential within $1300\ft1400\V$ has no
  visible impact on the drift field.
 
  \subsection{Gating grid}
  
\begin{figure}[ht]
    \centering
    \pgfimage[width=0.6\textwidth]{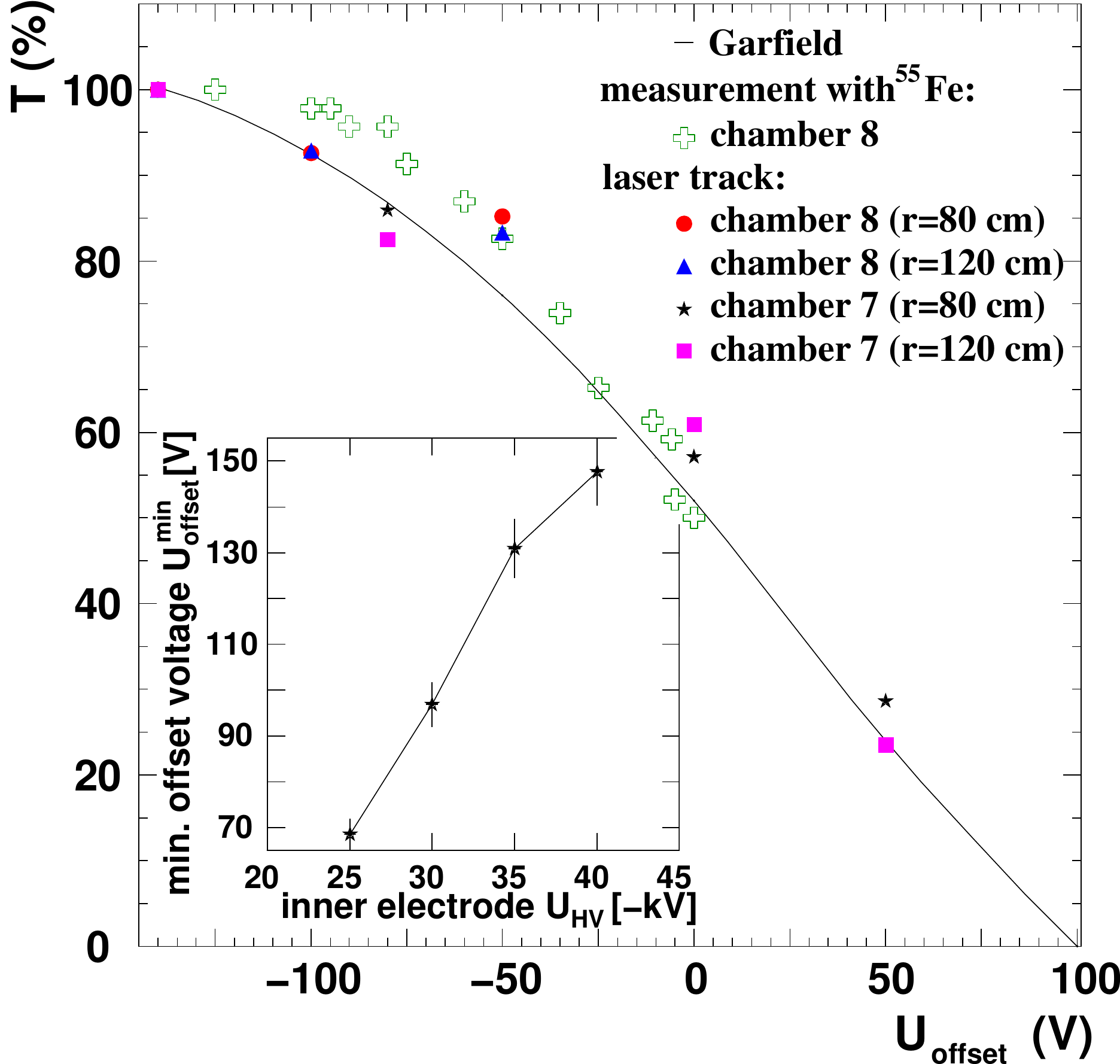}
    \caption{\label{fig:gatingGrid} {\small Measurement of the gating
	grid transparency as a function of the offset voltage for a
	gas mixture of Ne/CO$_{2}$ (80\perc/20\perc). Saturation is
	reached at an offset voltage of about $U_{\off} = -90\V$. The
	minimal offset voltage needed for full transparency depends
	linearly on the voltage of the HV-cylinder (figure inset).}}
  \end{figure}

  The performance of the gating grid is characterized by the
  transparency $T$, which is the ratio of electric flux lines crossing
  the gating grid plane to the total number of electric flux lines
  between gating grid wires and HV-cylinder. A measurement of the
  gating grid transparency as a function of the offset voltage
  $U_{\off}$ is shown in Fig.~\ref{fig:gatingGrid}, using a 
  $^{55}$Fe source and cross-checked with laser events
  \cite{Sch01}. The tests have been performed in static mode
  ($U_{bias} = 0$) without magnetic field using an anode wire voltage
  of $U_{a}=1.4\kV$ and a voltage of $U_{HV} = -29.2\kV$ for the
  HV-cylinder. The data shows a saturation of the transparency at an
  offset voltage of about $-90\V$. The minimal offset voltage needed
  for full transparency depends linearly on the voltage of the
  HV-cylinder, as shown by the inset in Fig.~\ref{fig:gatingGrid}.
  
\begin{figure}[ht]
    \centering
    \pgfimage[width=0.6\textwidth]{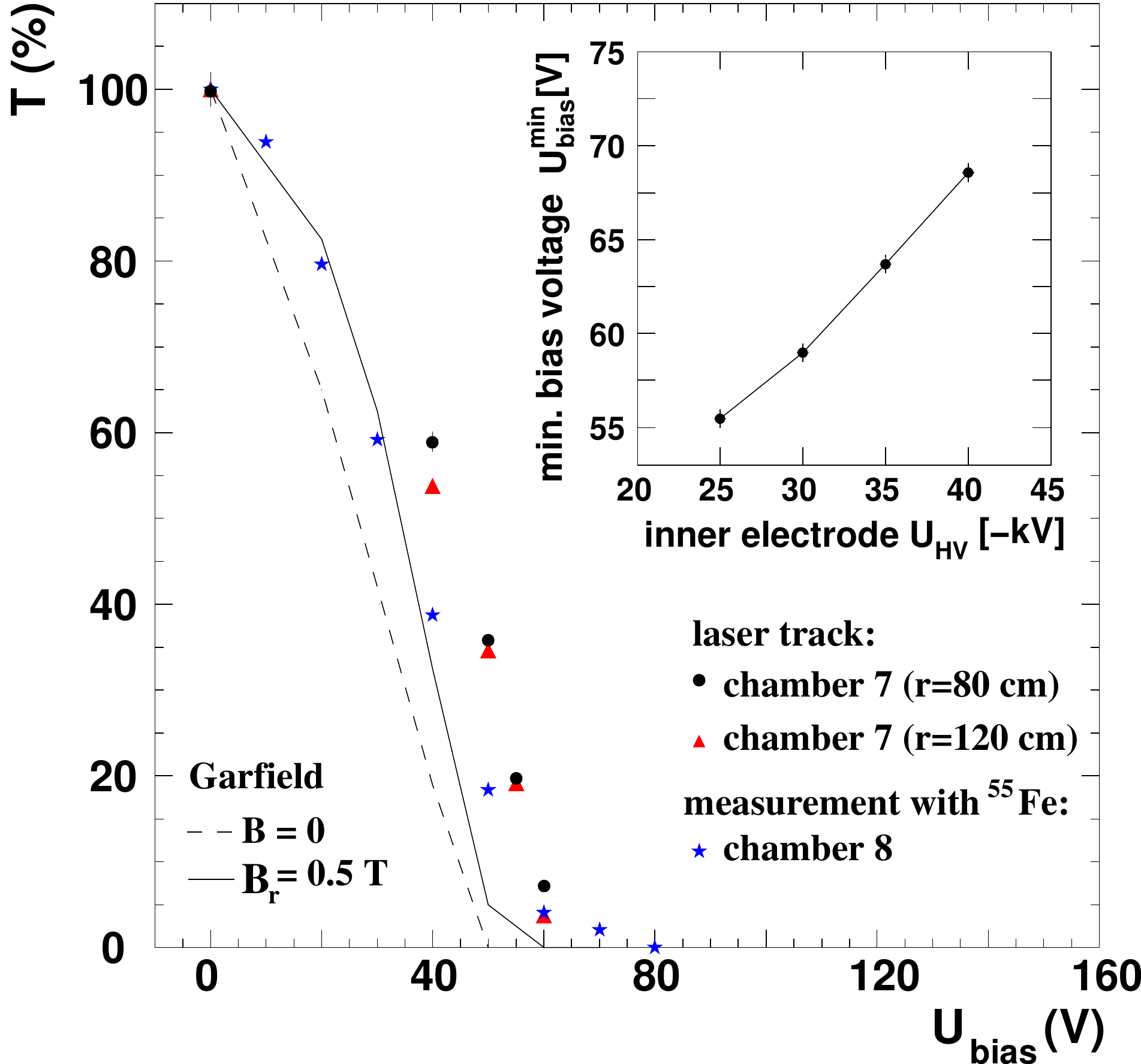}
    \caption{\label{fig:gatingGridBias} {\small Measurement of the
	gating grid transparency as a function of the alternating bias
	voltage for a gas mixture of Ne/CO$_{2}$
	(80\perc/20\perc). The gating grid is closed at about
	$U_{bias} = 60\V$. The bias voltage depends linearly on the
	voltage of the HV-cylinder (figure inset). GARFIELD
	calculations show the influence of the magnetic field and the
	CO$_2$ concentration on the efficiency of the closed gating
	grid.}}
  \end{figure}

  Fig.~\ref{fig:gatingGridBias} shows the transparency as a function
  of the alternating bias voltage $U_{bias}$ superimposed on the
  offset voltage \cite{Sch01}. The measurement was performed under the
  same conditions as described above using an offset voltage of
  $U_{\off}=-140\V$. The gating grid was closed with a
  bias voltage of about $60\V$. Also the alternating bias voltage
  depends linearly on the voltage of the HV-cylinder, as can be seen in the
  inset in Fig.~\ref{fig:gatingGridBias}.
  
  The magnetic field has an influence on the transparency of the
  closed gating grid. Due to the radial component $B_r$ of the
  magnetic field with respect to the drift direction, electrons
  precess around their drift line defined by the electric field. This
  precession, in addition to diffusion, enables some electrons to pass
  the gating grid in spite of it being closed. GARFIELD simulations
  depicted in Fig.~\ref{fig:gatingGridBias} show that a higher bias
  voltage has to be used in presence of a magnetic field to achieve
  the same transparency.

  For the CERES TPC the radial component $B_r$ of the magnetic field
  changes as function of the longitudinal coordinate $z$ (cf.
  Fig.~\ref{fig:bfield}). This implies a dependence of the closing
  efficiency of the gating grid along the beam axis. However, due to
  the weak magnetic field of at most $0.5\T$ used in the TPC and
  the cool properties of CO$_2$-based mixtures, the overall dependence
  of the closing efficiency of the gating grid on $B_r$ is small
  ($\Delta U_{bias}^{min} = 6.6\V\kern -0.1em /\kern 0.1em 0.5\T$), as
  confirmed by measurements. During operation a high value of
  $U_{bias}=70\V$ was chosen together with an offset voltage of
  $U_{\off}=-140\V$.

  \subsection{Gating grid settling time}

  The gating grid represents a capacitance C$_{gg}$. Thus, the
  opening happens with finite settling time $\tau_{gg}$ after
  receiving a trigger signal:
  \begin{equation}
    \label{eq:taugg}
    \tau_{gg} = f \cdot Z_0 \cdot C_{gg}.
  \end{equation}
  $Z_0=50\Ohm$ is the impedance of the RG58 cable providing the
  voltage to the gating grid. The factor $f = 4.6$ is applied to
  describe when the signal reaches $99\perc$ of its maximum value. The
  capacitance of the gating grid is given by:
  \begin{equation}
    C_{gg} = C_0 + 2 \cdot C_{M}.
  \end{equation}
  $C_0$ is the capacitance of the gating grid with respect to the
  cathode wire grid and $C_{M}$ is the capacitance between the
  individual wires of the gating grid, resulting in an expected value
  of $C_{gg} \approx 8.1\nF$ and, using Eq.~(\ref{eq:taugg}), in
  $\tau_{gg} = 2\us$. The capacitance between the gating grid and the
  HV-cylinder is small and can be neglected.

  A capacitance measurement of C$_0$ and C$_M$ resulted in
  $C_{gg} = 6.6\nF$. The settling time was measured with a noise
  analysis of the readout channels of events without target
  interaction (empty events), resulting in $\tau_{gg} = 2.4\us$,
  in good agreement with the  expectations.

  \subsection{Pad plane}
  \label{sec:PadPlane}

\begin{figure}[ht]
    \centering
    \pgfimage[width=0.5\textwidth]{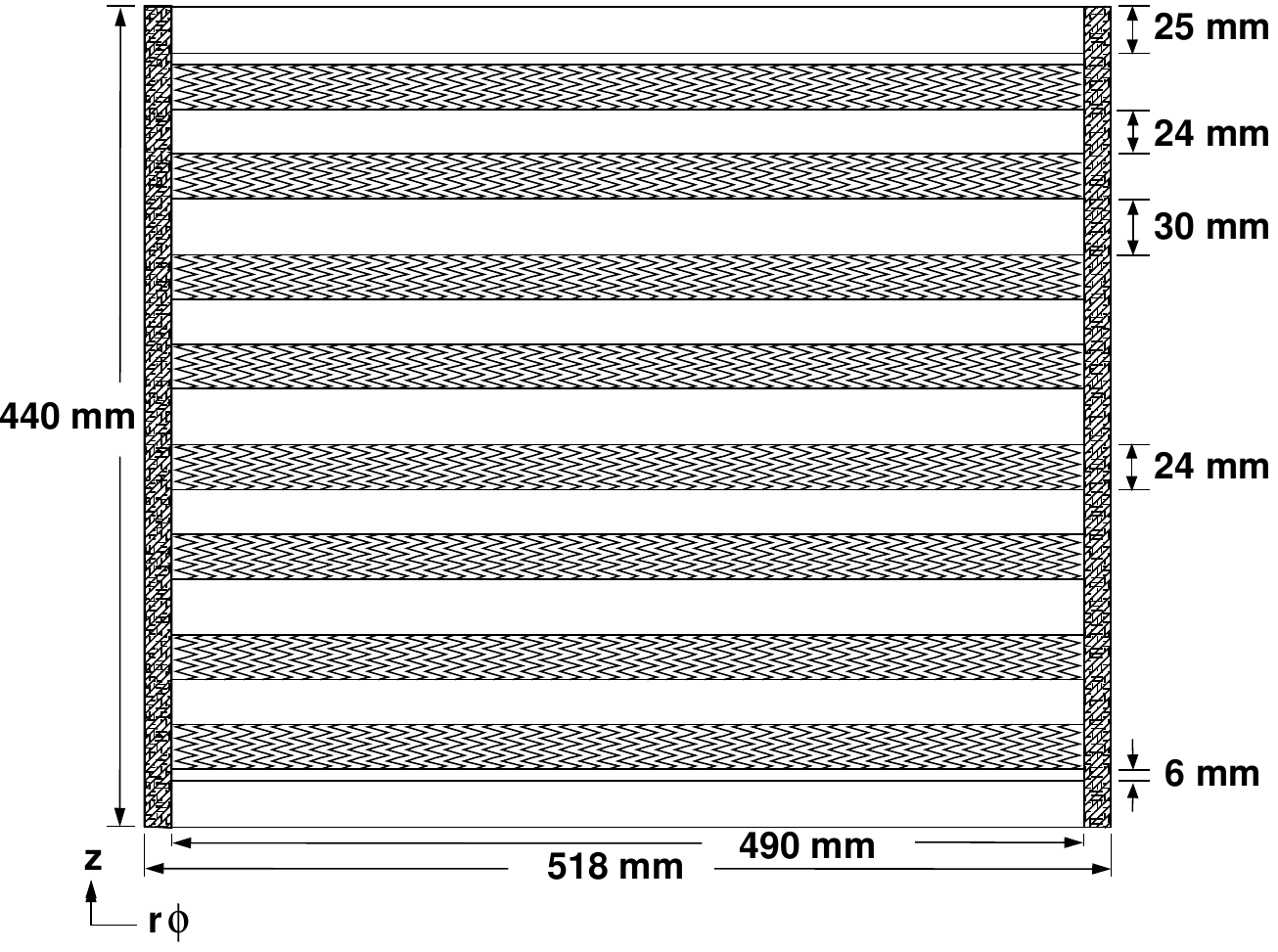}\\
    \pgfimage[width=0.5\textwidth]{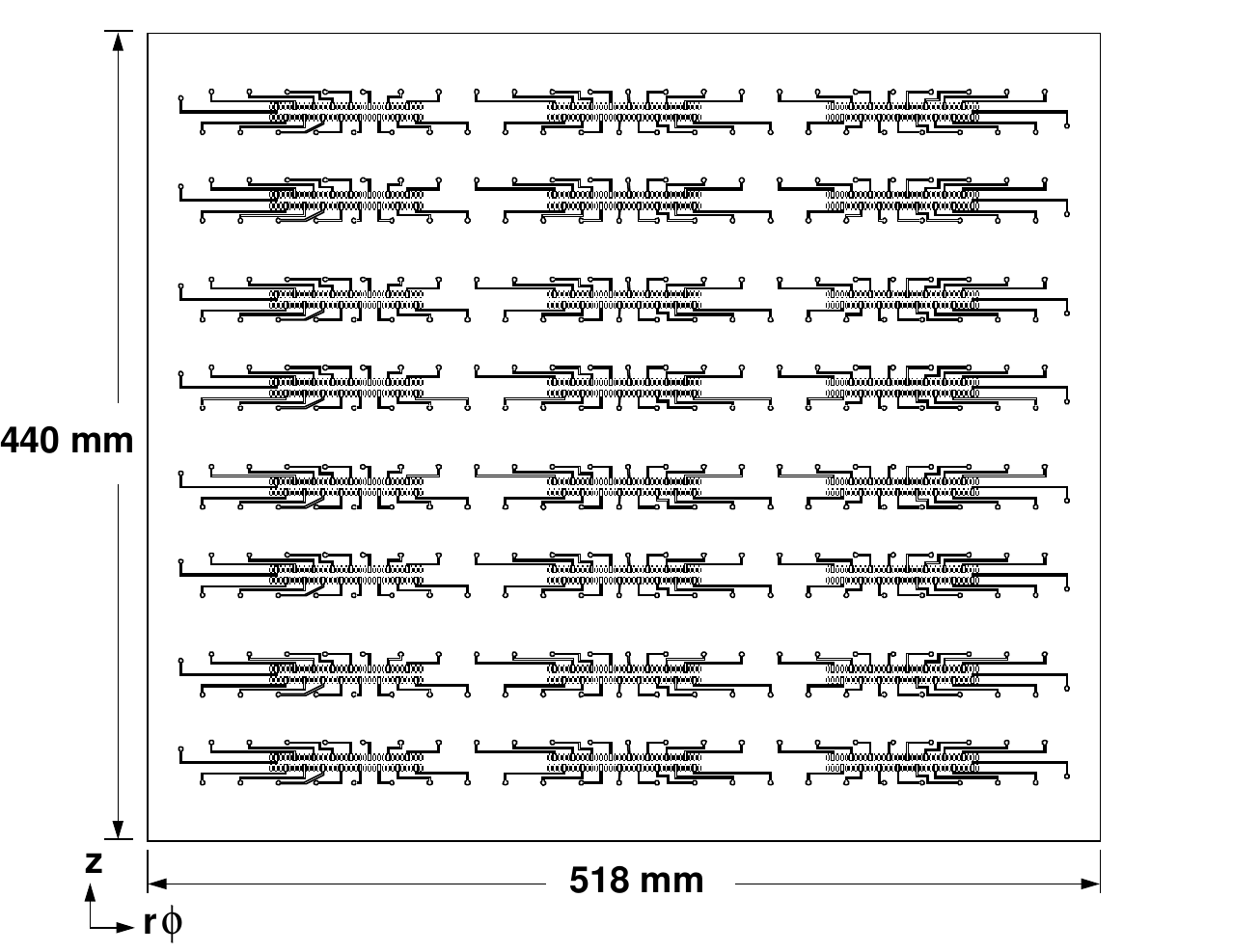}
    \caption{\label{fig:padPlane}
      {\small Top: Pad side of a pad plane. Bottom: Readout side of a pad plane.}}
  \end{figure}

  The TPC has 16 modular readout chambers, each having a total size of
  $2000\times518\mm^2$ and $48\times40$ readout channels. The pad
  planes are composed of five printed circuit boards shown in
  Fig.~\ref{fig:padPlane} with an etched readout pad structure on the
  front side connected to the readout traces on the rear side.

  The printed circuit board is made of $3.2\mm$ thick FR4 material
  with a gold plated ($0.09^{+0.1}_{-0.1}\um$ Au,
  $4.5^{+0.1}_{-0.2}\um$ Ni) copper surface ($18\um$ Cu) and a
  resulting tolerance of its total thickness of
  $^{+0.32}_{-0.30}\mm$. The readout pad structure is produced with an
  accuracy of $\pm50\um$ in each dimension. The gaps between
  conductive areas are etched with a width of $125\um$ at the pad side
  and $250\um$ at the readout side of the board. Each readout channel
  is connected via plated through holes ($\dia1.0\mm$) to the rear
  side of the printed circuit board to pick up its signal. To
  guarantee gas tightness, each hole is sealed with epoxy. The pad
  rows are separated from each other by ground strips of alternating
  thickness of $24\mm$ and $30\mm$. This large spacing between the pad
  rows was chosen due to financial constraints and concerns with the
  data volume and therefore readout speed. Furthermore, only half of
  the 40 pad rows were equipped with Front-End Electronics (FEE). The
  area of the three inner pad planes is $390\times518\mm^2$, and of
  the two outer ones is $415\times518\mm^2$.  For mechanical stability
  the printed circuit boards are glued onto a $5\mm$ thick
  G10-backplate. The backplate has cut-outs for the connectors of the
  readout electronics and is itself glued onto an Al frame. The ledges
  holding the wires are glued onto the pad side of the printed circuit
  boards.

  \subsection{Pad response}

\begin{figure}[h]
    \centering
    \pgfimage[width=0.5\textwidth]{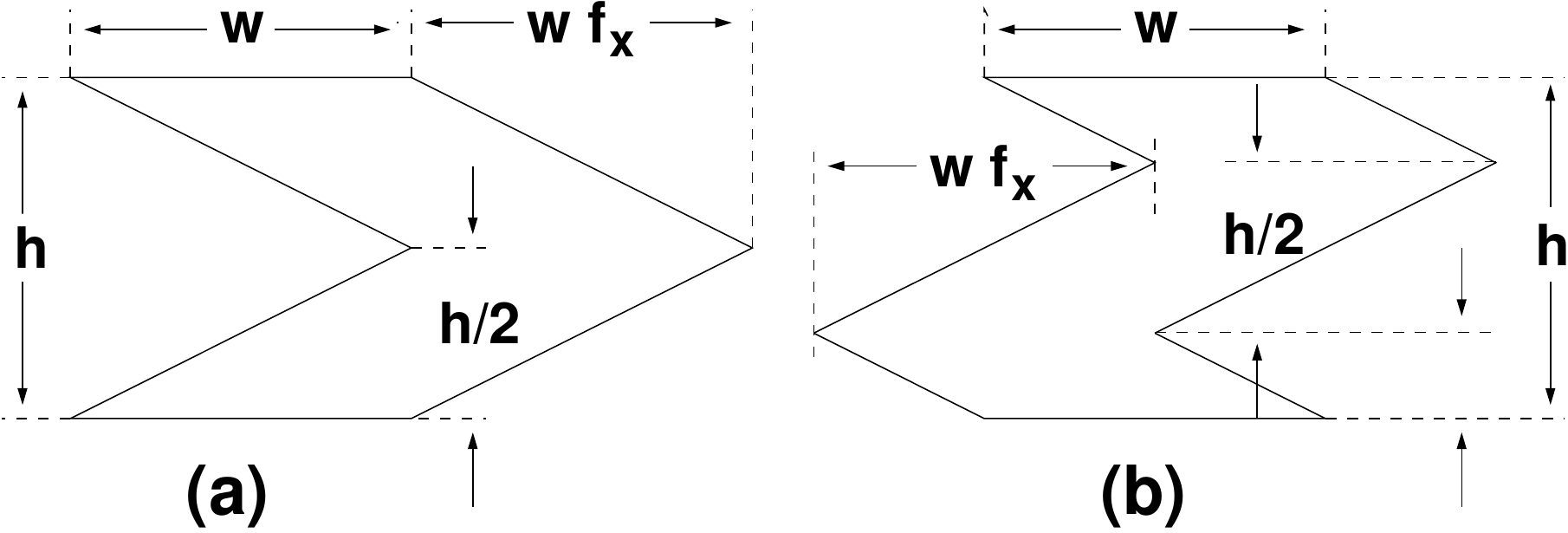}
    \caption{\label{fig:chevron}
      {\small Pattern of a centered (a) and displaced (b) single
      chevron pad. Chevron pads allow a larger
      readout node spacing than rectangular pads. For the CERES TPC the
      chevron version (b) was chosen.}}
  \end{figure}

  The CERES TPC employs chevron-shaped cathode pads to perform
  centroid determination of a spatially extended induced charge signal
  by geometric charge division (Fig.~\ref{fig:chevron}). Compared to
  the commonly used rectangular pads, chevron pads allow a
  substantially larger readout node spacing. Based on theoretical
  calculations and experimental measurements \cite{Yu91} the displaced
  single chevron version (Fig.~\ref{fig:chevron} (b)) was chosen.

\begin{figure}[h]
    \centering
    \pgfimage[width=0.4\textwidth]{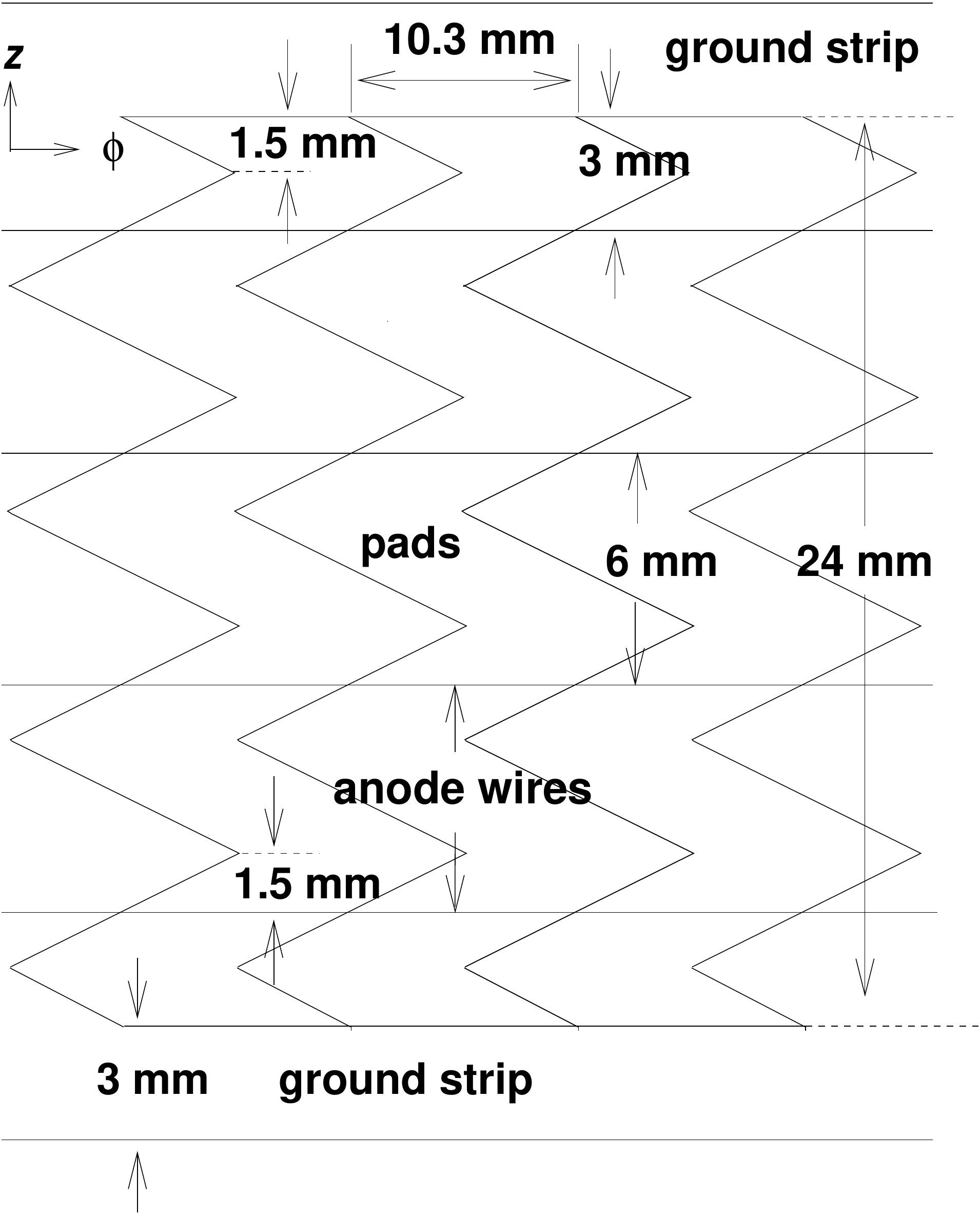}
    \caption{\label{fig:cathodePad} {\small Layout of the cathode
      pads. One readout channel is composed of four single chevron
      pads shown in Fig.~~\ref{fig:chevron} (b).}}
  \end{figure}

   The error of position measurement $\delta_{r\phi}$ has
   statistical and systematical contributions:
   \begin{equation}
     \delta_{r\phi} = \delta_{stat} + \delta_{syst}.
   \end{equation}
   Here, $r\phi$ refers to a point on the pad plane in
   $\phi$-direction. The statistical terms are given by the finite
   number statistics of the electrons created in the drifting electron
   cloud ($\delta_{diff}$), the track angle $\alpha$ with respect to
   the pad direction ($\delta_{\tan(\alpha)}$), and an additional smearing of
   the electron distribution close to the anode wire where the
   electron trajectory is crossing the radial component of the
   magnetic field ($\delta_{\vec{E}\times\vec{B}}$). The overall noise
   behavior is contained in $\delta_{0}$,
   \begin{eqnarray}
     \label{eq:statError}
     \delta_{stat}^2 = \delta_{diff}^2 + \delta_{\tan(\alpha)}^2 +
     \delta_{\vec{E}\times\vec{B}}^2 + \delta_{0}^2, \nonumber \\
     \delta_{diff}^2\sim 1/l; \;\; \delta_{\tan(\alpha)}^2\sim l; \;\;
     \delta_{\vec{E}\times\vec{B}}^2 \sim d; \;\; \delta_{0}^2\sim const.,
   \end{eqnarray}
   where $l$ refers to the length of the chevron pads and $d$ to the
   anode wire spacing. With the specific geometry of the CERES TPC and
   the Ne/CO$_2$ (80\perc/20\perc) gas mixture, calculations using
   Eq.~\ref{eq:statError} showed that for $l=24\mm$ and $d=6\mm$ the
   statistical position error can be minimized to $200\ft400\um$
   depending on the drift distance. Four of these displaced single
   chevron structures, with four anode wires running across, are
   connected to one single readout channel
   (Fig.~\ref{fig:cathodePad}). The width of a chevron pad has been
   fixed to $w = 10.3\mm$. Thus, each readout element contains 46
   complete chevron pads in $\phi$-direction and two half sized
   chevron pads at the edge.

   The finite granularity of the readout pads introduces an additional
   systematical error $\delta_{syst} = r\Delta\phi$ to the error of
   position measurement. This contribution can be minimized by
   optimizing the overlap factor $f_x$ as defined in
   Fig.~\ref{fig:chevron}. The induced cathode charge distribution $f$
   can be described by a two-dimensional empirical expression
   \cite{Mat89}:
   \begin{eqnarray}
     \label{eq:chargeDistr}
     f(\lambda) \simeq K_1 \cdot \frac{1-\tanh^2(K_2\lambda)}{1+K_3
     \tanh^2(K_2\lambda)}, \nonumber \\ {\mathrm{with}} \quad K_1 =
     \frac{K_2\sqrt{K_3}}{4 \arctan\sqrt{K_3}}; \; K_2 = \frac{\pi}{2}
     (1-0.5\sqrt{K_3}); \; \lambda = \frac{x}{L},\frac{y}{L}.
   \end{eqnarray}
   Here $x$ and $y$ are the coordinates in the pad plane, $L$ is the
   anode-cathode spacing. $K_3$ is a parameter depending on the
   chamber geometry (anode-cathode spacing and anode wire spacing) in
   $r\phi$- and $z$-direction. For the CERES-TPC the values are
   $K_3^{r\phi}=0.655$ and $K_3^z=0.805$. The distribution of
   Eq.~\ref{eq:chargeDistr} was used to integrate the induced charge
   over each pad.

   The uniform irradiation response UIR is defined as the inverse of
   the derivative of the reconstructed centroid $r\phi_{rec}$ as a
   function of the original position $r\phi$:
   \begin{equation}
    r\phi_{rec} = f(r\phi); \;\; UIR(r\phi) = 1/f'(r\phi).
   \end{equation}
   One of the commonly used measures of nonlinearity is the
   differential nonlinearity DFNL defined as:
   \begin{equation}
     DFNL = \frac{UIR_{max} - UIR_{min}}{(UIR_{max}+ UIR_{min})/2}.
   \end{equation}
   Calculations of the DFNL as a function of the overlap factor $f_x$
   for the CERES TPC chevron pattern showed that the nonlinearity can
   be minimized by choosing $f_x\simeq1.0$, reducing the remaining
   position error $r\Delta\phi$ to $\pm~6\um$. Previous
   simulations and measurements \cite{Yu91} showed that the ideal
   value of $f_x = 1.0$ has to be increased to $f_x= 1.05$ in order to
   take the finite gap between adjacent pads into account. It has to
   be noticed that the above considerations are ideal, i.e. noise
   present in the induced charge on a cathode pad is not considered as
   well as counteractive measures such as pedestal subtraction. The
   measured position error will be described in
   Sect.~\ref{sec:nonlincor}.

\begin{figure}[ht]
    \centering
    \pgfimage[width=0.5\textwidth]{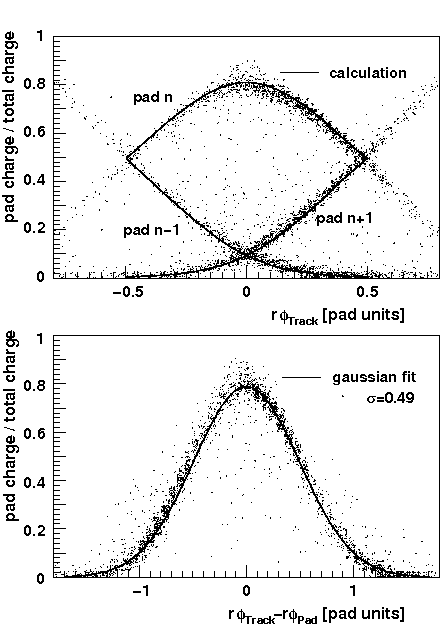}
    \caption{\label{fig:prfmeas} {\small Top: Measured charge
      sharing between three adjacent pads and the corresponding
      calculation. Bottom: Pad response function with a Gaussian width
      of 0.49 pad units, corresponding to $5.05\mm$.}}
  \end{figure}

  The pad response function describes the induced charge on a cathode
  pad as a function of its distance to the charge centroid.
  Fig.~\ref{fig:prfmeas} shows a measurement \cite{Sch01} from a
  readout chamber using an $^{55}$Fe source.
  It shows the
  charge sharing between three adjacent pads and the corresponding pad
  response function. The charge sharing was calculated integrating
  Eq.~\ref{eq:chargeDistr} and shows good agreement with the
  measurement. The pad response function can be described in good
  approximation by a Gaussian with $\sigma=0.49$ pad
  units (one pad unit $= 10.3\mm$).

  \subsection{Correction for the pad response nonlinearity}
  \label{sec:nonlincor}

\begin{figure}[ht]
  \centering \pgfimage[width=0.6\textwidth]{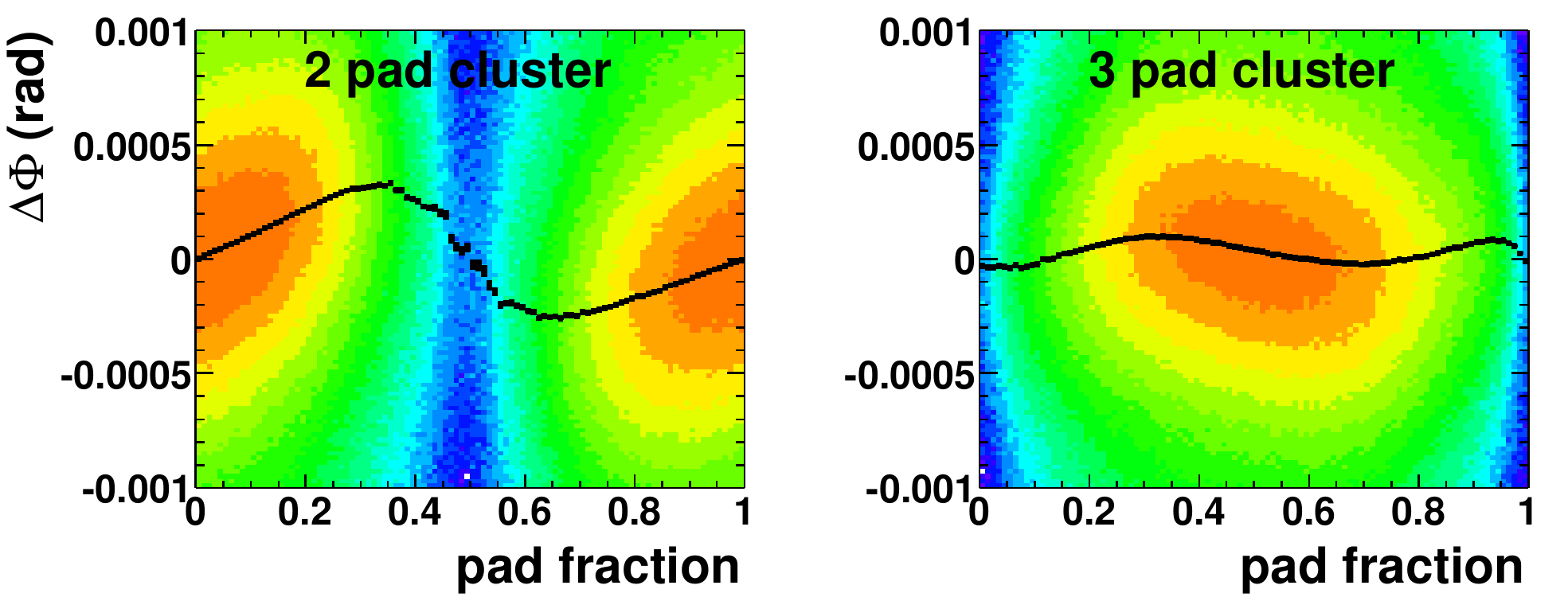}\\
  \pgfimage[width=0.6\textwidth]{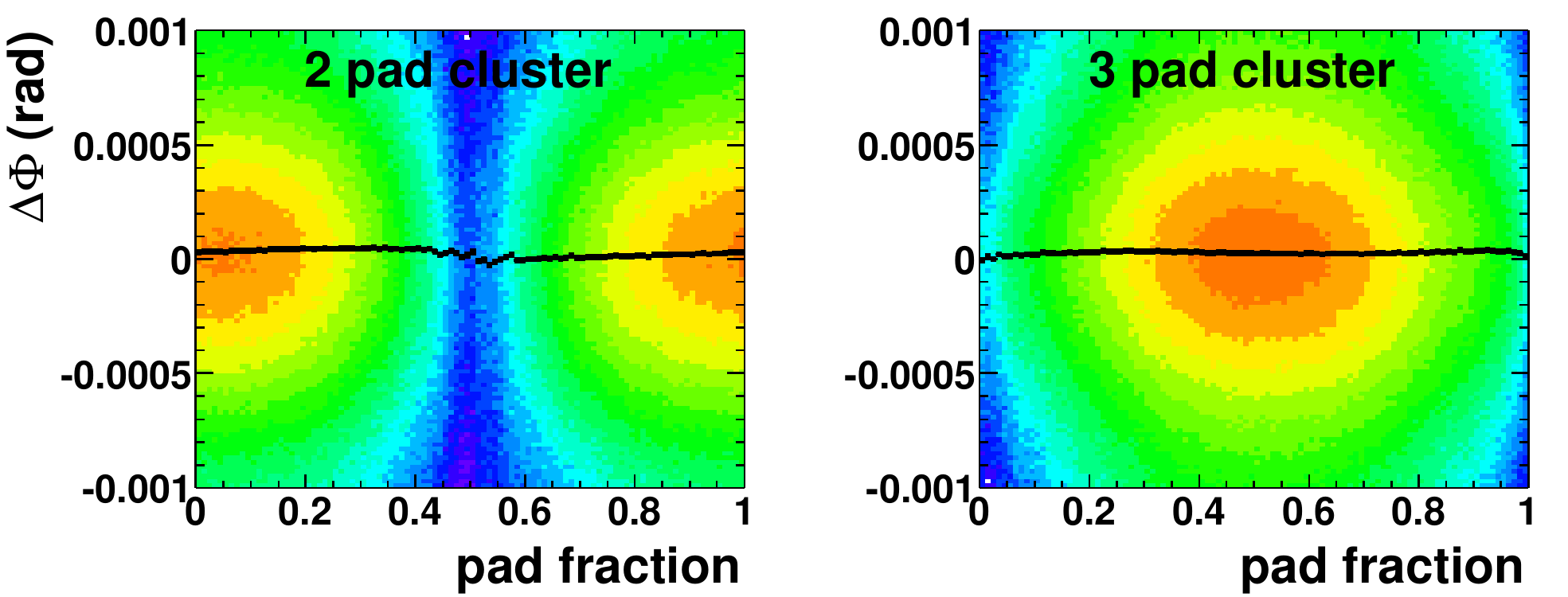}
  \caption{\label{fig:nonlinCorrection} {\small Top: Angular
    difference \mbox{$\Delta \phi = \phi_{track} - \phi_{hit}$} vs. a
    fraction of a pad. The remaining nonlinearity of the pad response
    is shown as bin-by-bin average of the entries in the color
    scatter plots. Bottom: Data after nonlinearity correction. The
    example is plotted for a positive magnetic field in the TPC.}}
\end{figure}

  To determine the remaining small nonlinearity in the pad response it
  was assumed that the true angle $\phi_{true}$ can be approximated by
  the azimuthal angle of a track $\phi_{track}$ at a given space
  point.  This assumption is justified by the fact that a track is
  fitted through $12\ft20$ space points and thus minimizes position
  uncertainties. The nonlinearity depends on the number of pads on
  which a signal is induced (Fig.~\ref{fig:nonlinCorrection},
  top). The angular difference between $\phi_{track}$ and the
  reconstructed space point $\phi_{hit}$ plotted vs. a fraction of a
  pad shows a maximum azimuthal position distortion of about
  $0.3\mrad$ for 2 pad clusters and about $0.1\mrad$ for 3 pad
  clusters \cite{Lud06}, respectively. The corresponding correction
  was performed separately for positive and negative magnetic fields
  via look-up tables containing the position uncertainties. The result
  of the correction is shown in the bottom of
  \mbox{Fig. \ref{fig:nonlinCorrection}}.

  \subsection{Correction for anode wire positions}

\begin{figure}[htb]
    \centering
    \pgfimage[width=0.5\textwidth]{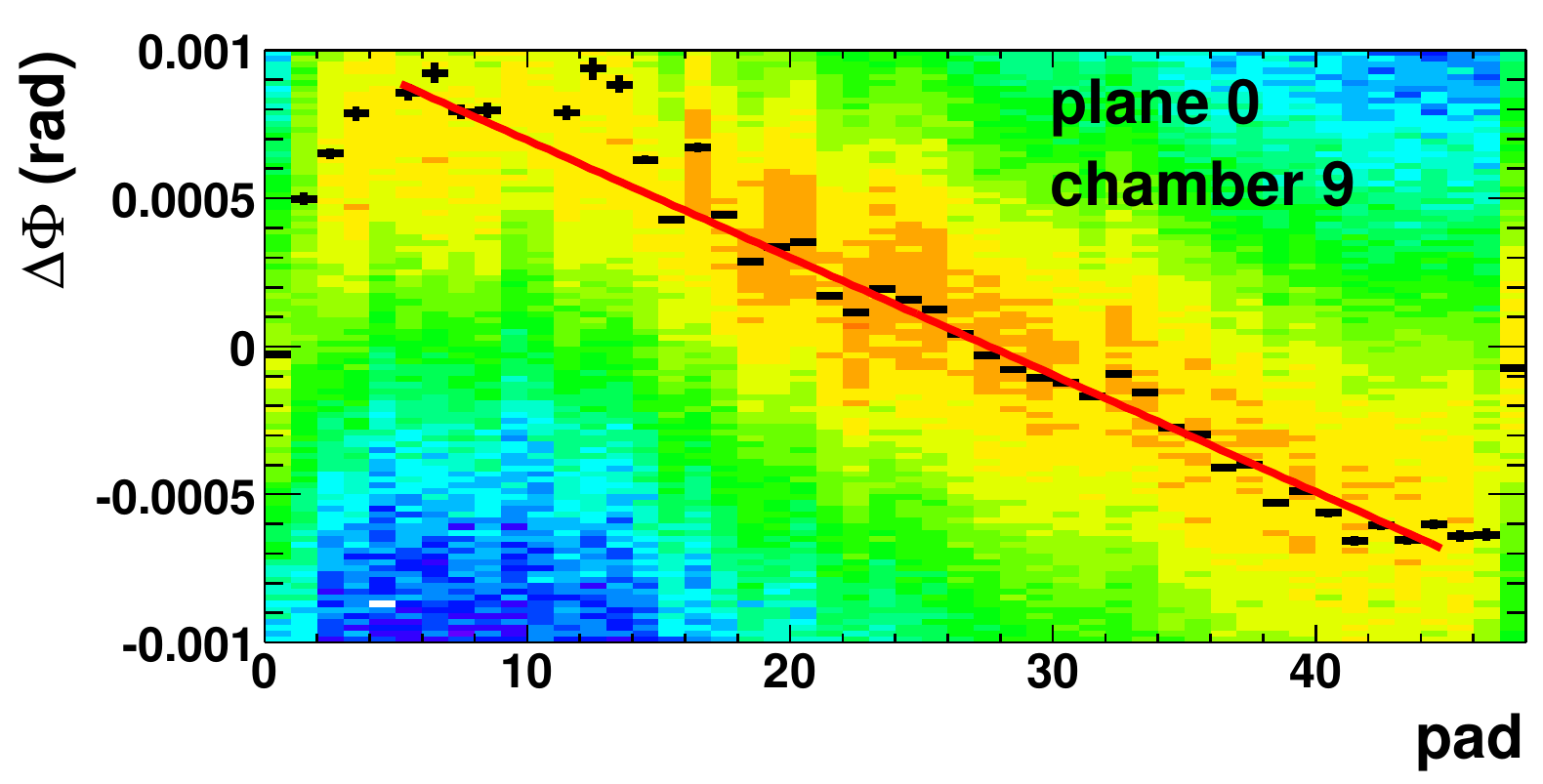}\\
    \pgfimage[width=0.5\textwidth]{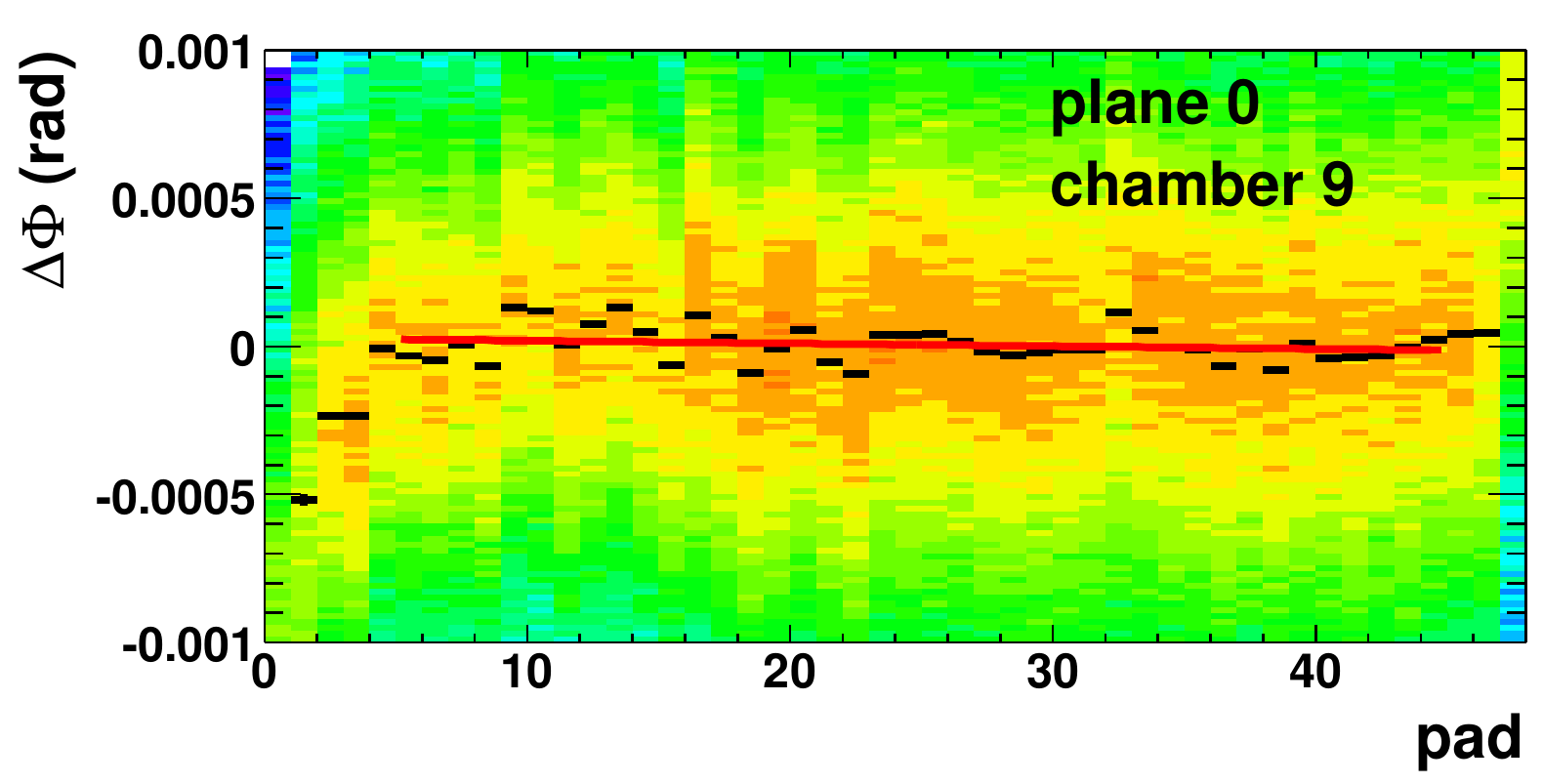}
    \caption{\label{fig:anodeWireCorrection} \small Top: Shifts of
      anode wires result in a linear dependence of \mbox{$\Delta \phi
      = \phi_{track} - \phi_{hit}$} vs. pad number. This is shown by
      the bin-by-bin average of the entries in the color
      plots. Bottom: The anode wire positions were corrected
      individually for each plane and each chamber via a look-up
      table.}
  \end{figure}

  The precision of the adjustment of the anode wires above the cathode
  pad planes is finite. Deviations from the nominal positions were
  calibrated using a sample of data taken in the absence of a magnetic
  field. For each plane and each chamber the difference of the
  azimuthal angle between the track and its hits \mbox{$\Delta \phi =
  \phi_{track} - \phi_{hit}$} was plotted versus the pad number
  (Fig.~\ref{fig:anodeWireCorrection}, top) \cite{Lud06}. Again $\phi_{track}$
  stands for the true azimuthal angle $\phi_{true}$.

  The observed offsets depend linearly on the pad number and were
  parametrized with a polynomial of first order. The linear
  dependence has its origin in the chevron-shaped structure of the
  cathode pads. If the position of an anode wire above the pads is
  shifted in beam direction (parallel to the $z$-coordinate) the
  charge sharing between the pads must necessarily change. This in
  turn influences the determination of the centroid of a hit. An
  offset of $\Delta \phi= 1\mrad$ corresponds to an anode wire shift
  of about $\Delta z = 0.37\mm$. The azimuthal hit position $\phi_{hit}$ was
  corrected via a look-up table (Fig.~\ref{fig:anodeWireCorrection},
  bottom).
  
  \section{Readout electronics}
  \label{sec:electronics}

  The data acquisition chain of the TPC starts with the Front-End
  Electronics (FEE) comprising a Preamplifier/Shaper (PASA) and a
  Switched Capacitor Array (SCA), an analog memory to record the
  analog output signal of the amplifier. Both chips were implemented
  in the $0.8\um$ AMS CMOS process \cite{Bau98}. The FEE-boards are
  directly mounted on the readout chambers of the TPC.

\begin{figure}[hb]
    \centering
    \pgfimage[width=1.\textwidth]{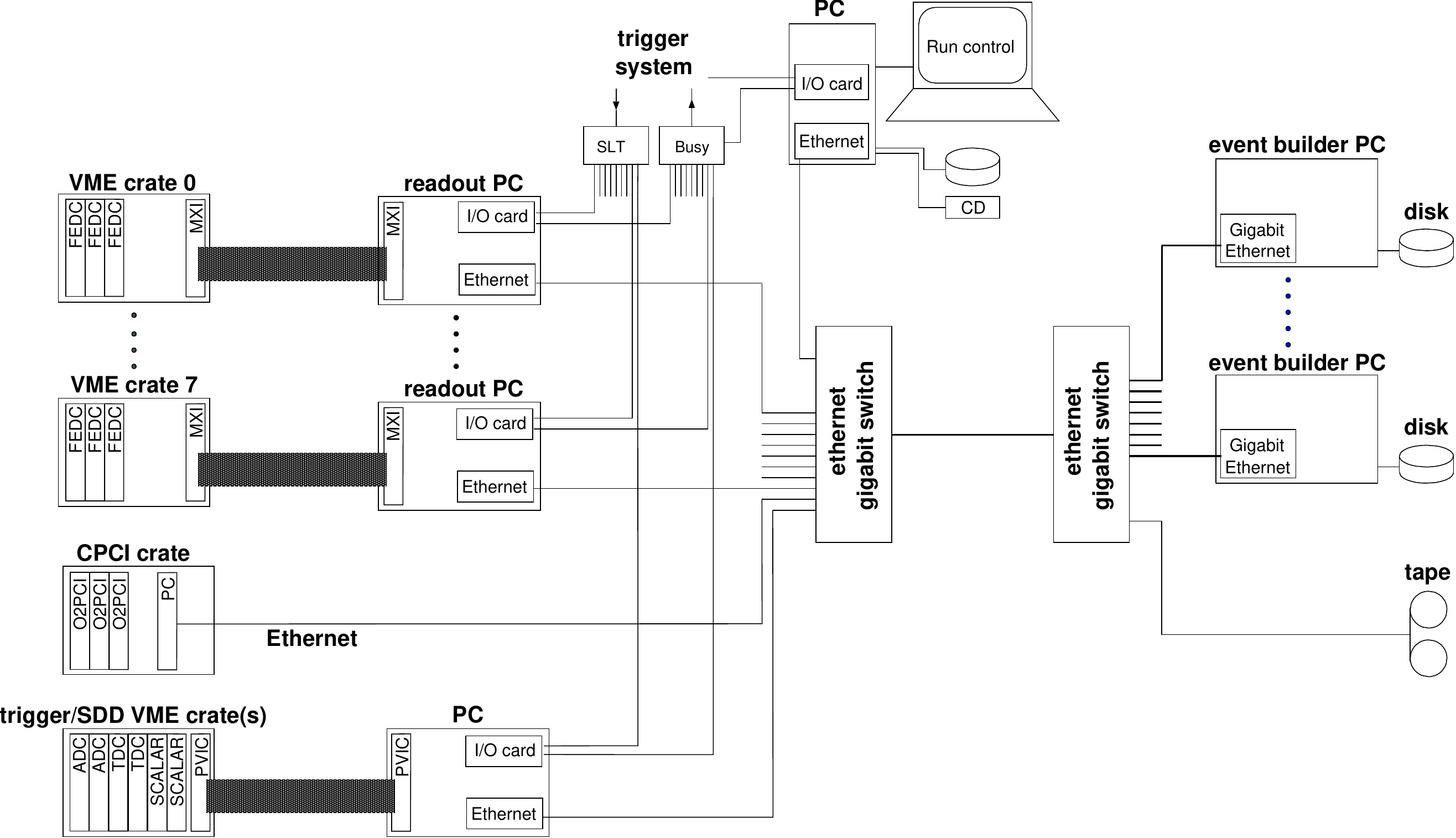}
    \caption{\label{fig:daq2000} {\small Overview of the CERES data
      acquisition in 2000. 24 FEDC modules handle the data of the 16
      TPC chambers. The FEDC VME crates are connected via MXI-buses to
      8 readout PCs, which in turn are connected via Gigabit Ethernet
      to event builder PC in the CERN Central Data Recording facility.}}
  \end{figure}

  An overview of the CERES readout system implemented for the beam
  time in 2000 is shown in Fig.~\ref{fig:daq2000} \cite{Til02}.  The
  analog output signals from the FEE-boards are sent via $14\m$ long
  coaxial cables to the Front-End Digital Cards (FEDC)
  \cite{Bae98,Eng00} where the signal is digitized. To handle the data
  of two TPC chambers three FEDC modules are necessary (one FEDC per
  640 cathode pads) and they are grouped in one 9U VME crate close to
  the TPC. The digitization process is clocked by an external clock
  signal which is provided by a central TPC Clock Module.

  The FEDC VME crates are connected via MXI-buses (Multisystem
  Extension Interface) to readout PCs. Eight readout PCs were needed
  to read the 16 TPC chambers. During the heavy-ion beam time in 2000
  the SPS beam was extracted and steered onto the target for $5\s$
  followed by a pause of $14\s$. During extraction all data were
  collected in the readout PCs. The interval between bursts was used
  to send the data via Gigabit Ethernet connection to one of the seven
  event builder PCs, located in the CERN Central Data Recording
  facility (CDR). There the data of one event coming from different
  detector systems were merged into one single data block and saved to
  disk. A tape daemon asynchronously archived the data on tape.
  
  The data of the SDD and the RICH detectors from the compact PCI
  modules (CPCI) are collected in the main memory of an embedded PC
  which is plugged into the CPCI crate. From there the data is sent
  via Ethernet to the event builder.

  The discriminators, coincidences, and downscalers used for the
  trigger logic, ADCs and TDCs of the beam related photomultiplier
  detectors, and counters of various beam and trigger signals are
  sitting in three VME crates located in the grillage. These crates
  are daisy-chained via VME extenders, and connected via a PVIC (PCI
  Vertical Interconnect) interface to another readout PC.
  
  The start of the readout was triggered by an external signal applied
  to an input channel of an I/O-card plugged in each readout PC. After
  receiving a trigger, a readout PC would set a busy signal on an
  output channel of its I/O-card. A logic OR of all busy signals was
  used to inhibit new triggers. After all data were sent to the
  readout PC the busy signal was removed.

  The typical event size for a central \mbox{Pb-Au} collisions at
  $158\gevc$ per nucleon was $500\kB$. From all detectors the average
  busy duration, i.e. the average time needed to get the event into
  the memory of a readout PC, was largest for the TPC with
  $5.7\ms$. With a beam intensity of $10^6$ per burst, with the
  requirement of no other beam particle within $\pm 1\us$, and with
  the centrality trigger of $7\perc$ the event taking rate was
  $300\ft400$ per burst.

  \subsection{Preamplifier/Shaper}
 
\begin{figure}[ht]
    \centering
    \pgfimage[width=0.6\textwidth]{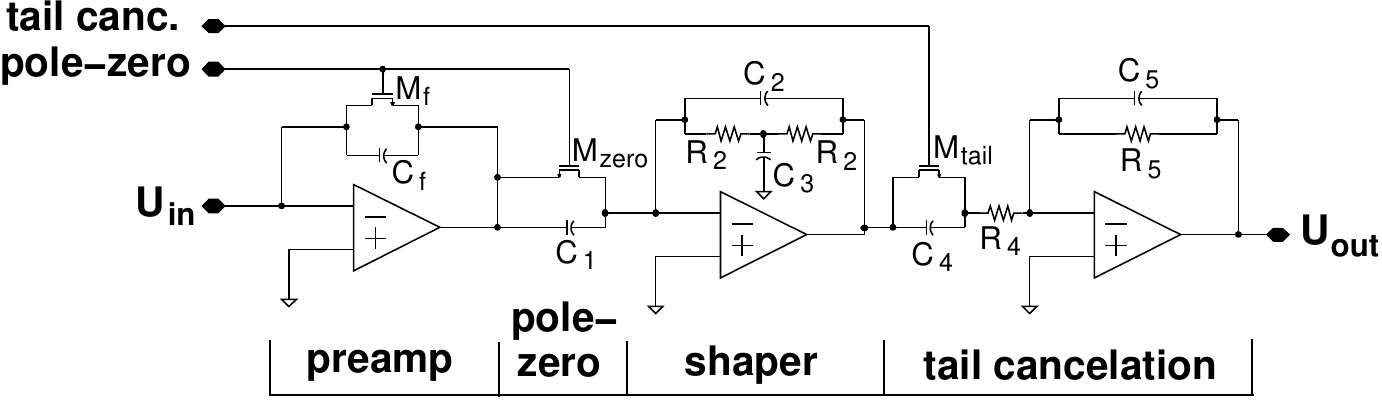}
    \caption{\label{fig:pasa}
      {\small Schematics of the PASA comprising a charge-sensitive 
	preamplifier with a
	semi-Gaussian shaper and tail suppression. The PASA is part of
      the Front-End Electronics of the TPC.}}
  \end{figure}

  A schematic drawing of the custom PASA chip is shown in
  Fig.~\ref{fig:pasa} \cite{Bau98}. It comprises a charge-sensitive
  preamplifier with a semi-Gaussian shaper and tail suppression.  In
  contrast to common designs which use a pulsed reset, this amplifier
  is continuously sensitive. This is important in a high track density
  environment typical for a heavy ion collision.

  A feedback resistor $\mosfetf$ continuously discharges the integration
  capacitor $\capf$ with a decay time $t_{decay}= \capf \mosfetf$. The value of
  $\mosfetf$ is a trade-off between noise performance and the capability to
  process events with high occupancy. For a peaking time $t_{decay}=
  400\ns$, noise considerations dictate a feedback resistance
  $\mosfetf>4\MOhm$. The only way to implement such a high resistance in CMOS
  technology is by using the associated drain-source resistance $R_{ds}$ of a
  MOSFET transistor. The value of $R_{ds}$ depends on the biasing conditions of
  $\mosfetf$. It decreases as the gate-source voltage $\Vgs$ of $\mosfetf$
  increases. This dependence is an advantage in our case as signal
  charges increase $\Vgs$: the integration of small charges results
  in small $\Vgs$ swings, thus $R_{ds} \approx R_{ds,DC}$. When this
  value is large enough, it prevents deterioration of the noise
  performance. Conversely, large charges are discharged with a faster
  decay time and the baseline of the preamplifier is quickly
  restored. Even more important is that undesired large signals (e.g. from
  $\delta$-electrons) collected on TPC pads are transferred onto $\capf$
  and quickly discharged, minimizing dead time.

  In case of a conventional pole-zero cancellation, the
  dependence of $R_{ds}$ on the injected charge $Q_{in}$ would
  deteriorate the linearity of the PASA. Here, an
  adaptive pole-zero cancellation scheme was used to suppress the pole
  associated with $R_{ds}$ and $\capf$. The transistor $M_{zero}$ is
  biased in the same way as $\mosfetf$ during the discharge of
  $\capf$. The zero associated to the network $M_{zero}\ft C_1$ adapts
  itself dynamically to accurately cancel the pole associated with the
  network $\capf \ft \mosfetf$.

  The peaking time of the shaper can be adjusted between $140$ and
  $580\ns$ and the tail suppression can be varied over a range of
  $0.1\ft1.5\us$ to cope with different input signals. The gain of the
  preamplifier can be adjusted from $35\ft110\mVperfC$.

  \subsection{Switched Capacitor Array}

\begin{figure}[ht]
    \centering
    \pgfimage[width=0.6\textwidth]{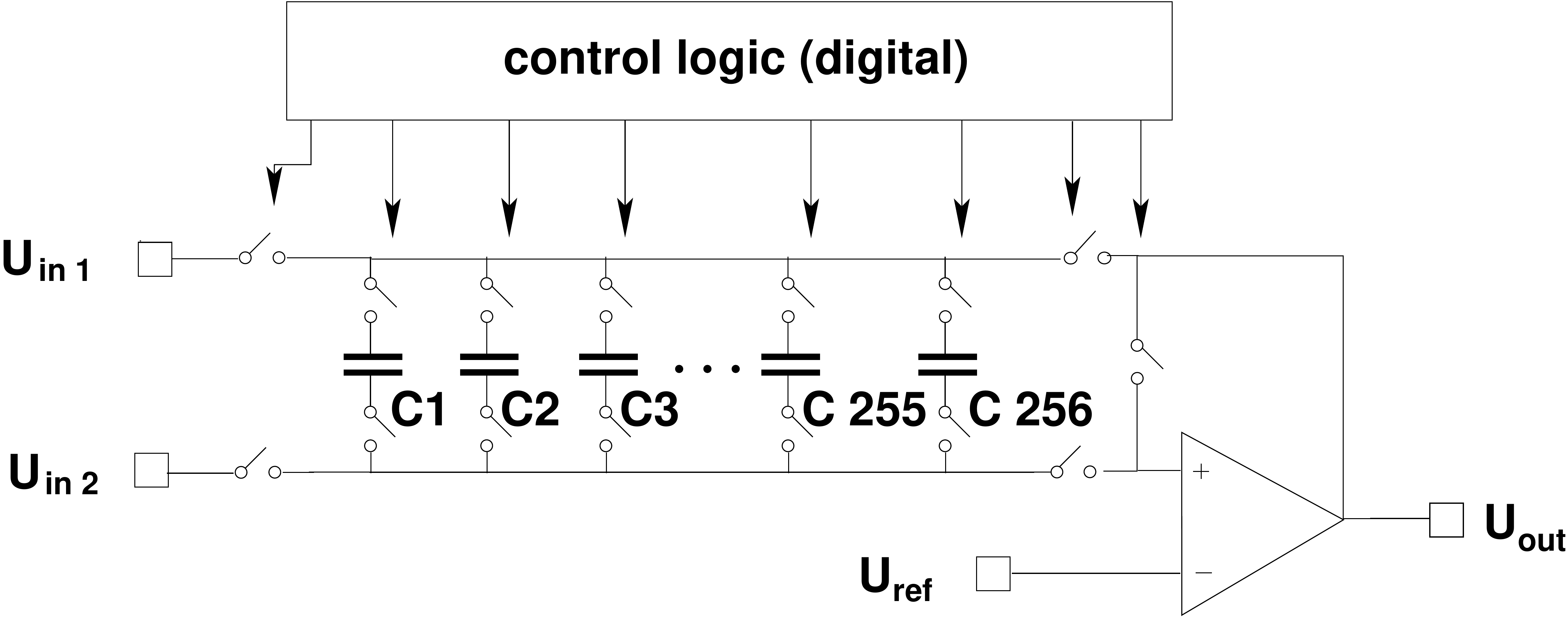}
    \caption{\label{fig:sca} {\small Schematics of the SCA. The SCA is
    an analog memory to record the analog output signal of the PASA.
    It is part of Front-End Electronics of the TPC.}}
  \end{figure}

  During the drift time of the TPC the output of the PASA is sampled
  and stored in an analog memory (SCA) at a rate of up to
  $14\MHz$. Fig.~\ref{fig:sca} shows the simplified schematics of this
  device. The SCA chip contains 16 channels, each with 256 individual
  samples. A 16-to-1 analog output multiplexer permits the use of a
  single external Analog-to-Digital Converter (ADC) per chip. The SCA
  is operated in voltage-read-voltage-write configuration which has
  the benefit that the output is independent of the exact value of the
  storage capacitor. Each memory cell consists of a $1.4\pF$ double
  poly capacitor connected by two transmission gates to the common
  lines. This scheme reduces parasitic capacitances during readout,
  when the storage cells are switched in the feedback loop of an
  operational amplifier. Without this second switch the read-amplifier
  would charge the total capacitance of all bottom plates to the
  substrate. Because the two transmission gates are inherently less
  sensitive than single transistor switches, the clock feed-through
  and charge injections are further reduced. Only the differences of
  the capacitor voltages between the top and the bottom plates are
  relevant and over a small range of input voltages the noise
  influence is the same on both plates. Symmetrical layout and only
  complementary signals running near analog cells ensure low switching
  noise caused by digital signals.

  The digital part includes a shift register to sequentially address a
  column of cells. A programmable clock window prevents overlapping
  addressing signals to inhibit charge sharing between adjacent
  cycles. Two registers on the chip can be used to store a delay value
  for the start of the sampling and the maximum number of readout
  channels. This allows to suppress data from tracks outside the conical
  acceptance of the tracks matching the other detectors.

  \subsection{Front-End Digital Cards}
  \label{sec:fedc}
  
  The digitization and further processing of the TPC signals is done
  on the FEDC-boards. These boards are realized as 9U VXI devices
  which can contain up to 48 readout channels. Each readout channels
  comprises a 10-bit ADC and a digital chip for signal processing (a
  modified version of the ALTRO chip for the ALICE TPC \cite{Mus03}) and
  processes the data from one FEE-board. For the CERES readout only 40
  channels per FEDC were used.

  Immediately after receiving a first-level trigger signal the SCA
  starts to sample the 16 outputs of the amplifier in parallel. This
  sampling phase is followed by the readout phase in which the stored
  analog values are dumped in a time-wise order. The data coming from
  the FEE-boards are digitized using the 10-bit ADC on the FEDC-board of
  which only the 9 most significant bits are used for further
  processing. Therefore, inside the ALTRO chip the signals are
  described by 9-bit codes ($0\ft511$). After converting the analog
  signal the data stream is demultiplexed according to the 16 channels
  of the preamplifier. The resulting 16 data streams are processed in
  parallel inside the ALTRO chip.

  In the following processing steps the polarity of the signal is
  changed and the baseline is subtracted. After this subtraction the
  signal should be contained in the first half of the 9-bit
  range. Therefore the most significant bit can be omitted reducing
  the signal representation to 8-bit codes. Finally the signal
  undergoes zero suppression. Samples with a value smaller than a
  constant threshold (8-bit) are rejected. This threshold is stored in
  one of the control registers of the ALTRO chip. When a sample is found to
  be above the threshold, it is considered as the start of a pulse and
  stored in the central memory of the FEDC.

  Because each ALTRO chip processes data coming from 16 readout
  channels with a maximum of 256 time samples, it was not possible
  to provide enough memory inside the chips to hold pedestal values
  for all samples. Instead, a scheme using a look-up table was
  implemented. The slight disadvantage of this scheme is the missing
  possibility to specify a threshold for zero suppression for all time
  bins. This restriction was evaded by adding the threshold
  (determined individually for each time bin) to the pedestal value
  and storing this combined value in the pedestal memory. The zero
  suppression value which is used to detect the start of a pulse was set
  to zero. With this scheme, not only the true baseline is subtracted
  from the signal but also the specific threshold. If the resulting
  value is above zero, the pixel is considered being part of the
  pulse. For the off-line analysis the additionally subtracted
  threshold had to be added again for each pixel.

  \subsection{Electronics response}
  
\begin{figure}[ht]
    \centering
    \pgfimage[width=0.5\textwidth]{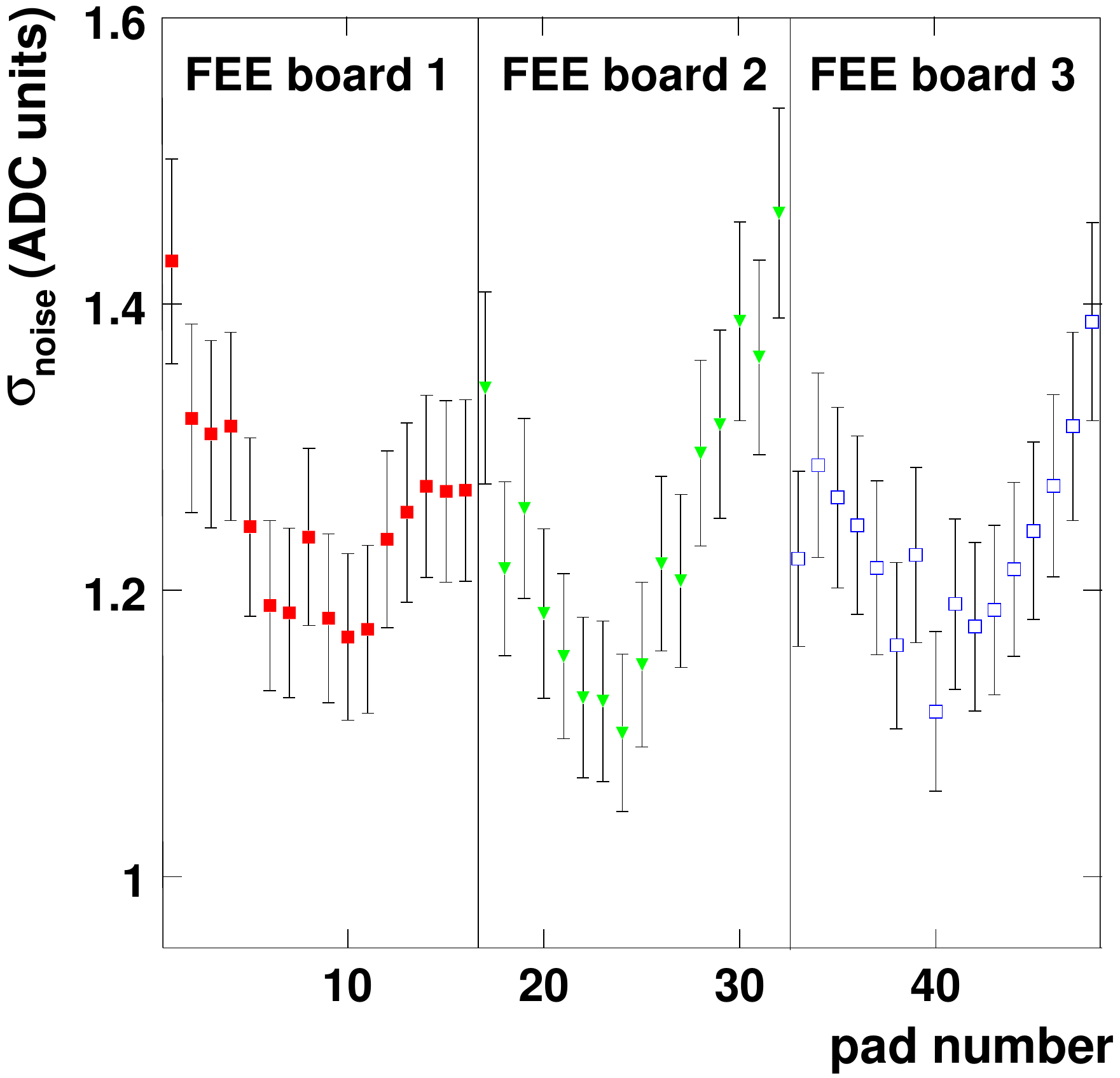}
    \caption{\label{fig:eNoise1} {\small Noise behavior of single
	readout channels in the three FEE-boards of one chamber
	caused by the different readout line capacitances.}}
  \end{figure}

\begin{figure}[ht]
    \centering
    \pgfimage[width=0.5\textwidth]{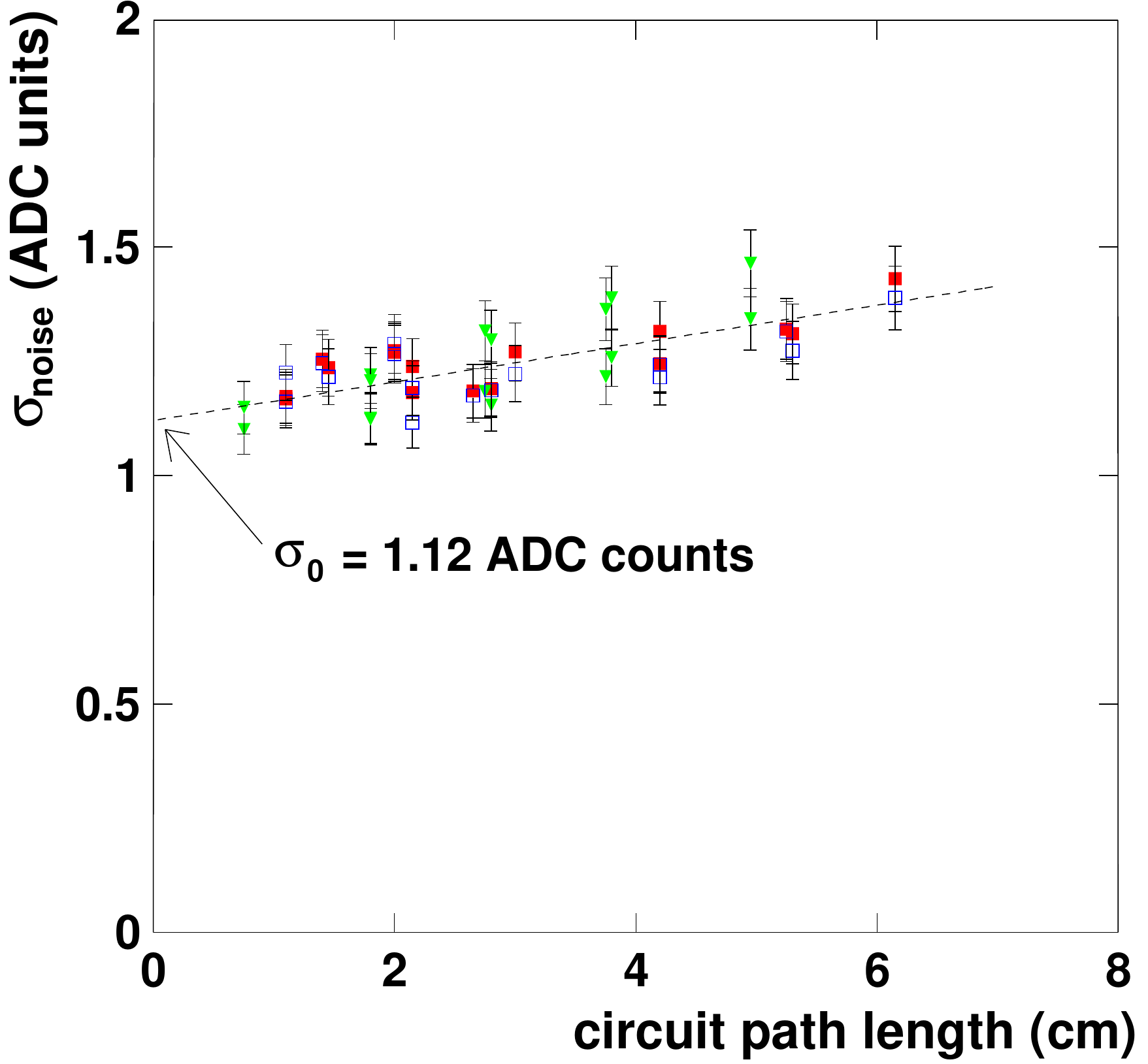}
    \caption{\label{fig:eNoise2} {\small Dependence of the noise
	behavior of a readout channel on its readout trace length. 
	An isolated readout channel with zero readout trace
	length would have a noise level of $\sigma_{0} = 1.12$
	ADC counts.}}
  \end{figure}

  Sixteen adjacent pads are processed in one FEE-board. Each pad row
  of a readout chamber is equipped with three
  FEE-boards. Fig.~\ref{fig:eNoise1} shows the electronic noise
  behavior of a row of 48 pads in one readout chamber \cite{Sch01}.
  This noise behavior is caused by the different readout line
  capacitances (cf. Fig.~\ref{fig:padPlane}, bottom). The correlation
  between circuit path lengths and noise behavior is shown in
  Fig.~\ref{fig:eNoise2}. A straight line fit results in a noise level
  of $\sigma_{0} = 1.12$ ADC counts for an isolated readout channel
  with zero readout trace length.

  The Equivalent Noise Charge (ENC) of the FEE-board
  electronics is given by:
  \begin{equation}
    ENC_{out} = g_{SCA} \cdot \sqrt{(g_{\pasa} \cdot ENC_{pad})^2 + ENC_{SCA}^2}.
  \end{equation}
  In this equation $g_{\pasa} = 1.5$ and $g_{SCA} = 1.6$ are the gains
  of the PASA and SCA, respectively. The measured $ENC_{out} =
  1.12\ADCcounts$ correspond to $1288\enc$ noise charge
  ($1\ADCcounts=1150\enc$), which in turn corresponds to a noise value
  of $ENC_{pad}=498\enc$.  The ENC is a function of the peaking time
  $\tau_{peak}$ where the signal reaches its maximum. Theoretical
  considerations of the serial ($\propto 1/\sqrt{\tau_{peak}}$),
  parallel ($\propto \sqrt{\tau_{peak}}$), and white (independent from
  $\tau_{peak}$) noise contributions to the ENC yield an expected
  cathode pad noise value of $ENC_{pad} \approx 445\enc$ at a peaking
  time $\tau_{peak} = 340\ns$ and a cathode pad capacitance $C_{pad} =
  12\pF$ \cite{Ern97}. This is in good agreement with the above
  measurement.

  \subsection{Effect of the trace length on the time measurement}
  \label{sec:padtopad}

\begin{figure}[ht]
    \centering
    \pgfimage[width=0.5\textwidth]{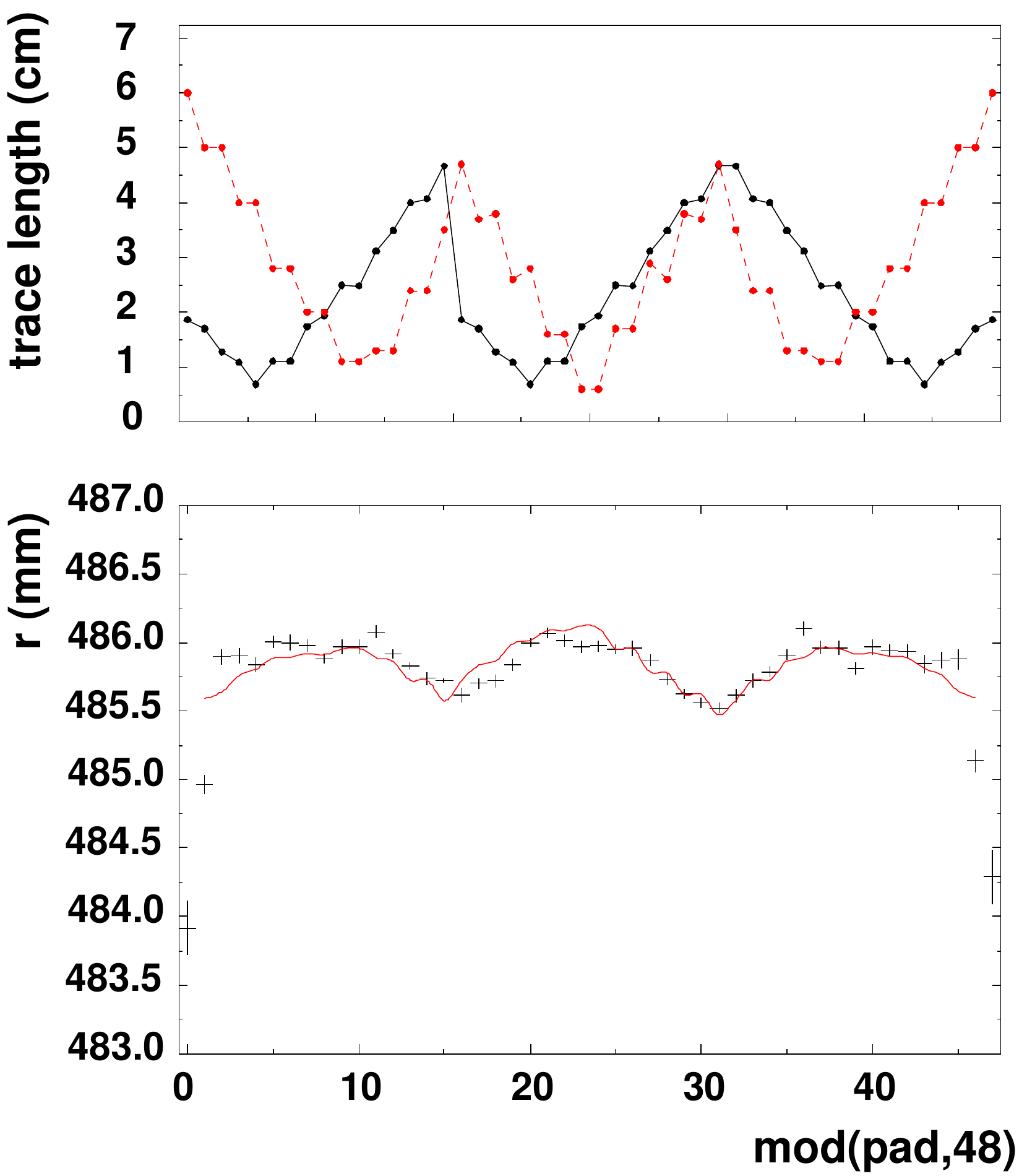}
    \caption{Top: Trace length on the FEE-board (solid) and on the pad
      plane (dashed).  Bottom: Periodic oscillation in the reconstructed
      radius of the HV-cylinder. The line is a linear combination
      of the pad plane and FEE-board trace lengths from the upper plot, with
      the two coefficients optimized to fit the data.}
    \label{fig:tribump}
  \end{figure}

  The reconstructed position of the high voltage cylinder, adjusted to
  its true value by shifting and tilting the chambers as described in
  Sect.~\ref{sec:corefieldmisalignment}, still shows a periodic
  oscillation when plotted vs. the pad number modulo 48, i.e. across a
  chamber (Fig.~\ref{fig:tribump}, bottom).
  The three bumps reflect the three FEE-boards of 16 pads each and
  originate from the pad-to-pad variation of the length of the trace
  within the pad plane $L_{ppl}(pad)$ and on the FEE-boards
  $L_{fee}(pad)$ (Fig.~\ref{fig:tribump}, top). The trace length
  differences are much too small to have any impact on the signal
  propagation time.  However, they affect the measured drift time via
  the capacitance and the signal shape.  As shown by the line in the
  bottom of Fig.~\ref{fig:tribump} the deviation can be
  reproduced by a linear combination of the two trace length
  distributions:
  \begin{equation}
    \label{eq:traceLenht}
    \Delta r(pad) = 0.10 \cdot L_{ppl}^{'}(pad) + 0.03 \cdot L_{fee}^{'}(pad)
  \end{equation}
  where the reconstructed radius correction $\Delta r$ is in mm and
  the trace lengths $L_{ppl}^{'},L_{fee}^{'}$ are in cm. The index
  $pad$ goes from 0 to 47 and measures the azimuthal position within
  one chamber. The trace length at the azimuthal center of the chamber
  has been subtracted:
  \begin{eqnarray}
    L_{ppl}^{'}(pad) &=& L_{ppl}(pad) - \frac{L_{ppl}(23) + L_{ppl}(24)}{2},\\
    L_{fee}^{'}(pad) &=& L_{fee}(pad) - \frac{L_{fee}(23) + L_{fee}(24)}{2}.
  \end{eqnarray} 
  For laser events the best fit is obtained with the two coefficients
  in Eq.~\ref{eq:traceLenht} being equal to 0.19 and 0.24, respectively. The
  difference between the nuclear collisions and the laser events is
  presumably caused by a different signal shape.
  
  \subsection{Anode induction}

\begin{figure}[ht]
    \centering
    \pgfimage[width=0.5\textwidth]{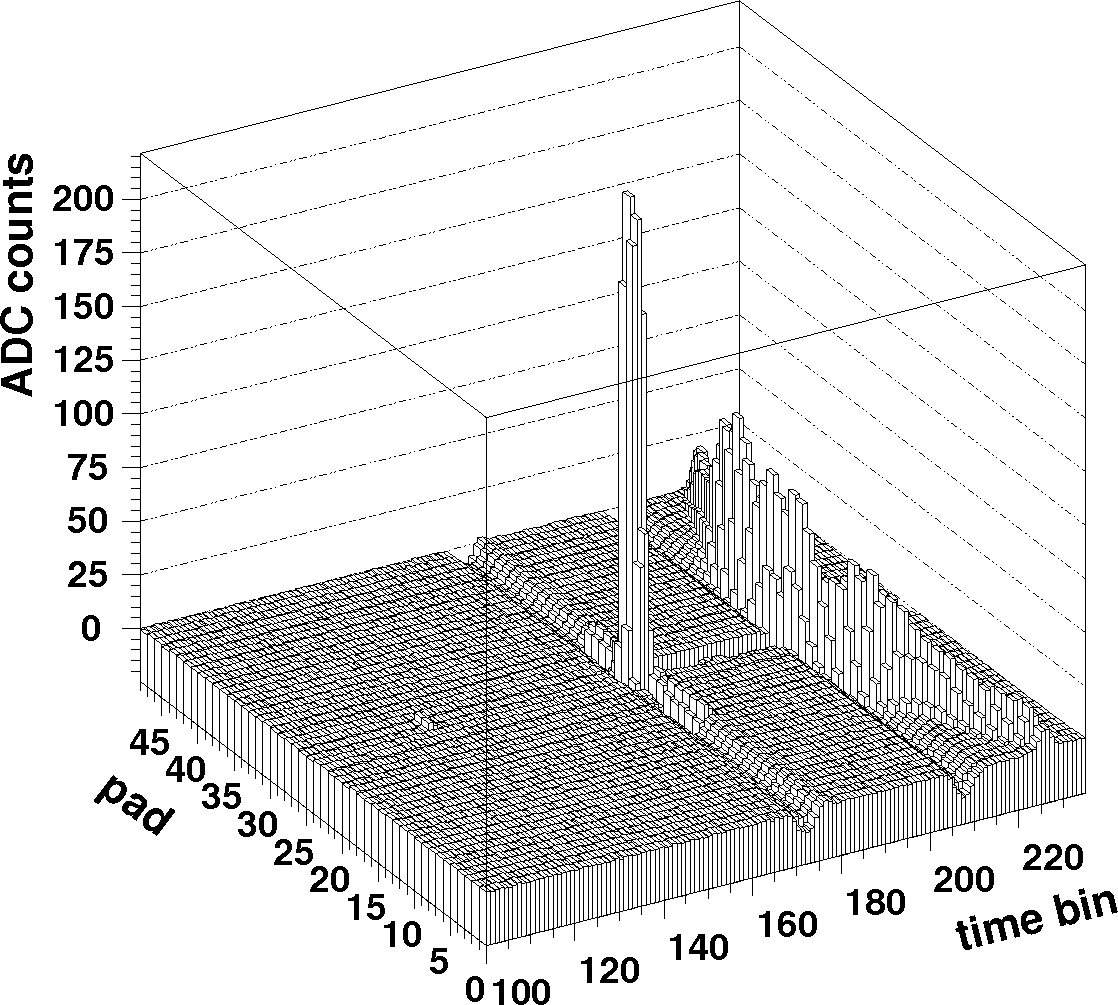}
    \caption{\label{fig:crosstalk} {\small Average of many laser
	events in pad-time space. The laser peak is followed by a
	baseline shift in temporal direction, as well as a sagging of
	the baseline on neighboring pads. The structure seen at large drift
	time is caused by UV stray light knocking out electrons from
	the inner cylinder.}}
  \end{figure}

  Fig.~\ref{fig:crosstalk} shows an average of a large number of
  pedestal subtracted laser events \cite{Sch01}. Clearly visible is
  the strong laser peak and the signals from electrons released from
  the inner cylinder by UV stray light at large time bins. Besides the
  baseline shift that follows the laser signal in temporal direction 
  there is also a sagging of the baseline of the neighboring pads. These
  are signals induced by the group of anode wires where the avalanche
  took place \cite{Ren97}.

  Whenever gas amplification happens at an anode wire, the resulting
  drop of the anode wire voltage is determined by the amount of charge
  $Q_{amp}$ that has been deposited during the amplification and by
  the capacitance $C_{aw}$ of the anode wire with respect to the
  surrounding electrodes:
  \begin{equation}
    \Delta U = \frac{Q_{amp}}{C_{aw}}.
  \end{equation}
  The measured pulse height $\PH_{signal}$ of the signal can be
  calculated from the actual charge at the anode wire:
  \begin{equation}
    \label{eq:phsignal}
    \PH_{signal}=\frac{256}{2.1 [\mV]} \cdot Q_{amp} [\fC] \cdot g_{electronics}
    [\mVperfC] \cdot \epsilon_{pad} \cdot \epsilon_{\pasa}.
  \end{equation}
  $Q_{amp}$ is given by the product of the primary ionization
  $Q_{prim}$ and the gas gain, $g_{electronics}$ is the gain of the
  electronics. The factor $\epsilon_{pad} = 0.3$ takes into account
  that only about $30\perc$ of the signal couples to the cathode pad
  plane, while the efficiency loss factor $\epsilon_{\pasa} = 0.5$
  occurs as a consequence of the finite integration time of the
  PASA. The factor $\frac{256}{2.1 [\mV]}$ converts the pulse height
  from mV to ADC counts.

  On the other hand, each single cathode pad has a capacitance
  $C_{pad-aw}$ with respect to the anode wire grid. Therefore, a drop
  of the anode wire voltage $\Delta U$ will induce a charge
  \begin{equation}
    \label{eq:qcrosstalk}
    q_{crosstalk} = C_{pad-aw} \cdot \Delta U =
    \frac{C_{pad-aw}}{C_{aw}}\cdot Q_{amp}
  \end{equation}
  on the cathode pad. This charge can be related to the measured pulse
  height signal $\PH_{crosstalk}$:
  \begin{equation}
    \label{eq:phcrosstalk}
    \PH_{crosstalk}=\frac{256}{2.1 [\mV]} \cdot q_{crosstalk} [fC]
    \cdot g_{electronics} [\mVperfC] \cdot \epsilon_{\pasa}.
  \end{equation}
  Note that the factor $\epsilon_{pad}$ does not enter because the
  coupling between anode wire grid and pad has already been explicitly
  taken into account by the capacitance $C_{pad-aw}$.

  According to Eq.~\ref{eq:qcrosstalk}, the
  magnitude of the induced signals scales with the amount of charge
  that has been created by gas amplification in the same time
  bin. Making use of Eq.~\ref{eq:phsignal} to \ref{eq:phcrosstalk},
  the measured dip can be related to the sum of the visible positive pulses:
  \begin{equation}
    \label{eq:phcrosstalk_phsignal}
    \PH_{crosstalk} = \frac{1}{\epsilon_{pad} \cdot \epsilon_{signal}} \cdot
    \frac{C_{pad-aw}}{C_{aw}} \cdot \PH_{signal}.
  \end{equation}
  The factor $\epsilon_{signal} = 0.24$ takes into account that in the case of
  the CERES TPC the pulse height corresponds to only $24\perc$ of the
  total charge deposited on the anode wires. This corresponds to the
  geometric fraction of the total signal which is read out (3 pad rows of
  $2.4\cm$ length, each in an anode wire group of $33\cm$ length). The
  capacitance $C_{pad-aw}$ has been measured to be $0.27\pF$. The
  anode wires of each chamber are powered by six HV-sectors, which are
  accessible from the backplate of the TPC. The capacitances $C_{aw}$
  of the six HV-sectors have also been measured and are listed in
  Table~\ref{tab:HVSectors}.
  \begin{table}[h]
    \centering
    \caption{\label{tab:HVSectors}{\small Capacitances of the
    individual sense wire groups (HV-sectors).}}
    \vspace{1mm}
    \begin{tabular}[h!]{|c|c|c|}
      \hline
      HV sector & cable length [cm] & capacitance $C_{aw}$ [$\pF$]\\
      \hline \hline
      1 & 250 & 812 \\ \hline
      2 & 200 & 722 \\ \hline
      3 & 170 & 668 \\ \hline
      4 & 144 & 622 \\ \hline
      5 & 116 & 571 \\ \hline
      6 &  74 & 492 \\ \hline
    \end{tabular}
  \end{table}

\begin{figure}[htb]
    \centering
    \pgfimage[width=.5\textwidth]{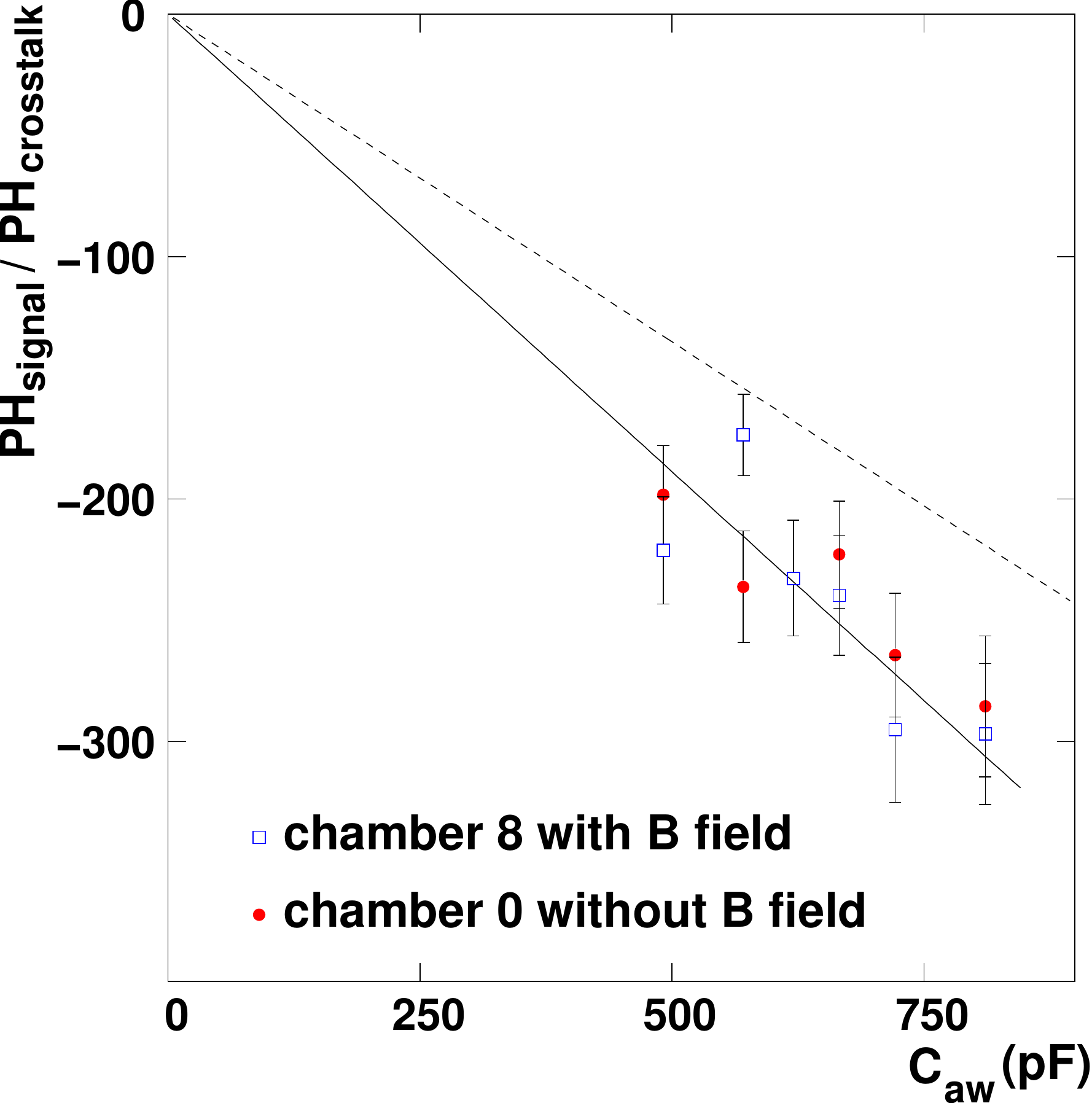}
    \caption{\label{fig:caw} {\small Signal to crosstalk pulse height
	ratio as a function of the anode wire group. Also shown is the
	expectation from Eq.~\ref{eq:phcrosstalk_phsignal} (dashed
	line).}}
  \end{figure}

  Fig.~\ref{fig:caw} shows the measured ratios
  $\PH_{signal}/\PH_{crosstalk}$ as a function of the capacitances of
  the individual anode wire groups. If
  Eq.~\ref{eq:phcrosstalk_phsignal} holds, one would expect a linear
  dependence in this representation, which is indeed suggested by the
  measurement. However, the magnitude of the effect is somewhat
  underestimated by Eq.~\ref{eq:phcrosstalk_phsignal}, as indicated by
  the dashed line. The crosstalk effect was reduced by a factor $2.5$
  by placing additional capacitors of $5\pF$ in parallel to each
  individual anode wire group.

  \subsection{Signal undershoot}

  The baseline shift in temporal direction seen in
  Fig.~\ref{fig:crosstalk} is caused by
  high-pass filters which are used to suppress leakage currents. High-pass
  filters work as differentiators for frequencies below the threshold
  frequency. Thus, a trailing edge of an incoming
  pulse will cause a negative outgoing pulse. This decreases the
  amplitude of pulses following in time.

  The shape of the signal undershoot was parametrized with an
  exponential function and included as correction \cite{Die01}. The
  correction assumes that the baseline shift is additive for
  subsequent clusters on a pad.

  \subsection{Ion signal}
  
  The electron avalanche occurs within less than $50\um$ of the anode
  wires \cite{Sau77}. The process typically lasts for about
  $1\ns$. The produced electrons are collected on the anode wires
  where they induce a negative signal. The remaining ions induce a
  positive signal on the readout pads. The ions drift towards the
  cathode wires about 1000 times slower than the electrons. The
  electric field strength is strongly weakened due to space charge and
  the drift velocity of the ions is further reduced. However,
  depending on the geometry of the chamber it may happen that the ions
  feel a second acceleration in proximity of the cathode wires where
  the electric field strength is again increasing. This late ion
  signal manifests itself in a negative induced signal on the cathode
  pads. According to GARFIELD simulations of the CERES TPC, the late
  ion signal is expected with a time delay of about $42\us$ after the
  main signal.

\begin{figure}[ht]
    \centering
    \pgfimage[width=.5\textwidth]{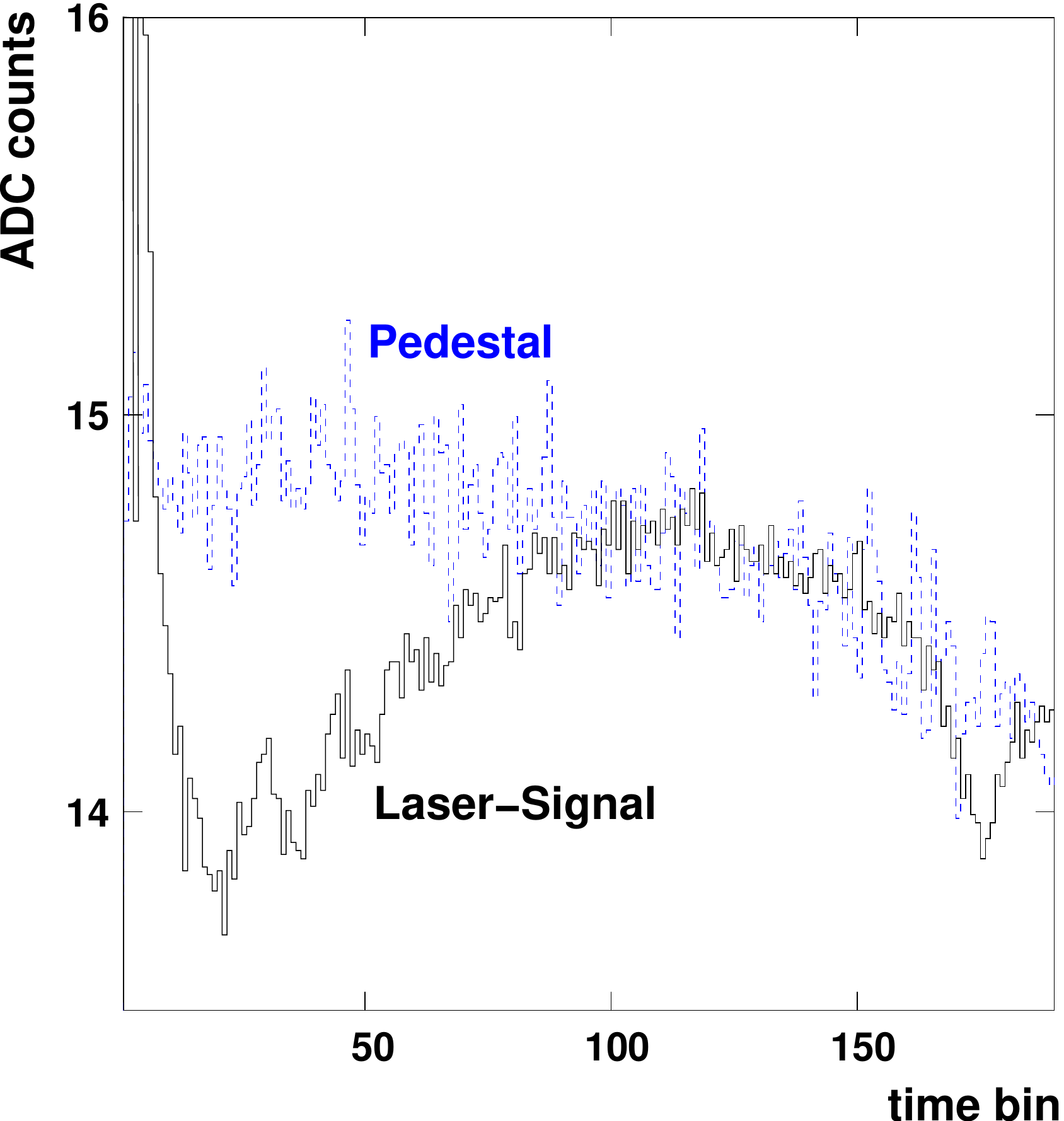}
    \caption{\label{fig:ionsignal} {\small Temporal development of the
	overlay of many laser events (solid line) and the
	corresponding pedestals (dashed line). The late ion signal is
	delayed by 175 time bins ($1\tb = 300\ns$).}}
  \end{figure}

  A measurement of the late ion signal is shown in
  Fig.~\ref{fig:ionsignal} \cite{Sch01}.  The figure shows an average
  of many laser events taken without zero suppression and not corrected
  for pedestals. Taking into account that the ion drift velocity is
  not well known for electric fields prevailing in the vicinity of the
  cathode wires, the measured time delay of about $52\us$ for the late
  ion signal is in good agreement with expectations. The lowering of
  the pedestals during the late ion signal is less that $0.8\perc$ as
  compared to the maximum signal and thus the correction of this
  effect was neglected.
  \subsection{Pad-to-pad gain variations}

\begin{figure}[ht]
    \centering
    \pgfimage[width=0.7\textwidth]{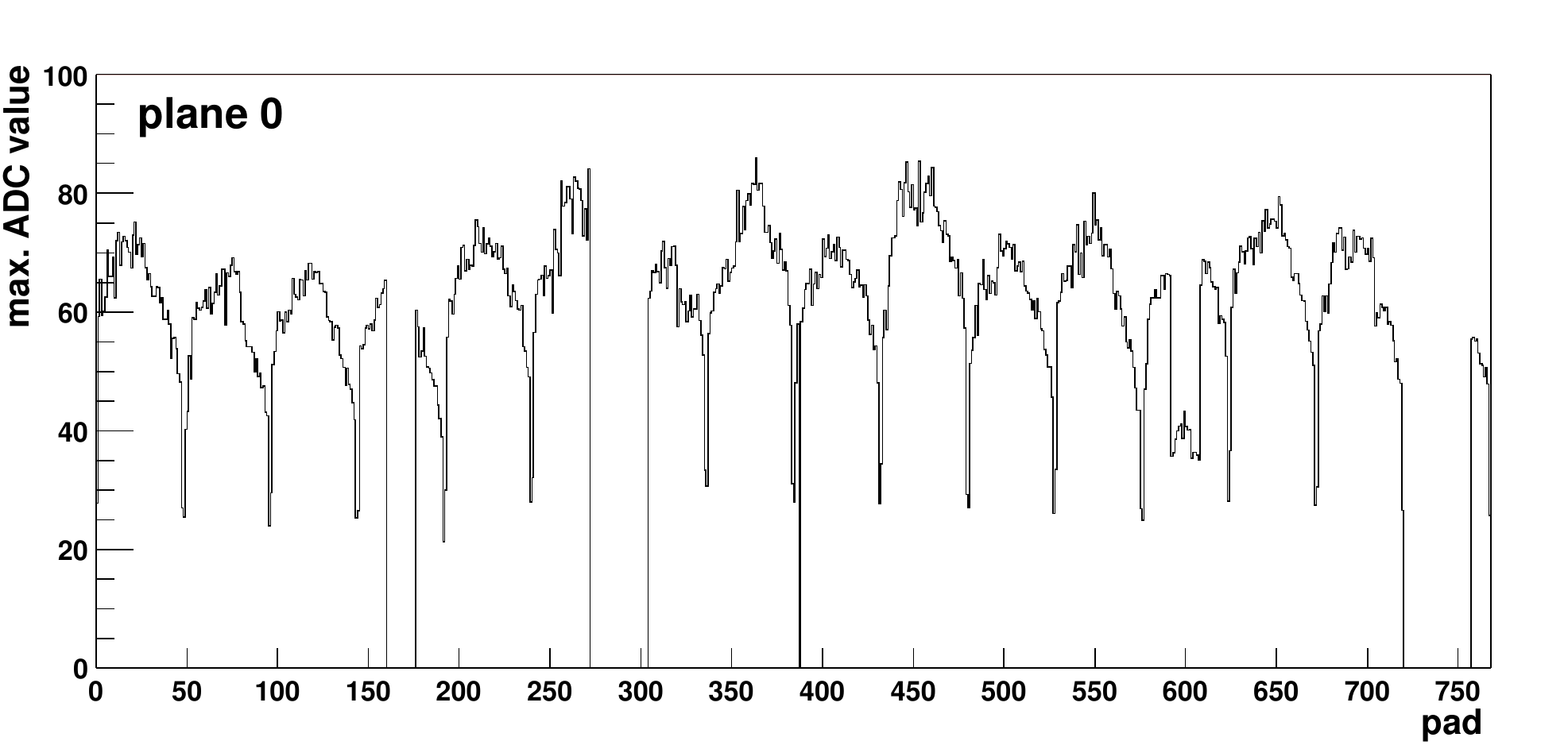}\\
    \pgfimage[width=0.7\textwidth]{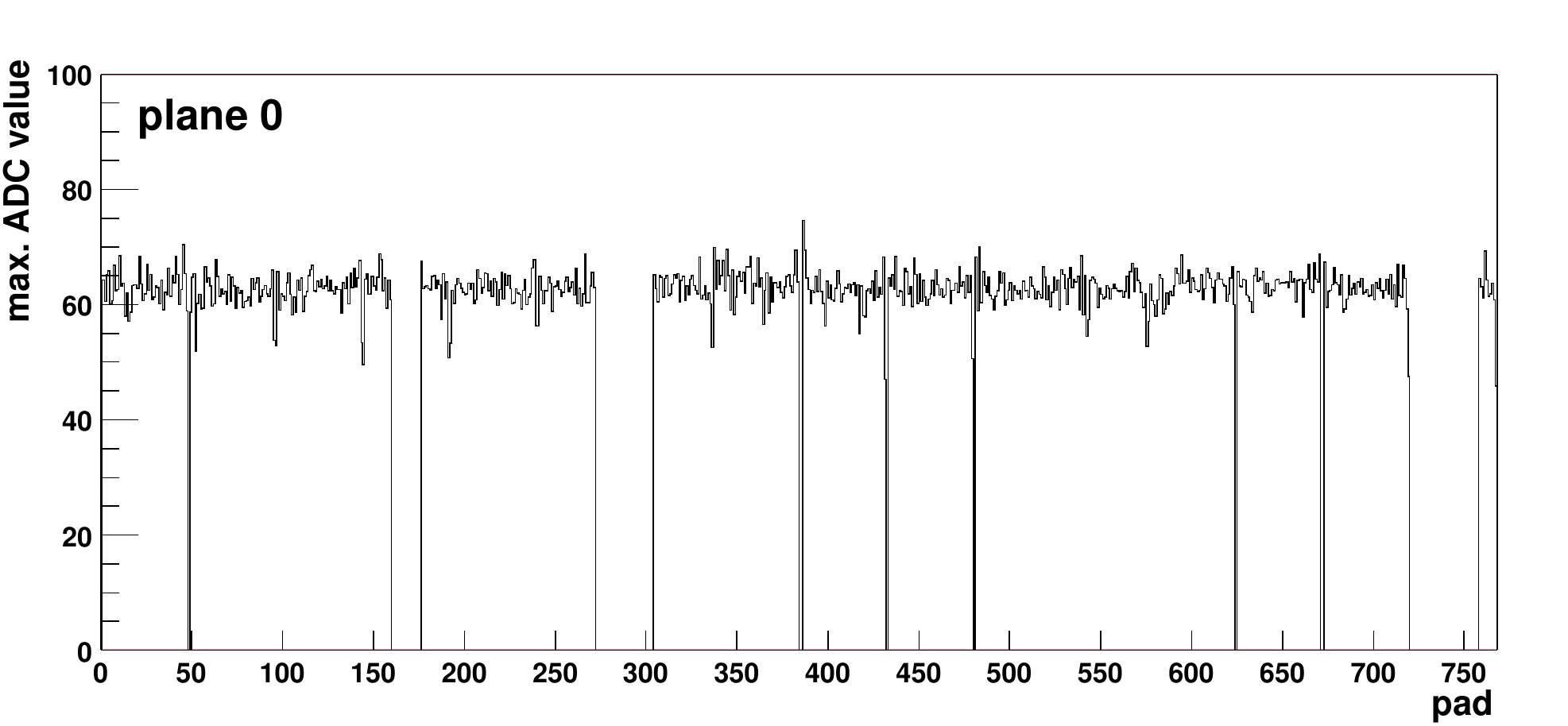}
    \caption{\label{fig:padToPadCor} {\small Top: Uncorrected
	pad-to-pad gain variations. For the pad-to-pad gain correction
	the maximum amplitude of a hit is used, which is localized on
	a single pad. The maximum amplitude is corrected for the polar
	angle $\theta$ of the track. Bottom: Pad-to-pad gain variations
	after correction.}}
  \end{figure}

  The pad-to-pad correction comprises all effects which cause gain
  variation from pad to pad, but also characteristics which extend
  over a whole electronic device. The top of
  Fig.~\ref{fig:padToPadCor} shows, as an example, the uncorrected
  pad-to-pad gain variation of the first plane of the TPC.

  The most noticeable structure extends over groups of 48 pads
  corresponding to the chambers of the TPC. The periodic peaks can partly
  be explained by the sag of the anode wires. The wires are glued
  to the edge of a chamber. In between these two fixed points the
  wires bend due to the electrostatic attraction towards the pad
  plane. This decreases the distance between the anode wires and the
  pads and thus a stronger signal is induced. The effect is strongest
  in the center of a chamber. It has also been observed that the gain
  drops at the end of the wires. Here the electric field differs from
  that of an infinite wire. Furthermore, the closing pads at the
  border of a chamber are smaller and have a different shape than the
  rest of the pads. This can be responsible for the dip at the edge of
  each chamber.  Finally, the three-fold structure inside a chamber
  reflects the individual behavior of the front-end-boards.

  The pad-to-pad gain variations have been studied in
  \cite{Die01}. The correction is implemented in form of look-up
  tables and determined for each calibration unit. The effect of the
  correction is demonstrated in the bottom of Fig.~\ref{fig:padToPadCor}.

  \section{Laser system}
  \label{sec:laser}
 
\begin{figure}[htb]
    \centering
    \pgfimage[width=1.\textwidth]{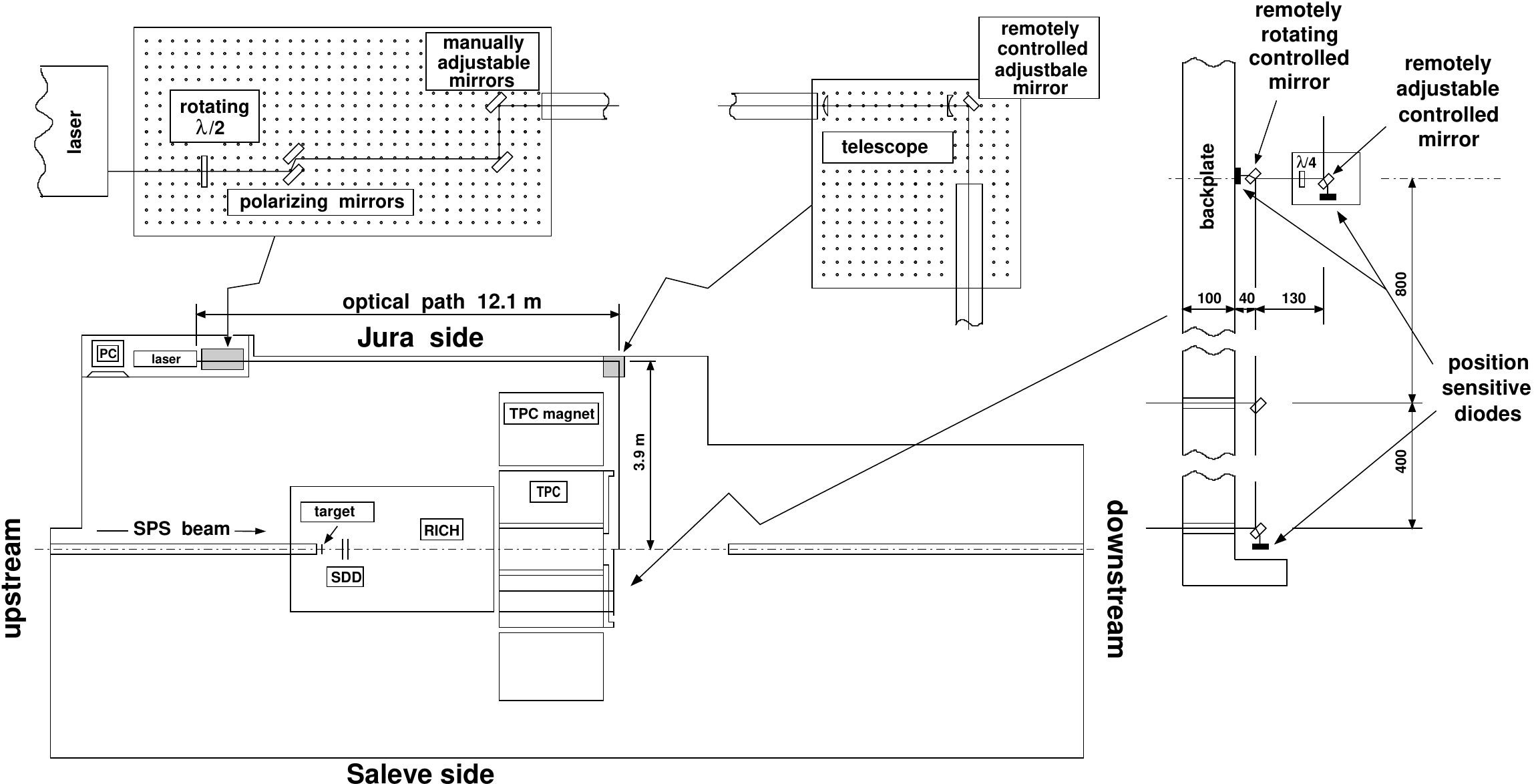}
    \caption{\label{fig:laser} {\small Laser system of the CERES
	TPC. The laser light from a Nd:YAG laser is frequency doubled
	twice and sent into the TPC using a mirror system mounted on
	the backplate of the TPC.}}
  \end{figure}

  For calibration and monitoring purposes the CERES TPC is equipped
  with a laser system, shown in Fig.~\ref{fig:laser}.  A Nd:YAG laser
  with a small beam diameter of $2\mm$ and low divergence of below
  $0.5\mrad$ was chosen.  With two frequency doublers the wavelength
  is converted from $\lambda = 1064\nm$ to $266\nm$.  With a pulse
  duration of $4\ns$ and an energy of $10\ft20\uJ$ per pulse an
  ionization similar to the one caused by charged particles can be
  achieved inside the TPC with a repetition frequency of up to
  $10\Hz$.

   The laser light is sent into the TPC, parallel to its axis at
   different radii and azimuthal angles, using a mirror system mounted
   on the backplate of the TPC. The position of each laser ray is
   monitored with position sensitive diodes located behind
   semi-transparent mirrors. The position of the laser ray and its
   angle at the entrance of the TPC are known within $0.25\mm$ and
   $0.5\mrad$, respectively. A detailed description of the laser
   system can be found in \cite{Mis06}.
   
  \section{Data analysis}
  \label{sec:analysis}

  The raw data recorded with the TPC during a beam period were
  processed into a suitable format for the subsequent data analysis. This
  section describes the basics of the reconstruction strategy and
  final calibration techniques.

  \subsection{Hit finding}
  \label{sec:hitfind}
 
  The TPC has 20~planes with 768~pads arranged in azimuthal direction.
  The drift time, and thus the radial coordinate, is digitized in
  256~time bins. The $20 \times 768 \times 256 \approx 4$ million
  pixels contain the linear amplitude information of the deposited
  charge in an 8-bit representation (cf. Sect.~\ref{sec:fedc}). 
  
  In order to reconstruct particle trajectories the first task to
  accomplish is to localize the passages of particles in the planes of
  the TPC, the so called hits. The hits are charged clusters,
  i.e. conglomerations of pixels with amplitudes $A>0$. It is also
  possible that several hits overlap in one charged cluster. The hit
  finding procedure scans the time-pad pixel grid of each plane to
  find absolute maxima $A_{max}$ within each cluster by first
  searching for local maxima in time direction and subsequently in pad
  direction \cite{Lud06}. In cases where the position of an absolute
  maximum within a cluster is ambiguous (i.e. several adjacent
  saturated pixels), the amplitude information from neighboring pixels
  is considered in the decision.
  
  A study based on simulated and real events has shown that the area
  of deposited charge does not extend over more than five time bins
  and three pads. This defines an area of 15 pixels arranged around an
  absolute maximum within a cluster. Hits containing only one time bin
  row are discarded. The same applies for one pad clusters, with the
  exception of a hit being localized at the edge of a chamber or
  adjacent to a dead FEE-board.

\begin{figure}[ht]
    \centering
    \pgfimage[width=1.\textwidth]{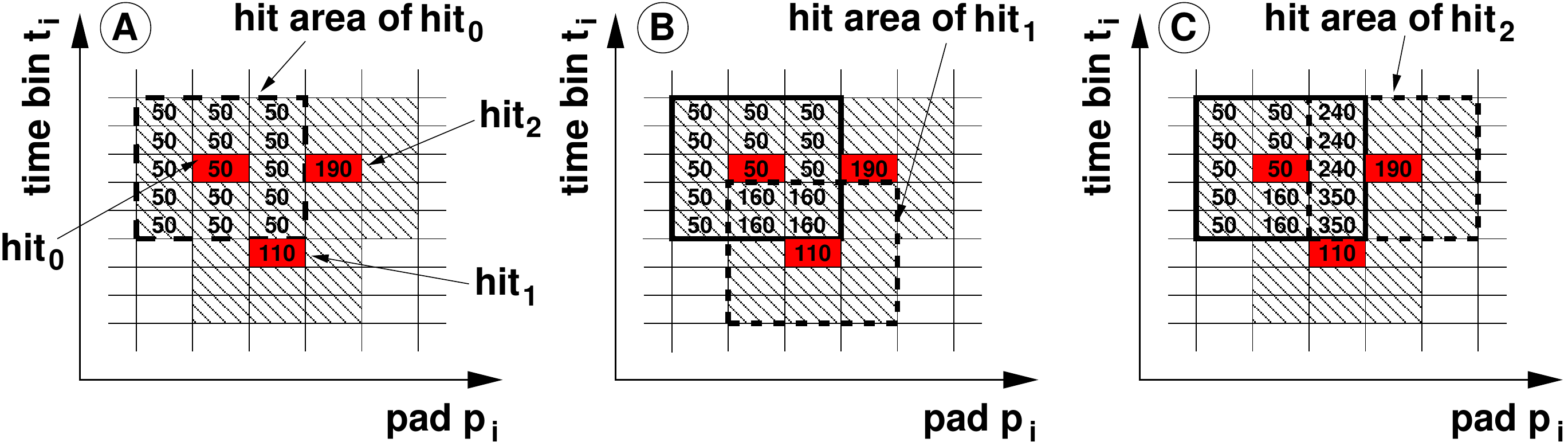}
    \caption{\label{fig:overlappingHits} {\small Reconstruction of
      overlapping hits in the TPC. The counter variables memorize the
      value of the absolute maximum of the hit to which the pixel
      belongs. If a pixel is assigned to several hits, its counter
      variable is augmented by the absolute maxima of the overlapping
      neighbors.}}
  \end{figure}

  To cope with the problem of overlapping hits, a counter variable
  $f_{i}$ is allocated for each pixel $i$. It memorizes the sum
  amplitudes of absolute maxima from those hits which share the same
  pixel. The method is sketched in Fig.~\ref{fig:overlappingHits}.  In
  this way it becomes possible to assign weights to individual pixels
  by the ratio of the absolute maximum $A_{max}$ of a certain hit and
  the counter variable $f_{i}$. For a given plane of the TPC the
  centroids for the time coordinate $\overline{t}$, and respectively
  for the pad coordinate $\overline{p}$, are given by:
  \begin{equation}
    \overline{t}  =  \frac{\sum_{i}A_{i} \cdot \frac{A_{max}}{f_{i}} \cdot
      t_{i}}{\sum_{i} A_{i}} \quad {\rm and} \quad
    \overline{p}  =  \frac{\sum_{i}A_{i} \cdot \frac{A_{max}}{f_{i}} \cdot
      p_{i}}{\sum_{i} A_{i}}.
  \end{equation}

  For a hit multiplicity between 17500 and 20000, about $80\perc$ of
  the hits are isolated, $17\perc$ are overlapping with one other hit
  and $3\perc$ are overlapping with two other hits. The advantage of
  calculating the centroid with weighted pixels as compared to a
  minimization procedure is a substantially decrease of computing time
  necessary for hit finding.

  \subsection{Coordinate transformation}
  \label{sec:coordTrans}

  The detector specific hit coordinates $(pad,time,plane)$ are
  transformed to spatial coordinates $(x,y,z)$ via look-up tables. The
  transformation contains the information about the transport process
  of the charged clusters in the electric and magnetic fields inside
  the TPC.

  The look-up tables are calculated using a Runge-Kutta method
  \cite{Pre92} that calculates the drift trajectory using in each
  point the drift velocity vector (Eq.~\ref{eqDriftVel}), starting at
  the cathode plane. The drift between the cathode plane and the pad
  plane is absorbed in a $\phi$-dependent time offset due the fact
  that each Front-End Electronic (FEE) channel had a slightly
  different capacitance. The difference between the MAGBOLTZ Monte
  Carlo drift \cite{Bia99} and the actual drift velocity vector is
  accounted for with a $z$- and $r(E)$-dependent correction for the
  drift velocity component parallel to the electric field and the one
  parallel to $\vec{E}\times\vec{B}$.

  \subsection{Track finding}
  
  The tracking starts in one of the middle planes 5 to 15 where the
  hit density is lowest \cite{Yur05}. The hits in this region are
  subsequently combined with their closest matches in the two
  upstream and two downstream planes to determine the sign of the
  track curvature in $\phi$-direction. The track curvature is linearly
  extrapolated to the remaining planes in which further hits are
  searched for within a narrow window of \mbox{$\Delta\phi=5.3\mrad$} and
  \mbox{$\Delta\theta=1.4\mrad$}.  The extrapolation contains
  $z$-dependent correction factors obtained from a GEANT \cite{GEA93}
  simulation of the TPC. The procedure stops if no further hit is
  found.

  In a second step the track is fitted in several iterations with a
  second order polynomial using Tukey weights~\cite{Tuk87} to predict
  the position of missing hits. The window to search for hits is
  widened to \mbox{$\Delta\phi=15$} $(10)\mrad$ for upstream
  (downstream) hits. The complicated track trajectory due to the
  strong inhomogeneity of the magnetic field leads to a less accurate
  prediction of the fit in the upstream direction for the low momentum
  (soft) tracks.  This is taken into account by applying again correction
  factors in the last five planes, now depending on both the
  $z$-position of the plane and the curvature of the track.
  The search for hits is continued in higher planes in the case of
  missing hits.
  
  A hit is only assigned to a single track,
  i.e. the sharing of hits between tracks is not allowed. The maximum
  number of hits per track is given by the \mbox{20 planes} in the
  TPC. The minimum number is limited to \mbox{7 hits} in order to
  reduce the contribution of imperfectly reconstructed or fake tracks.

  \subsection{Track fitting}
  \label{sec:trackfit}

\begin{figure}[ht]
    \centering
    \pgfimage[width=0.5\textwidth]{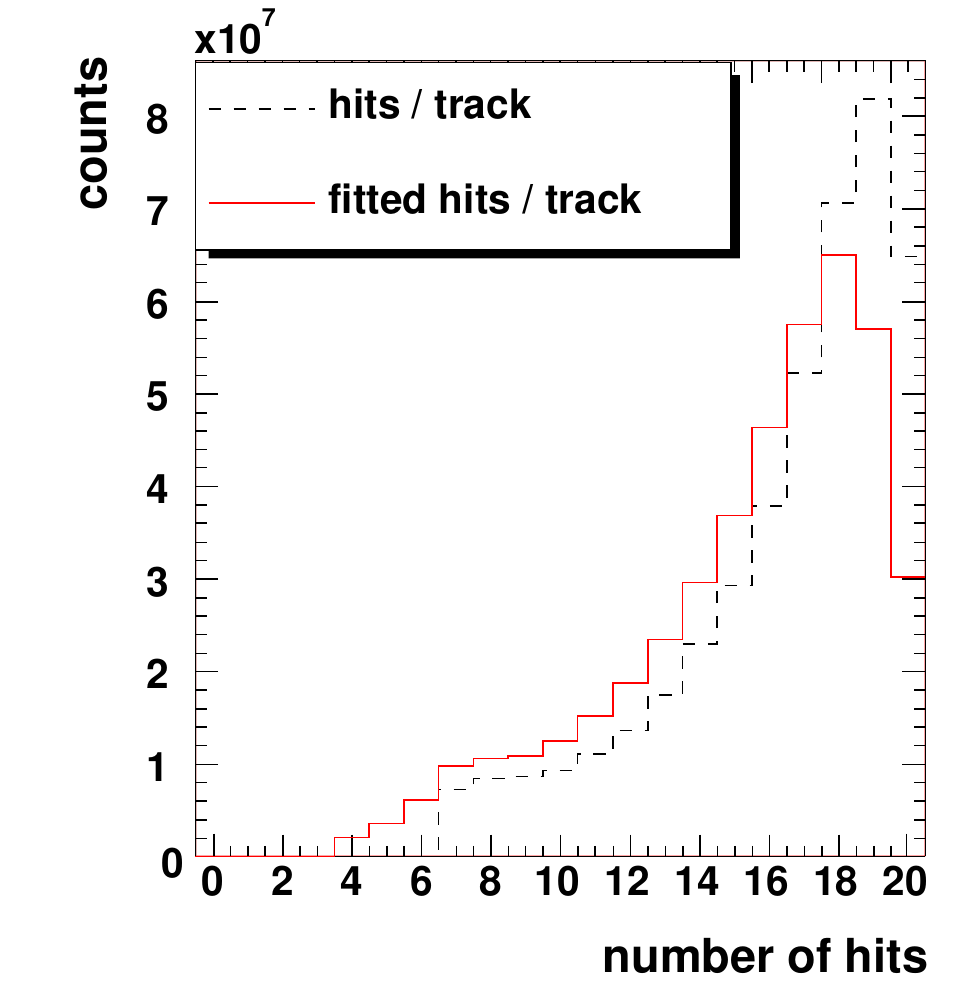}
    \caption{\label{fig:hitsPerTrack} {\small Number of hits per track
      and fitted hits per track. On average one hit is excluded from
      the track fit due to its large residual. Tracks with less than 7
      fitted hits are discarded.}}
    \end{figure}

  Because of the inhomogeneous magnetic field in the TPC an analytical
  description of a trajectory is impossible. This problem is handled
  by using reference tables for the track fit in the $\phi$-$z$ and
  $r$-$z$~planes. The reference tables contain the hit coordinates of
  Monte Carlo tracks from a GEANT \cite{GEA93} simulation of the TPC
  including the measured magnetic field configuration.  The Monte
  Carlo tracks are generated in bins of 32 different $\phi$ angles in
  the range \mbox{$-\pi < \phi < \pi$}, \mbox{18 different} $\eta$
  angles in the range \mbox{$2.05 < \eta < 2.95$} and 8~different
  inverse momentum values in the range \mbox{$-2 < q/p < 2$
  ($\gevc$)$^{-1}$} , where $q$ is the charge of the simulated
  particle. The track segments in the TPC are fitted with a
  two-parameter fit assuming that the tracks originate from the
  vertex, and with a three-parameter fit taking into account multiple
  scattering which happens mainly in the mirror of the RICH2
  detector. After several iterations, hits with large residuals
  $\Delta r > 0.4\cm$ and $r\Delta\phi > 0.2\cm$ are excluded from
  the fit. The distribution of initial and final ('fitted') number of
  hits per track is shown in Fig.~\ref{fig:hitsPerTrack}. The average
  number of hits (fitted hits) per track is $16.5$ ($15.3$) with a
  most probable value of $19$ ($18$) hits. This means that on average
  one hit per track is excluded from the fit.

\begin{figure}[ht]
    \centering
    \pgfimage[width=0.6\textwidth]{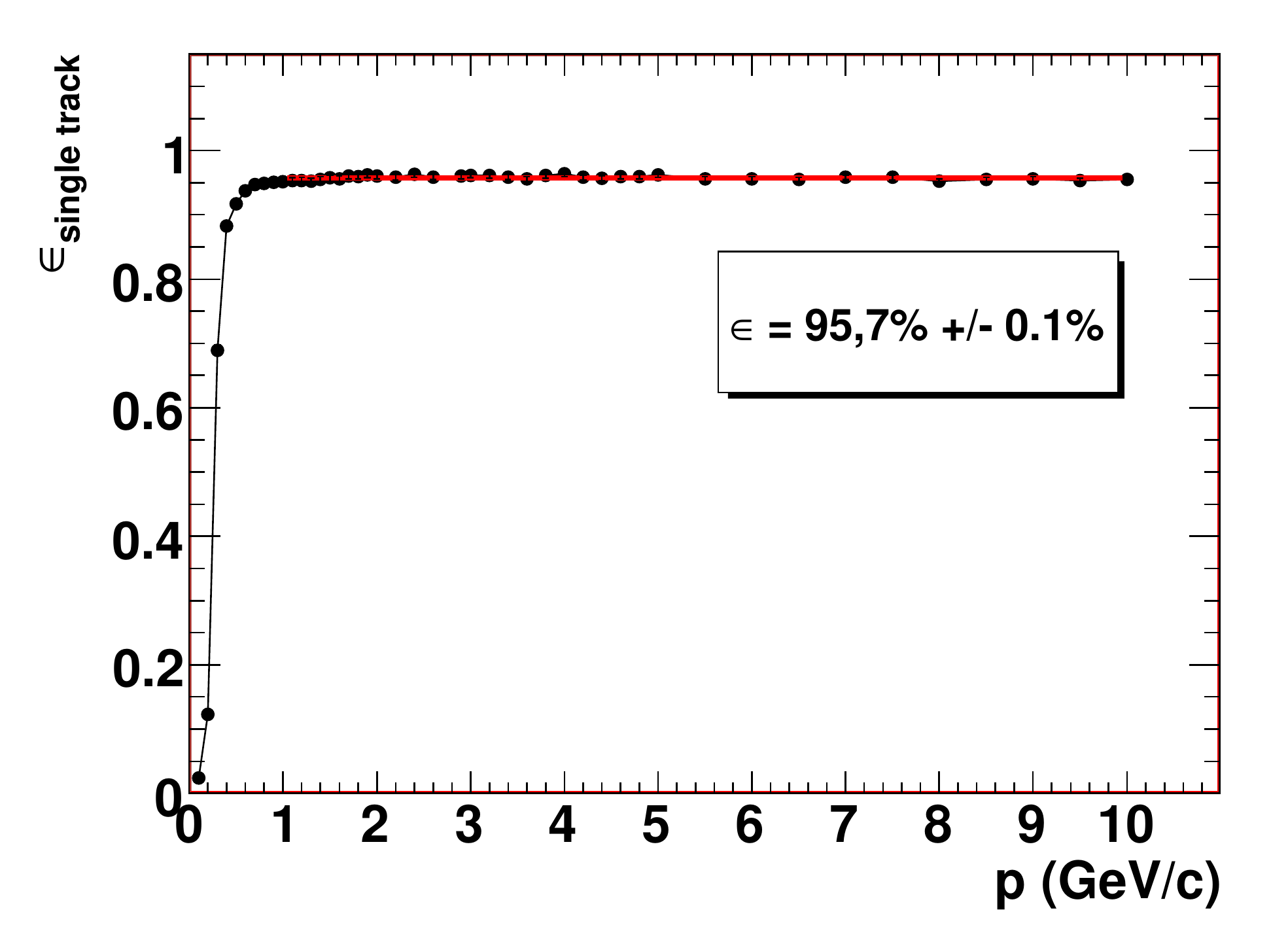}
    \caption{\label{fig:trackEff_mom} {\small Single track efficiency
      of the TPC as a function of momentum, obtained by embedding
      Monte Carlo tracks in experimental central Pb-Au events.}}
      \pgfimage[width=0.6\textwidth]{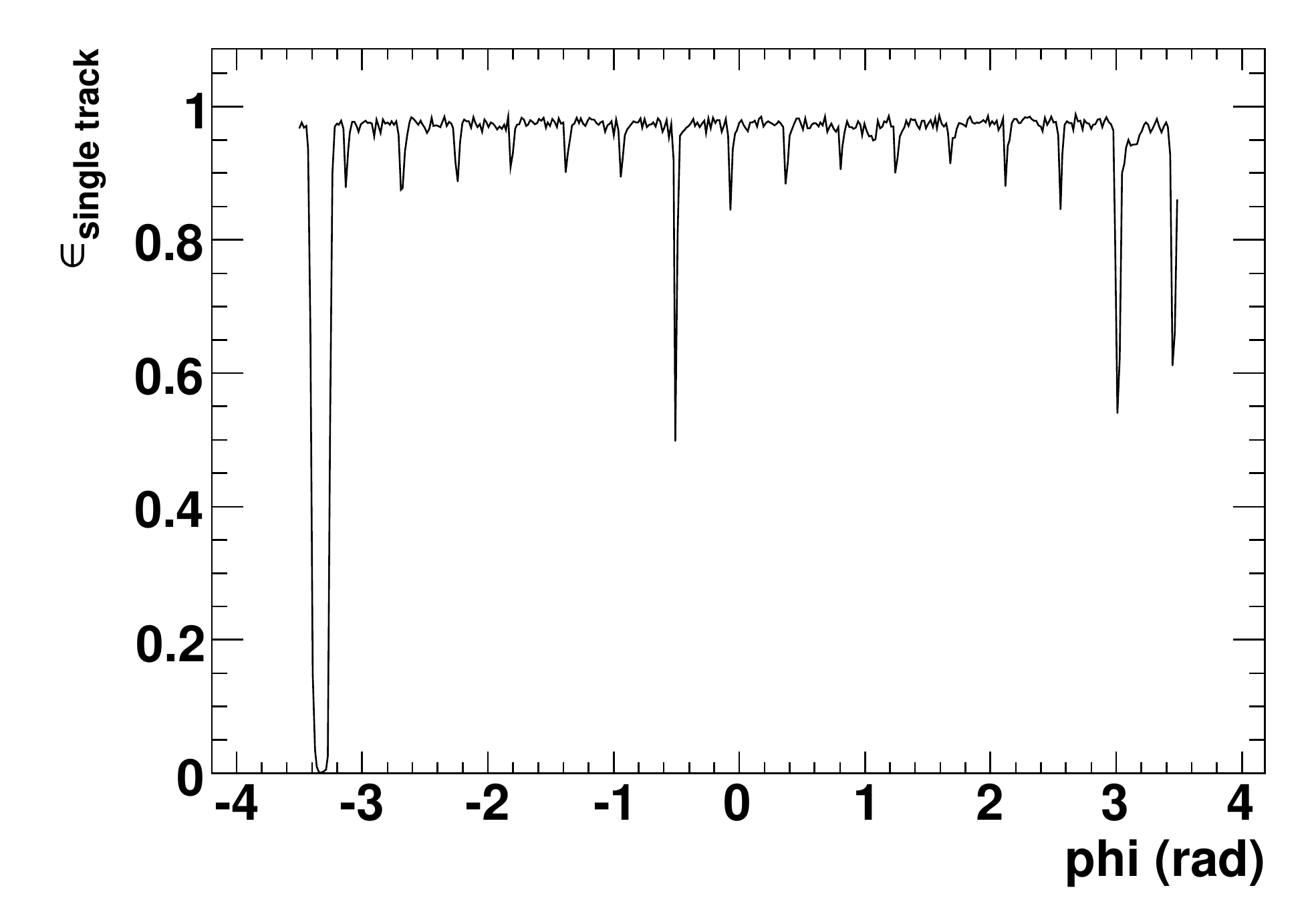}
    \caption{\label{fig:trackEff_phi} {\small The single track
      efficiency of the TPC as a function of azimuth.}}
  \end{figure}

  The single track efficiency of the TPC for positive pions traversing
  the detector at $\theta=0.18\rad$, i.e. at the center of the
  acceptance, is shown in Fig.~\ref{fig:trackEff_mom} as a function of
  momentum \cite{Kal07}. The efficiency was determined by embedding
  Monte Carlo tracks in experimental central Pb-Au events. The
  efficiency drops steeply for tracks with a momentum smaller than
  $0.6\gevc$. Tracks with $p>1\gevc$ are reconstructed with an
  efficiency of $96\perc$. About half of the inefficiency there is
  caused by a row of not powered FEE-boards at $\phi\approx-3.3\rad$
  (cf. Fig.~\ref{fig:trackEff_phi}) \cite{Kal07}. The finite hit
  density and the inefficiencies at the chamber boundaries account for
  $1/4$ each.  The single track efficiency is rather independent on
  the polar angle except when approaching the edge of the acceptance
  at low theta. At $\theta=0.12\rad$ the high momentum efficiency is
  dropping by $5\perc$ because of the finite hit density. Further work
  to improve the track efficiency at low momentum is still ongoing
  \cite{Bus06}.
 
\begin{figure}[h]
    \centering
    \pgfimage[width=0.5\textwidth]{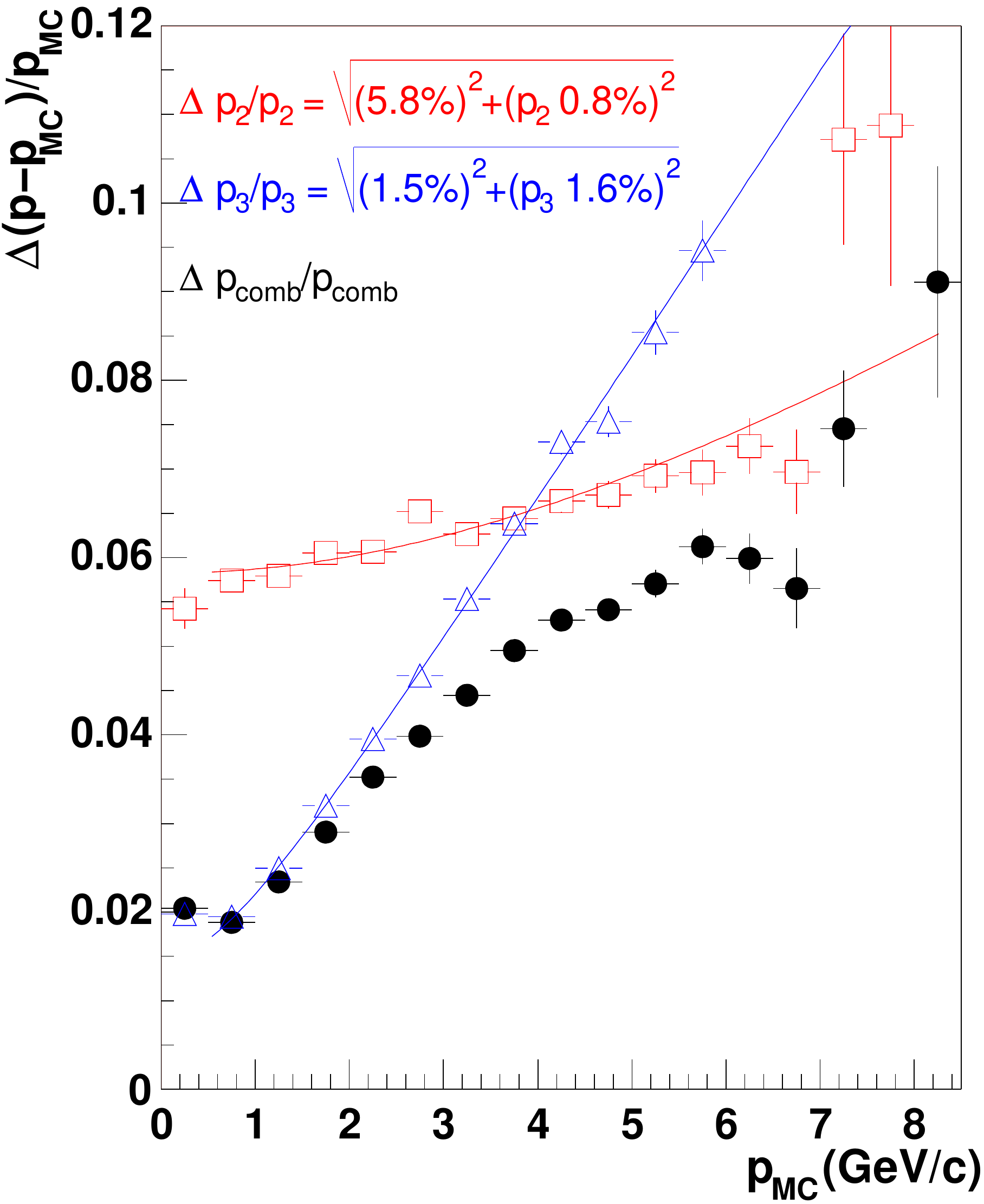}
    \caption{\label{fig:momentumRes}
	      {\small Relative momentum resolution. The two-parameter
	    fit (open squares) includes an additional
	    vertex constraint and is optimal for high
	    momentum tracks. The three-parameter fit (open triangles)
	    allows an
	    azimuthal inclination of the track at the entrance of the
	    TPC and yields the best result for low momentum tracks
	    which often suffer multiple scattering. The combined
	    momentum fit (filled circles) combines the advantages of both.}}
  \end{figure}

  The momentum of a track is determined from the deflection in
  $\phi$-direction. The deflection in $\theta$-direction caused by a
  second order field effect is taken into account by applying a small
  correction. In addition to a two-parameter fit in the $\phi$-$z$
  plane, a three-parameter fit allows an azimuthal inclination of the
  track at the entrance of the TPC caused e.g. by prior multiple
  scattering. Fig.~\ref{fig:momentumRes} shows
  the relative momentum resolution determined with a Monte Carlo
  simulation as a function of the momentum. The three-parameter fit
  yields an optimal result for low momentum tracks which often suffer
  multiple scattering. In contrast, high momentum tracks are better
  described by a two-parameter fit due to the additional vertex
  constraint. To exploit the positive aspect of both, a combined
  momentum
  \begin{equation}
    \label{pcomb}
    p_{comb} =
    \left(\frac{p_{2}}{\sigma_{2}^{2}}+\frac{p_{3}}{\sigma_{3}^{2}}\right)
    / \left( \frac{1}{\sigma_{2}^{2}} +
    \frac{1}{\sigma_{3}^{2}}\right),
  \end{equation}
  is used, where $p_{2}$ and $p_{3}$ denote the two-parameter and
  three-parameter fits, respectively, and $\sigma_{2} = \Delta
  p_{2}/p_{2}$ and $\sigma_{3} = \Delta p_{3}/p_{3}$ are the
  corresponding resolutions. The relative momentum resolution of the
  TPC, which is determined by the resolution of the detector and
  multiple scattering, was determined by GEANT simulations and is
  given by:
  \begin{equation}
  \label{momentumResolution}
  \frac{\Delta p_{comb}}{p_{comb}} = \sqrt{\left(1\% \cdot p_{comb}
    \right)^2 + \left(2\%\right)^2},
  \end{equation}
  with $p_{comb}$ expressed in $\gevc$.

  \subsection{Calibration of hit positions}
  \label{sec:calhit}

\begin{figure}[h]
  \centering
  \pgfimage[width=0.5\textwidth]{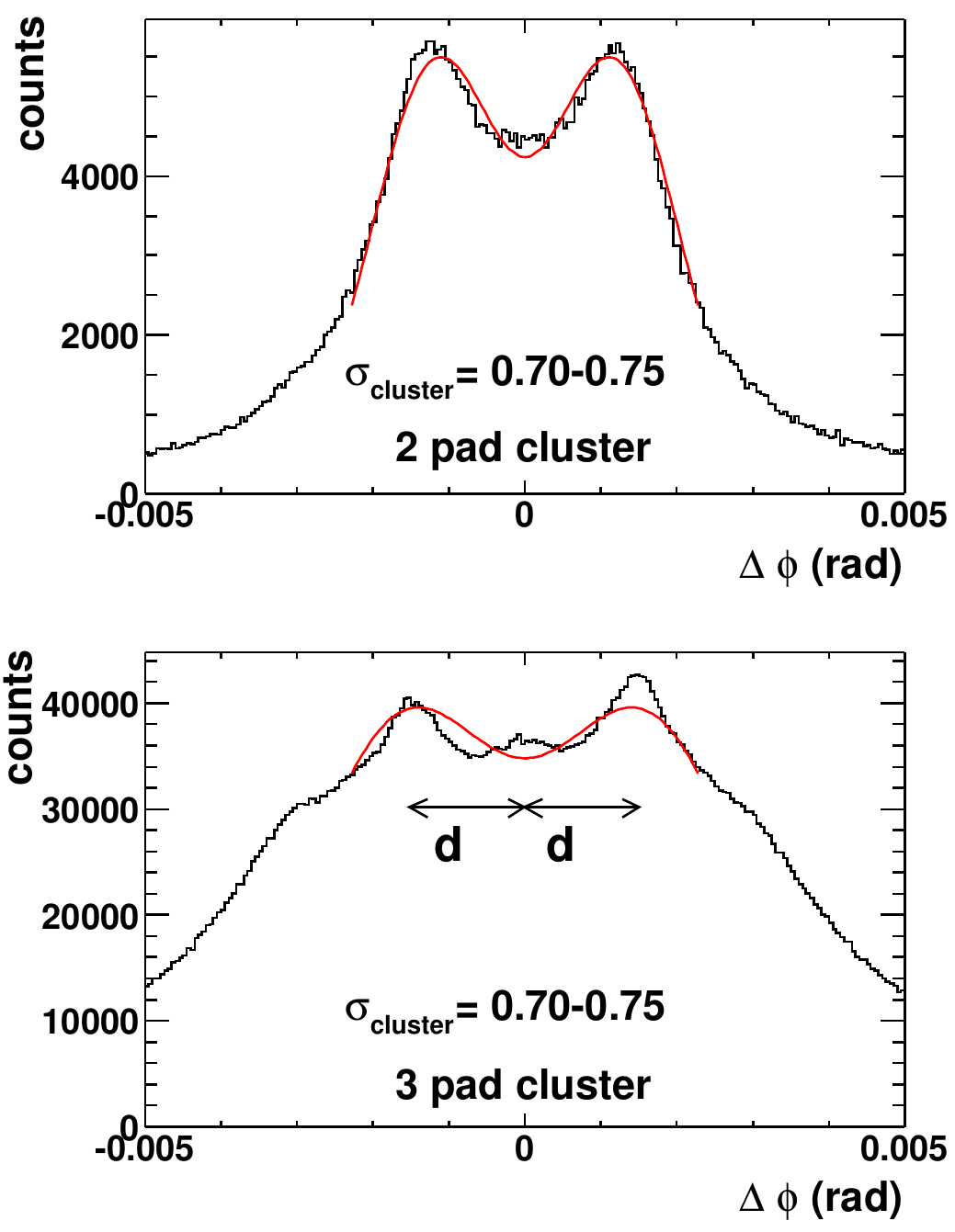}
  \caption{\label{fig:overlapHitCor} \small The splitting of the
    \mbox{$\Delta \phi = \phi_{track} - \phi_{hit}$} distribution in
    several peaks is due to unrecognized overlapping hits. The
    distance $d$ of the peak to the center as a function of the
    cluster width is used to derive a correction for the hit
    position. The cluster width $\sigma_{cluster}$ is given in pad units.}
  \end{figure}

  The hit reconstruction (cf. Sect.~\ref{sec:hitfind}) is able to
  separate overlapping hits as long as their absolute maxima are at
  least one pixel apart. If this is not the case, the merged clusters
  are assigned to a single hit. These hits are recognizable by their
  unusually large cluster width of $\sigma_{cluster}>0.6$~pad units
  ($\approx6.2\mm$). A double or even triple peak structure becomes
  visible in the $\Delta\phi=\phi_{track}-\phi_{hit}$ distribution
  with increasing cluster width (Fig.~\ref{fig:overlapHitCor})
  \cite{Lud06}. To correct for unrecognized overlapping hits the
  double peak structures were parametrized as a function of the
  distances $d$ from the center and included as corrections in the
  track fitting procedure. The triple peak structures were neglected.
 
\begin{figure}[h]
    \centering
    \pgfimage[width=0.8\textwidth]{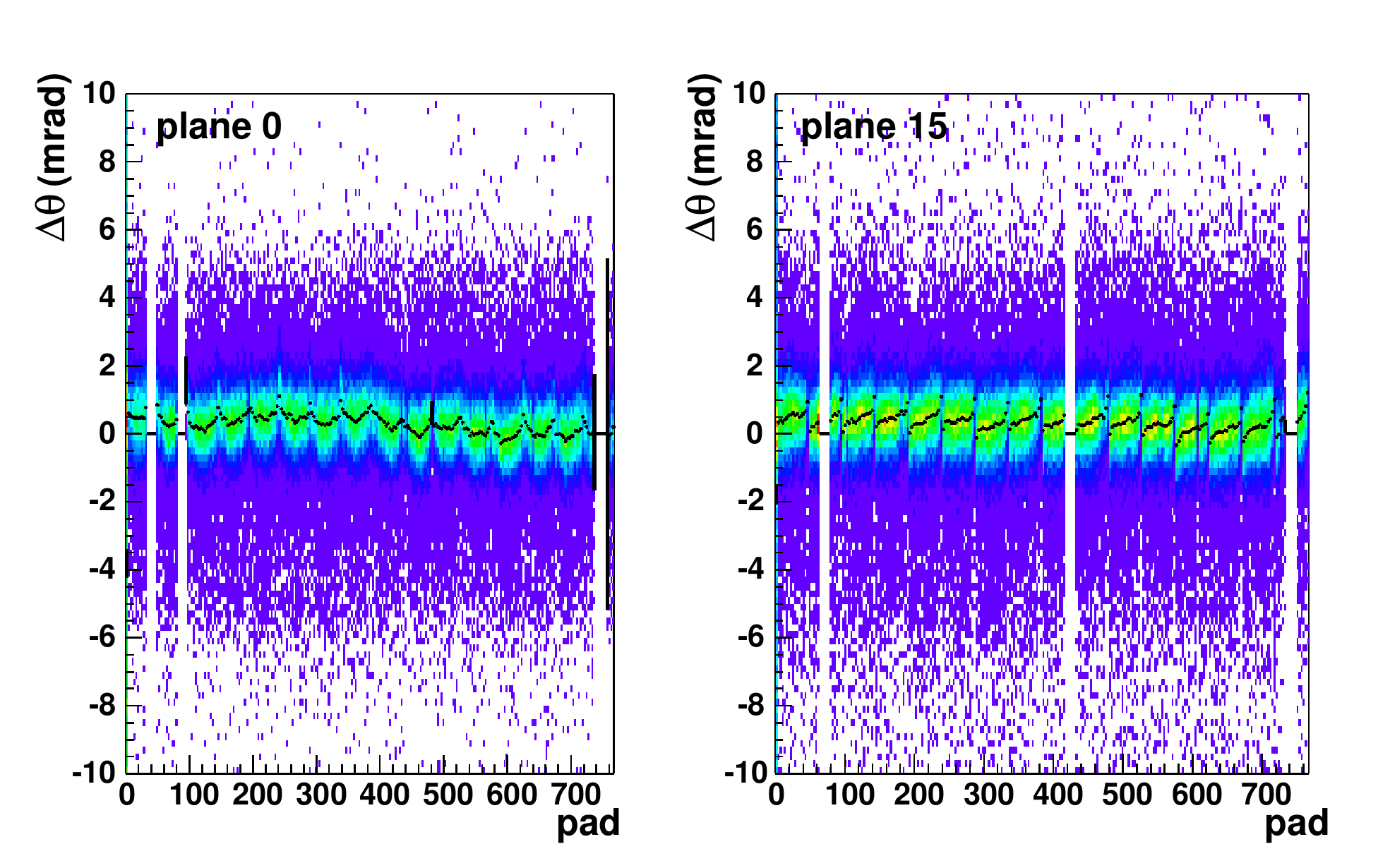}\\
    \pgfimage[width=0.8\textwidth]{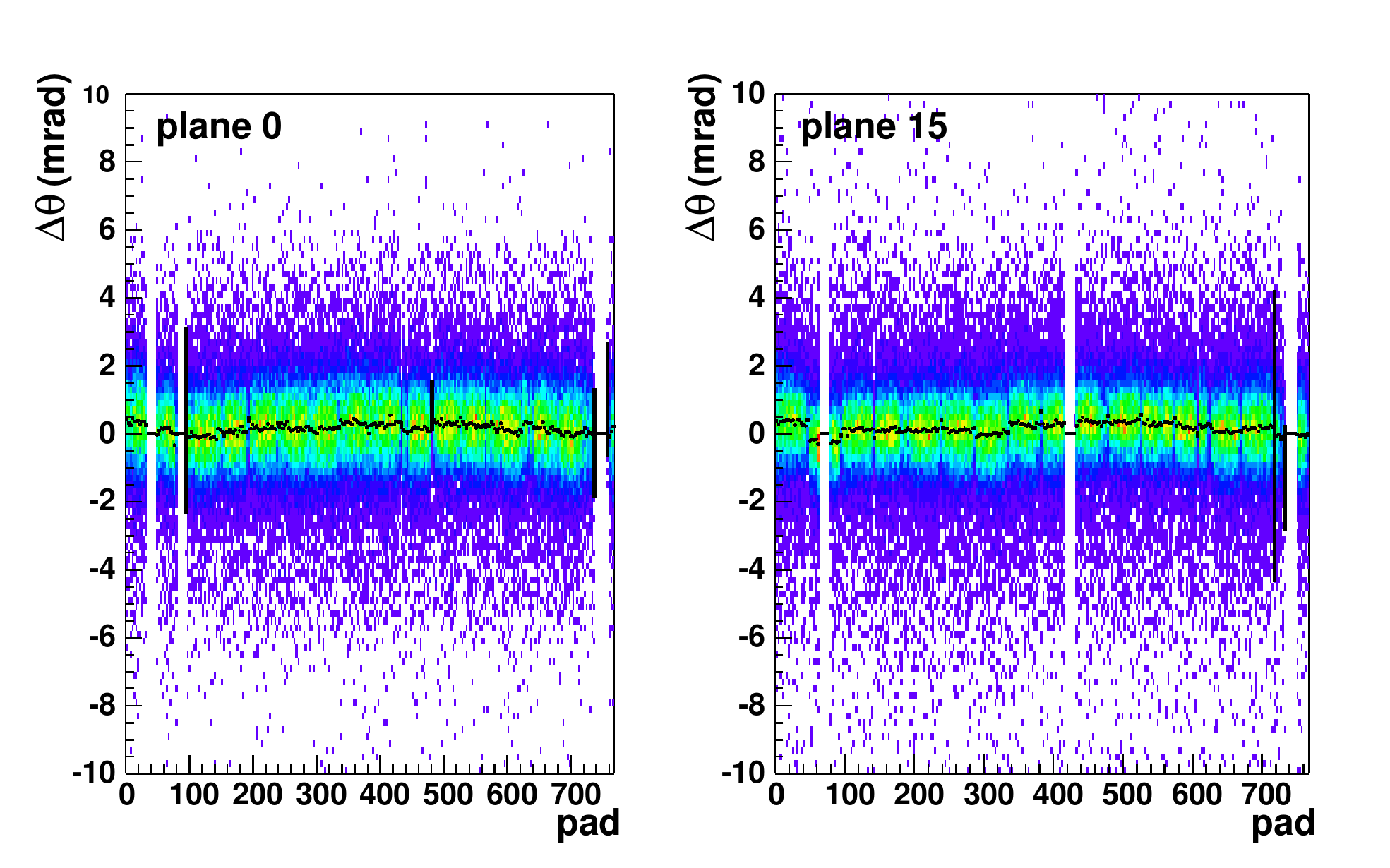}
    \caption{\label{fig:hitCorrectiona} {\small Top: Polar angle
	difference $\Delta \theta = \theta_{SDD,track} -
	\theta_{TPC,hit}$ between the SDD track segment and the TPC
	hits plotted vs. the pads in the TPC. The black lines
	represent the bin-by-bin averages of the entries in the color
	scatter plots. The tiny structures reflect the 16~readout chambers of
	the TPC. Bottom: The same distribution after correction of the
	hit positions in the TPC.}}
  \end{figure}

  Further corrections of the hit positions in the TPC were obtained by
  using the additional information from other detectors. The tiny
  structures seen in the polar angle difference
  $\Delta\theta=\theta_{SDD,track}-\theta_{TPC,hit}$ between the SDD
  track segment and the TPC hits plotted vs. the pads of the TPC
  (Fig.~\ref{fig:hitCorrectiona}, top) reflect the 16~readout
  chambers.  No periodic structures were observed in the distribution
  of $\theta_{SDD,track}$. Using this knowledge the positions of the
  hits in the TPC were shifted by the amount $\Delta \theta_{cor}$
  with respect to the mean value $\Delta \theta$ in each each
  chamber. The corrections were applied as a function of the polar
  angle $\theta$ to the individual pads and planes
  (Fig.~\ref{fig:hitCorrectiona}, bottom).

\begin{figure}[ht]
    \centering
    \pgfimage[width=1.\textwidth]{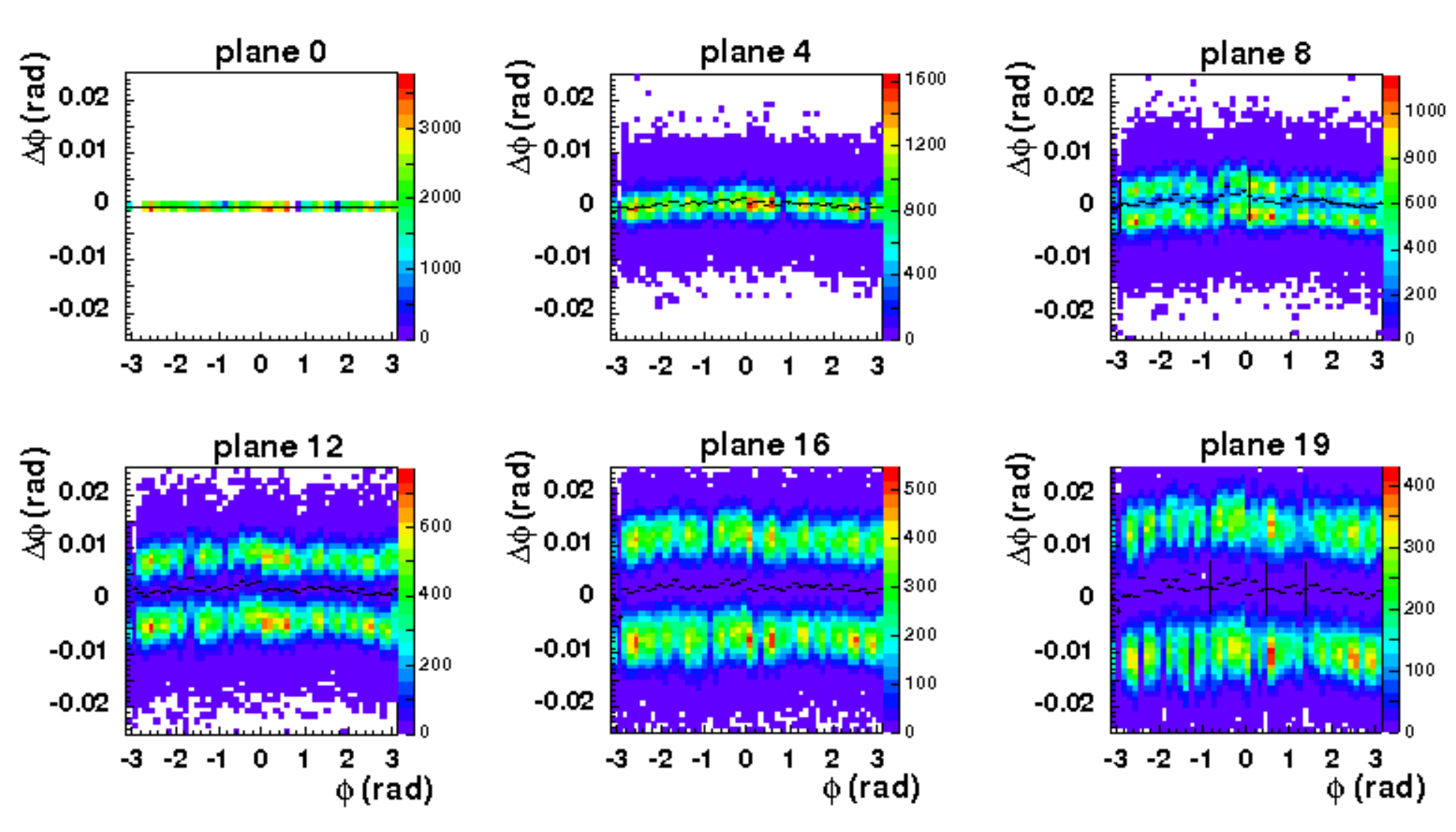}
    \caption{\label{fig:hitCorrectionb} {\small Remaining apparent
	deflection for infinite momentum tracks. For a clean pion
	sample the azimuthal angle difference $\Delta \phi =
	\phi_{R2M} - \phi_{TPC,hit}$ is plotted vs. $\phi_{R2M}$.  The
	black lines represent the bin-by-bin averages of the entries
	in the color scatter plots. The valley between the positive
	and negative particles should, ideally, be at
	$\Delta\phi=0.$}}
  \end{figure}

  A clean sample of high momentum pions ($p > 4.5$ GeV/c), selected
  with the RICH detectors, was used to study the azimuthal angle
  difference $\Delta\phi=\phi_{R2M}-\phi_{TPC,hit}$ between the TPC
  track segment as measured at the mirror of the RICH2 detector and
  the hits in the TPC (Fig.~\ref{fig:hitCorrectionb}). The advantage
  of using a clean pion sample for calibration purposes is given by
  the similar multiplicities of $\pi^+$ and $\pi^-$ differing by only
  $10\%$ \cite{Sta05}. The deflection of oppositely charged particles
  with increasing plane number can clearly be seen in
  Fig.~\ref{fig:hitCorrectionb}. With good approximation, the minima
  of the distributions are expected to be centered at zero because
  tracks with infinite momentum are not deflected
  (cf. Sect.~\ref{sec:momcalib}). The deviations from zero were used
  to correct the hit positions of the TPC as a function of the
  azimuthal \mbox{angle $\phi$} and the plane number. The same
  calibration was performed for the polar angle $\theta$.
  \subsection{Hit position resolution}
  \label{sec:difres}

  The result of the track fitting procedure can be improved by
  assigning individual weights to the hits according to their
  resolution. The weights contain the information about the hit
  positions in 3-dimensional space and about special hit
  characteristics. In this way remaining inaccuracies in the
  determination of the drift velocity are taken into account, as well
  as other dependencies like the hit amplitude or the hit multiplicity
  of the event.
  
  As the momentum is determined from the deflection of a particle
  trajectory in the magnetic field, the azimuthal coordinate $\phi$
  has a dominant influence on the momentum resolution. For this
  coordinate the hit position resolution $\Delta\phi$ was
  parametrized as a function of the radius, the 20 planes in the TPC,
  the hit amplitude, the hit multiplicity in the TPC, the number of
  responding pads, and the hit being isolated or not, and included as
  weights ($w=1/{\Delta\phi}^2$) in the track fitting
  procedure.

\begin{figure}[ht]
    \centering \pgfimage[width=1.\textwidth]{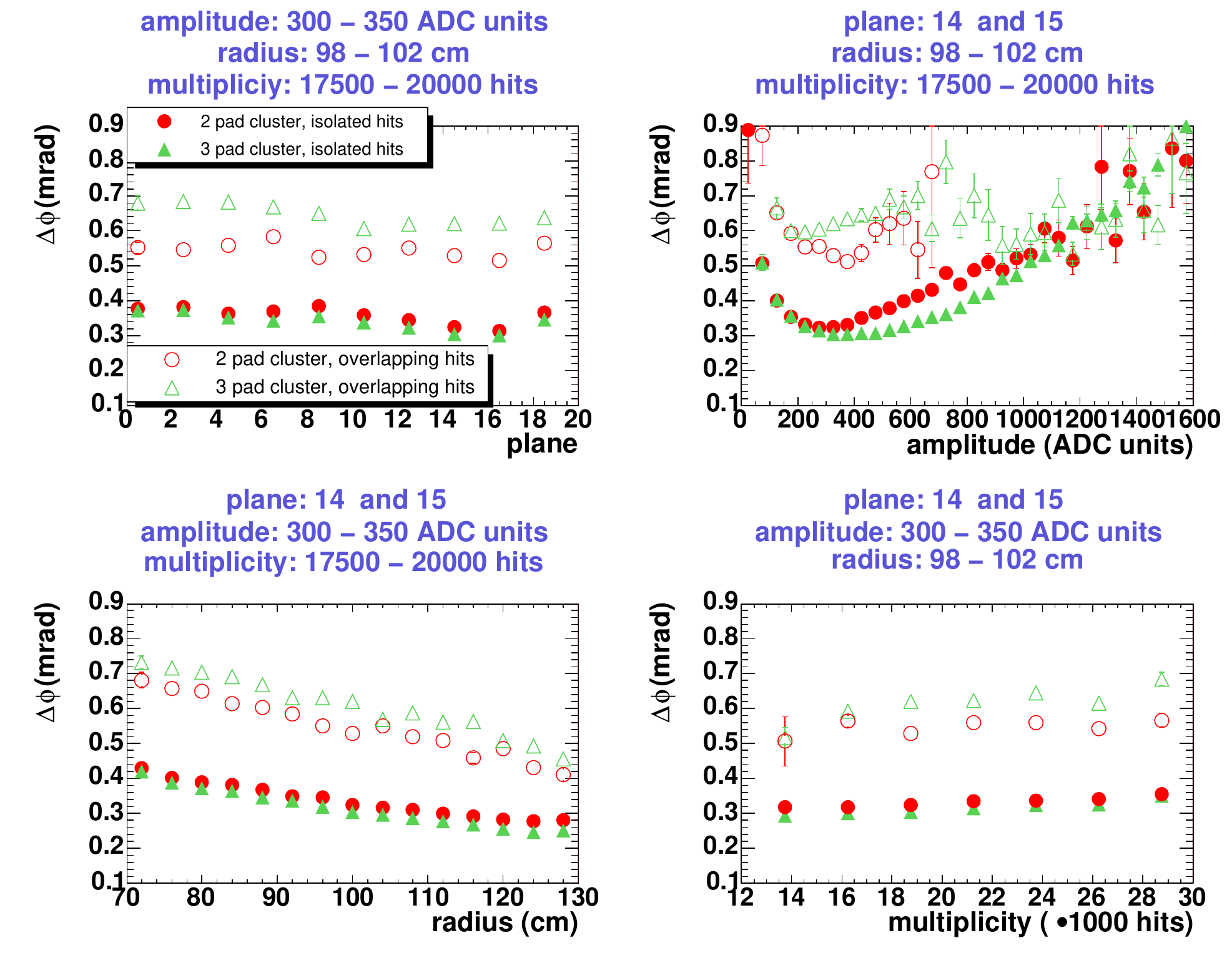}
    \caption{\label{fig:difRes} \small The hit position 
      resolution was parametrized as a function of the radius,
      plane, amplitude, hit multiplicity, number of responding pads and
      hit isolation status. This allowed to weight each hit individually
      in the track fitting procedure according to its characteristics.}
  \end{figure}

  Fig.~\ref{fig:difRes} shows a subset of the 6-dimensional resolution
  matrices.
  For isolated hits, the 3~pad clusters have a somewhat better
  resolution than the 2~pad clusters due to the more favorable charge
  sharing. The resolution of overlapping hits deteriorates about a
  factor of two compared to isolated hits. Remarkable is also the
  strong dependence of the resolution on the amplitude. Firstly, the
  resolution improves with the amplitude. This effect can be ascribed
  to a better signal-to-noise ratio. Due to saturation effects the
  resolution deteriorates again at higher amplitudes. As expected, the
  resolution deteriorates with the number of hits in the TPC, though
  the effect is small. The linear improvement of the resolution with
  the radius has its origin in the increasing influence of the
  diffusion with the drift length.  The dependence on the plane number
  of the TPC is due to the higher occupancy in the first planes, but
  also due to remaining uncertainties in the knowledge of the electric
  or magnetic field.

  \subsection{Calibration of the momentum}
  \label{sec:momcalib}

\begin{figure}[ht]
    \centering
    \pgfimage[width=0.5\textwidth]{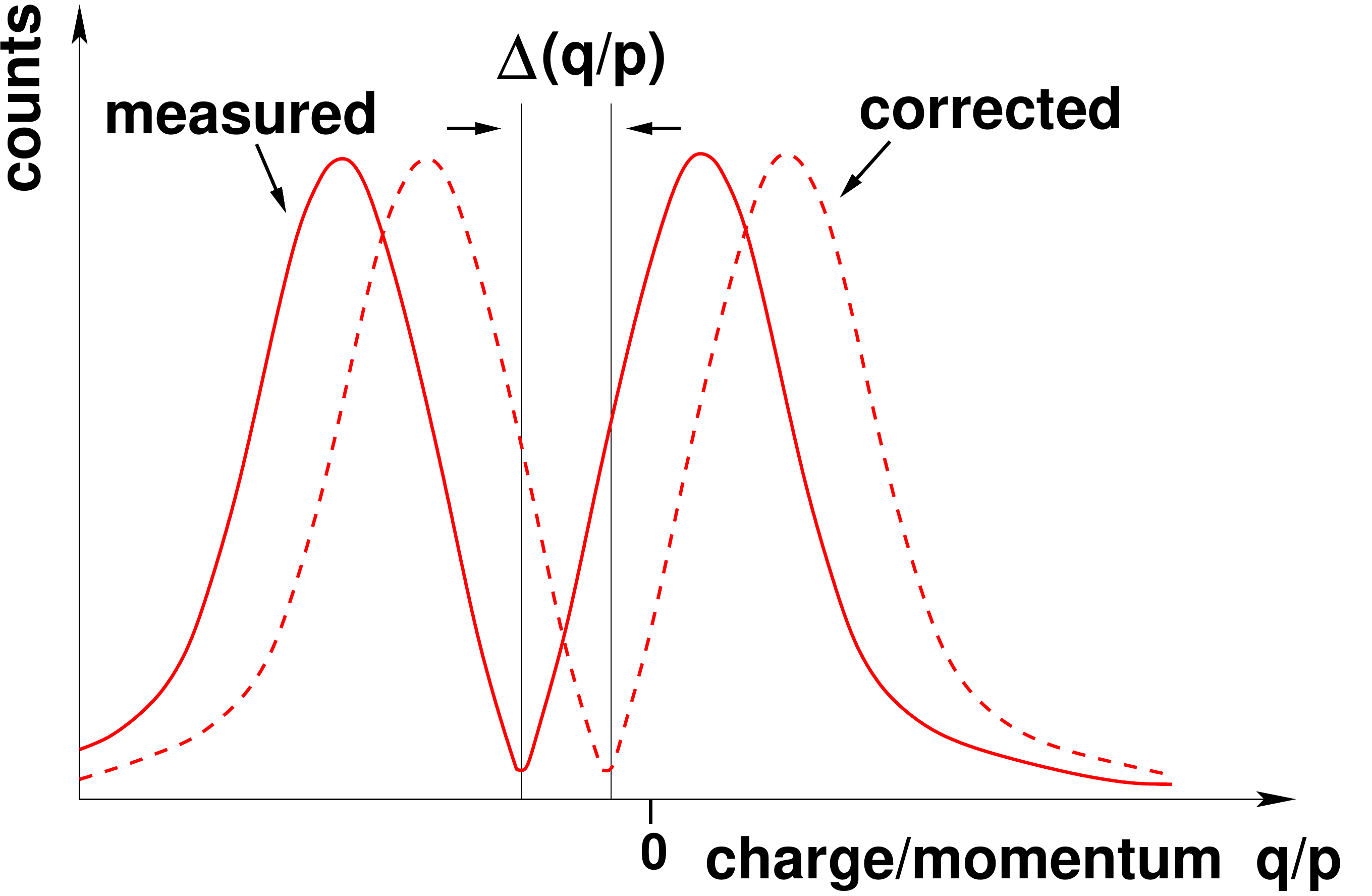}
    \caption[Inverse Momentum Correction]{\label{fig:invPCor} \small A
      shift of inverse momentum distribution of positively and
      negatively charged particles can be used to determine the
      correction.}
  \end{figure}

  The absolute measured value of a particle momentum can vary during
  the data taking period. It is caused by the fluctuation in the
  electric and the magnetic field, the change in the gas composition,
  temperature, and ambient pressure. The calibrations described in the
  previous sections removed most of these effects. The remaining
  variations are taken out phenomenologically by looking at the
  position of the minimum in the $q/p$-distribution
  (Fig.~\ref{fig:invPCor}). Assuming similar abundances of positively
  and negatively charged particles and the infinite spatial and
  temporal resolution of the detector, the distribution of the charge
  over momentum ($q/p$) should have a minimum at zero
  (cf. Sect.\ref{sec:calhit}).

  The shift can be used to correct remaining deviations in the
  momentum. A convenient particle choice for this measurement is
  pions. As already discussed in Sect.~\ref{sec:calhit} the
  multiplicities of $\pi^{+}$ and $\pi^{-}$ are similar and effects
  related to a different abundance of particles with opposite charge
  are reduced.

  To determine the proper position of the minimum with high accuracy,
  the reconstructed masses of $\Lambda$- and
  $\overline{\Lambda}$-hyperons were compared as a function of
  momentum, while applying different shifts $\Delta(q/p)$. Only at the
  position of the proper minimum the reconstructed masses coincide.
  
\begin{figure}[ht]
    \centering
    \pgfimage[width=1.\textwidth]{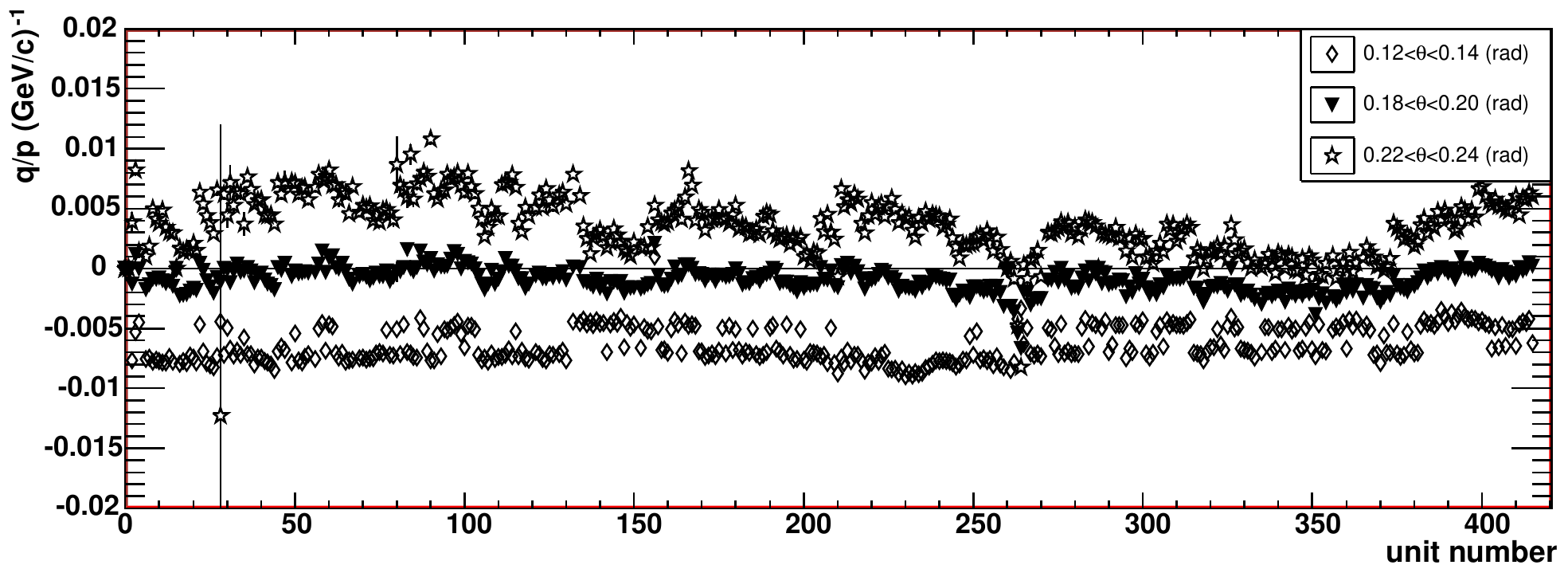}\\
    \pgfimage[width=1.\textwidth]{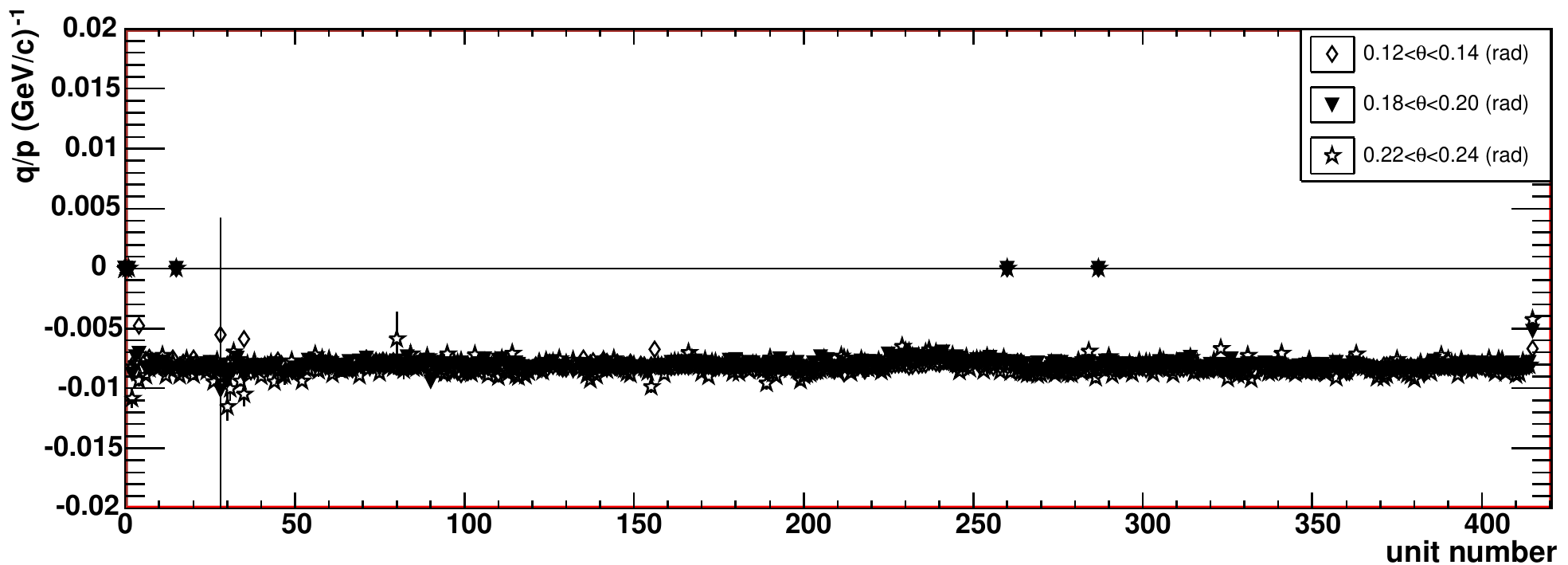}
    \caption{\label{fig:momCalib} \small Top: The minimum of the
    $q/p$-distribution plotted vs. calibration units (1 calibration
    units $\approx 1$ hour) for different polar angles $\theta$
    without applying any momentum calibration. Bottom: After momentum
    calibration the position of the minimum remains stable.}
  \end{figure}

  For the calibration of the data a clean pion sample was selected
  with the RICH detectors by the ring radius. The determination of
  the shift of the minimum with respect to its proper position was
  performed in four steps in order to optimize statistics
  \cite{Ant06}. First, the data were divided in three groups:
  \begin{itemize}
  \item positive magnetic field at the beginning of the beam time,
  \item negative magnetic field, and
  \item positive magnetic field at the end of the beam time,
  \end{itemize}
  and a coarse correction was calculated as a function of the
  azimuthal and polar angle $\phi$ and $\theta$. Second, a finer
  correction was derived in intervals of one hour data taking. 
  The next calibration steps were
  performed on a pion sample identified via the differential energy
  loss $\dedx$ in the TPC. This allowed to increase significantly
  statistics, but decreasing the purity of the pion samples. The
  corrections were applied as a function of $\theta$ integrating over
  $\phi$ in finer time scales. Finally, a last correction was
  determined in even finer entities of 10~bursts, but integrated over
  $\phi$ and $\theta$. The effect of the momentum calibration on a
  pion sample selected via the $\dedx$ in the TPC can be
  seen in Fig.~\ref{fig:momCalib} where the minimum of
  the $q/p$-distribution is plotted as a function of time. 

\subsection{Specific energy loss}
\label{sec:SpecificEnergyLoss}

\begin{figure}[htb]
\centering
\vspace{0.1cm}
\pgfimage[width=0.7\textwidth]{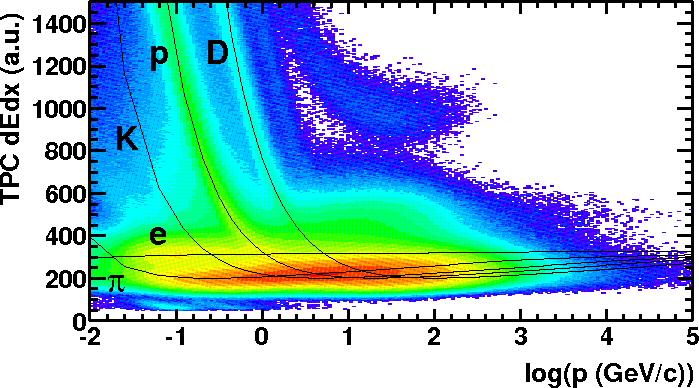}
\caption{\label{fig:tpc_dedxvsp}
\small The truncated-mean energy loss as a function of momentum for different 
particle species.}
\end{figure}

The specific energy loss of charged particles at a given momentum
carries information about the particle mass.  In the CERES TPC the
energy deposited along the track is sampled 20 times with each sample
corresponding to a path length of approximately $2.4\cm$ (cf
Fig.~\ref{fig:padPlane} in Sect.~\ref{sec:PadPlane}).  The signal in
each sample follows the Landau distribution with a long tail due to
energetic delta electrons.  The truncated-mean method, in which hits
with lowest (one hit) and highest (40\% of all hits) amplitudes are
excluded, leads to a significantly improved $\dedx$ resolution.  The
truncated-mean energy loss measured in the CERES TPC for electrons,
pions, kaons, protons, and deuterons is shown in
Fig.~\ref{fig:tpc_dedxvsp}.

\begin{figure}[ht]
  \centering
  \pgfimage[width=0.5\textwidth]{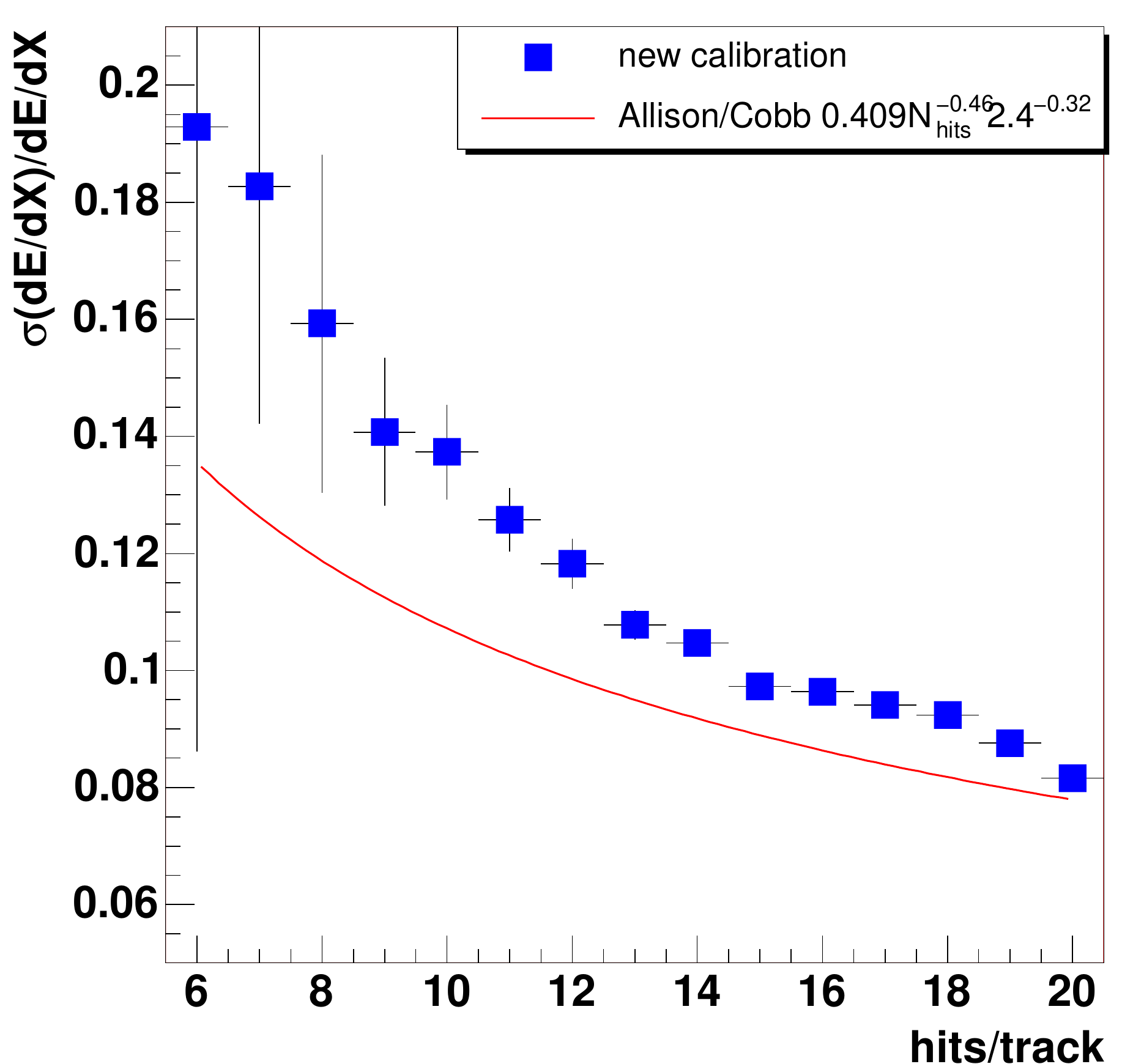}
  \caption{\label{fig:dedxresvshits}
    Truncated mean $\dedx$-resolution vs. number of hits per
    track (squares) compared with the parametrization by Allison and
    Cobb \cite{All80} (line). For a number of hits above 15 the $\dedx$-resolution is
    better than 10\perc.}
\end{figure}

After pad-to-pad calibration based on total amplitude in the pad of maximum 
amplitude (cf. Sect.~8.9), correction of undershoot after each pulse 
(cf. Sect.~8.7), and attachment correction taking into account 
different particle composition at different $\theta$ (cf. Sect.~6.7)
the d$E$/d$x$ resolution was improved from 18\% [35] to less than 10\% 
[10,35] for tracks with more than 15 hits. 
A comparison between the calibrated truncated mean $\dedx$-resolution for 
electrons and the Allison/Cobb parametrization \cite{All80} 
is shown in Fig.~\ref{fig:dedxresvshits} as a function of the number of 
hits per track. For the maximum number of hits the $\dedx$-resolution 
approaches the parametrization. With the number of hits per track peaked 
at 18 (cf. Fig.~\ref{fig:hitsPerTrack}) most of the electrons are measured
with a $\dedx$-resolution of better than $10\perc$.

  \section{Performance of the TPC}
  \label{sec:summary}

  For a track multiplicity between 350 and 400, the global position
  resolution achieved with the calibration of the TPC is
  $r\Delta\phi\approx340\um$ and $\Delta r\approx640\um$ \cite{Lud06}
  to be compared with the aimed design resolution of
  $r\Delta\phi^{design}=250-350\um$ and $\Delta r^{design}=400-500\um$
  \cite{App98}. The design value was met for the azimuthal coordinate
  which has a crucial influence on the momentum resolution. Less
  effort was taken for the radial coordinate, where a detailed weight
  determination for the individual hits (cf. Sect.~\ref{sec:difres})
  was disregarded.  The improvements due to the complex calibration
  presented in this paper can be judged by comparing the position
  resolution with former values from a preliminary calibration, namely
  $r \Delta \phi \approx 500\um$ and $\Delta r \approx 800\um$
  \cite{Sch01}.

\begin{figure}[ht]
    \centering
    \pgfimage[width=1.\textwidth]{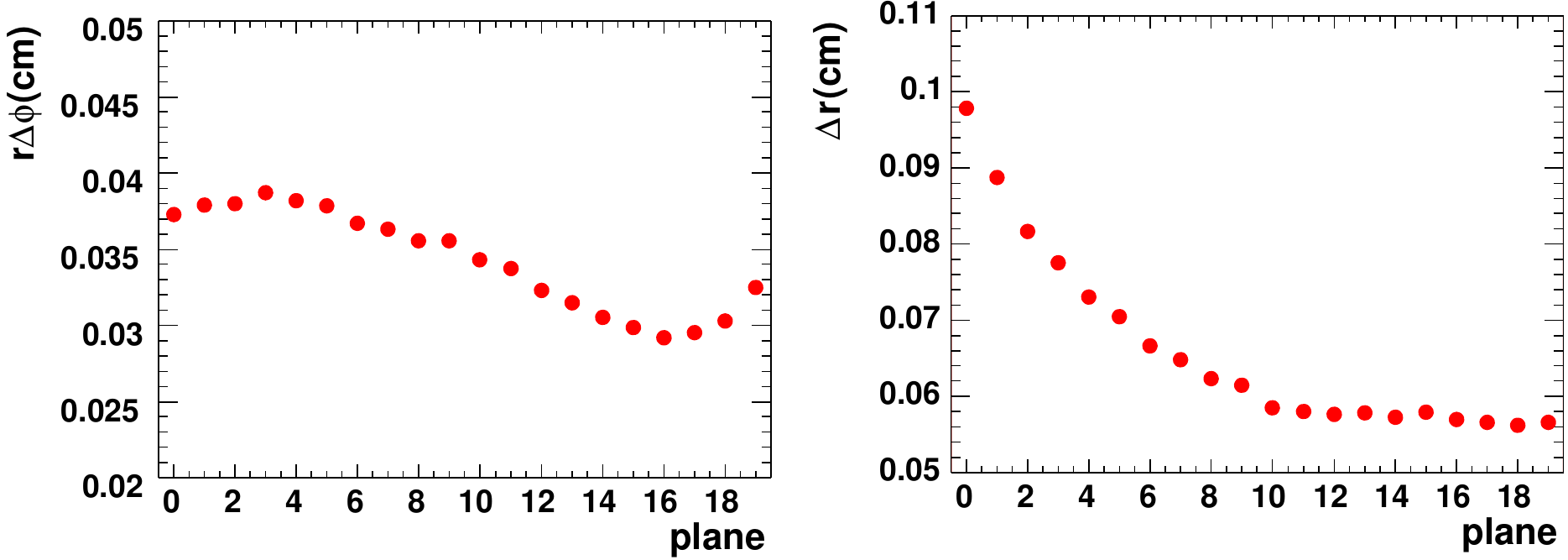}
    \caption{\label{fig:residuals}
      \small Azimuthal (left) and radial (right) position resolution
      as a function of the plane number in the TPC.}
  \end{figure}

  Fig.~\ref{fig:residuals} shows the width of the residuals as a
  function of the plane number in the TPC. The azimuthal resolution
  varies between $0.4\mm$ at the point where the radial magnetic field
  is strongest and $0.3\mm$ close to the end of the TPC. The higher
  occupancy in the first TPC planes is responsible for the
  deterioration of the radial resolution. Furthermore, the drift
  length is on average longer in the first planes and thus diffusion
  plays a larger role.

\begin{figure}[ht]
    \centering
    \pgfimage[width=0.9\textwidth]{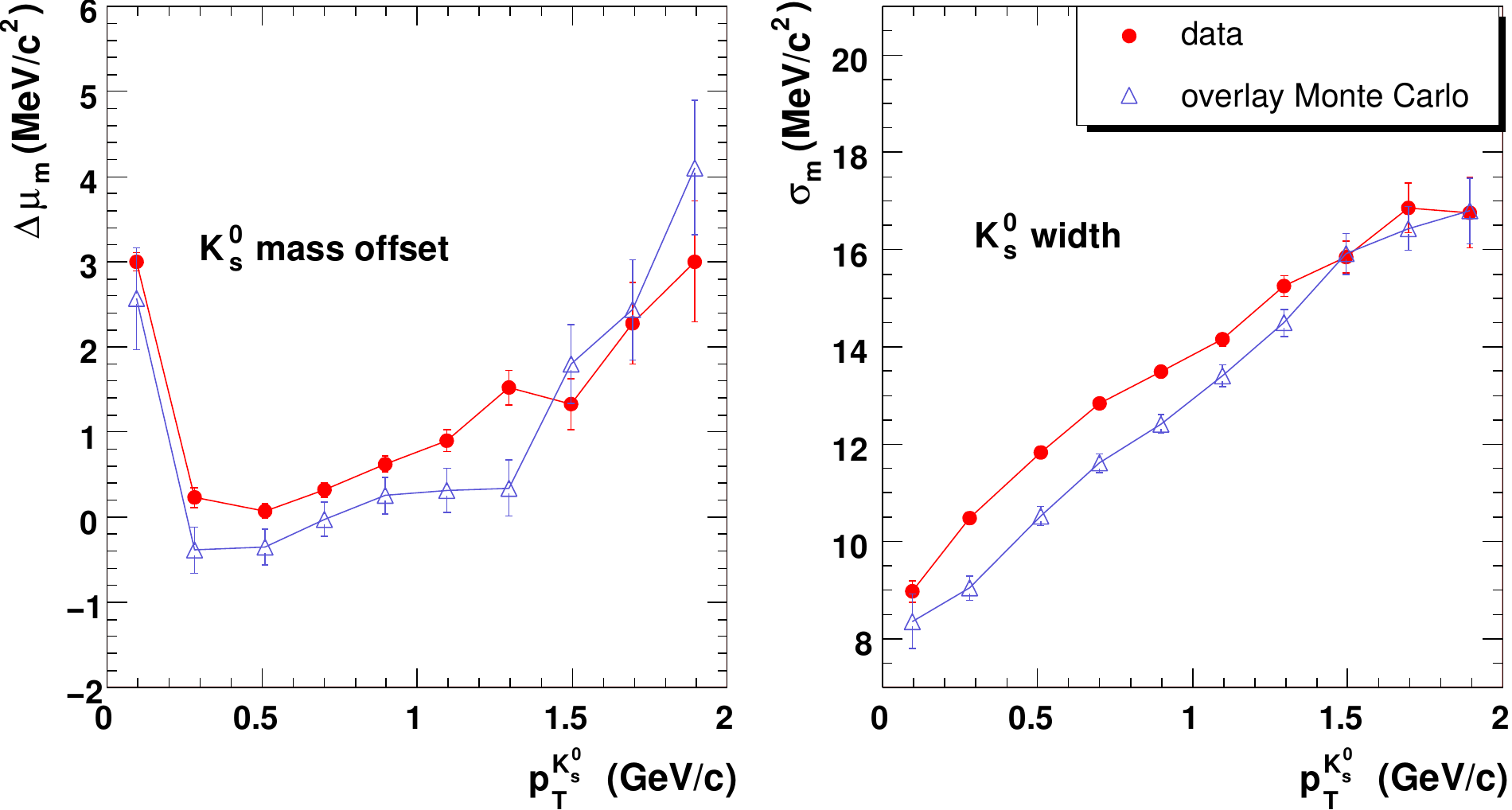}
    \caption{\label{fig:K0_massOffset_width}
      \small Detailed simulations confirm the
      invariant mass shift (left) and width (right) of the
      $K^{0}_{S}$ meson over the whole
      range of transverse momentum. The points are connected
      by a line to guide the eye.}
  \end{figure}

  Another informative quantity is the mass resolution of $K^{0}_{S}$
  meson in the decay mode $\pi^{+}\pi^{-}$.  The contribution of the
  pion masses to the mass of the $K^{0}_{S}$ is small. Thus, the mass
  resolution of the $K^{0}_{S}$ is sensitive to the momentum
  resolution. The calibration improved the width of the reconstructed
  $K^{0}_{S}$ mass from $22\mevcc$ \cite{Sch01} to $13\mevcc$
  \cite{Lud06}.  Furthermore, the performance of the TPC was confirmed
  in detailed Monte Carlo simulation over a wide range of
  parameters. Fig.~\ref{fig:K0_massOffset_width} shows the agreement
  between the mass shift and width of the reconstructed $K^{0}_{S}$
  meson in data and simulation over the whole range of transverse
  momentum. The mass shift indicates a bias in the momentum
  determination. However, the effect is small as compared to the
  average mass resolution of the $K^{0}_{S}$ of $13\mevcc$, indicating
  that the momentum bias is small as compared to the momentum
  resolution $\Delta p$.

\begin{figure}[ht]
    \centering
    \pgfimage[width=0.5\textwidth]{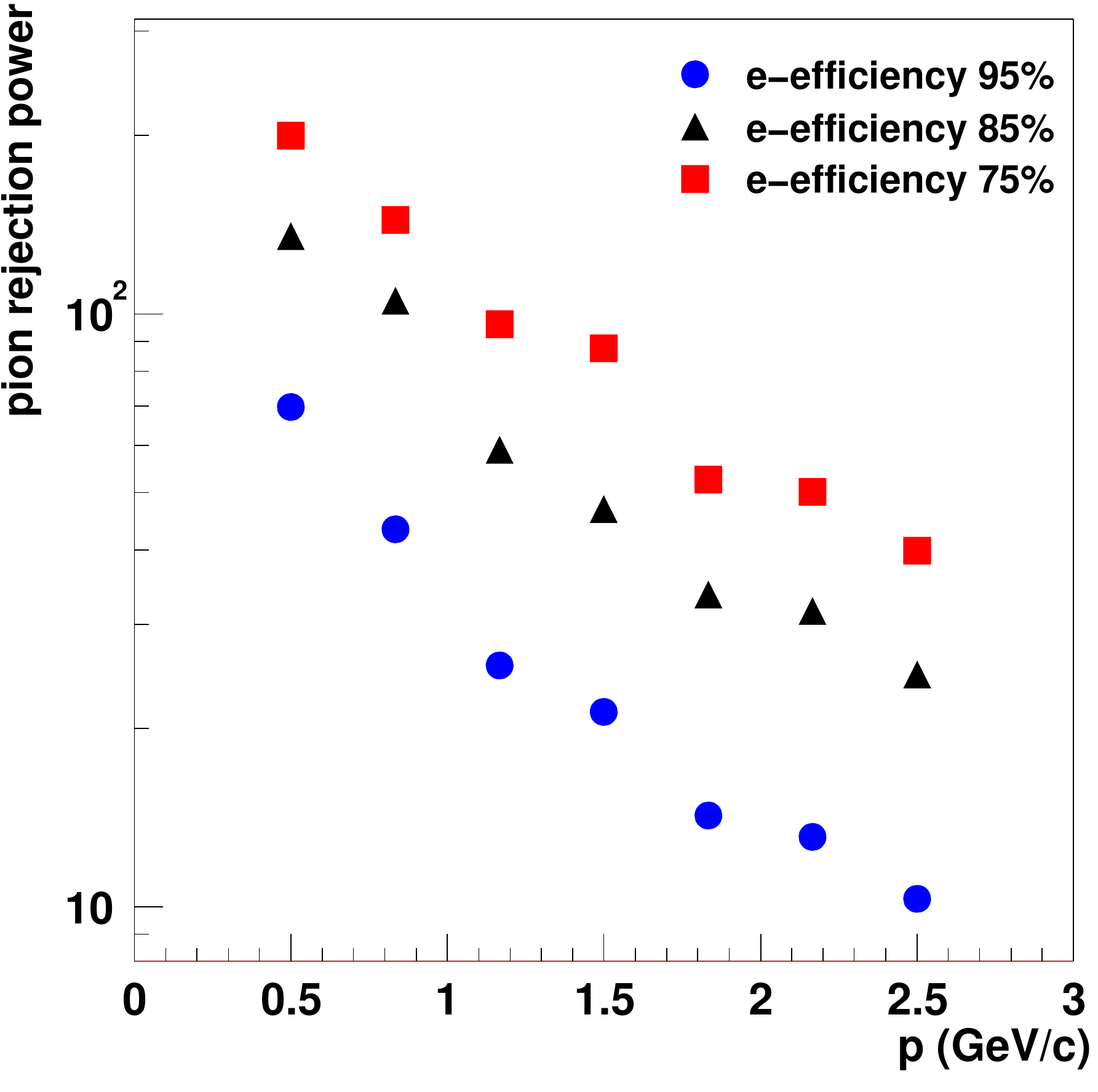}
    \caption{\label{fig:tpc_rejpowervsp}
      \small Pion rejection power as a function of momentum for
      different electron track efficiencies.}
  \end{figure}

  The particle identification capability and the rejection power
  depends on the $\dedx$-resolution. The pion rejection power of the
  TPC, given by the inverse of the pion misidentification rate, is
  shown in Fig.~\ref{fig:tpc_rejpowervsp} as a function of the
  particle momentum for three different electron efficiencies
  \cite{Yur05}. For a particle momentum of $1\gevc$ and an electron
  efficiency of 0.97 in the TPC (not shown in
  Fig.~\ref{fig:tpc_rejpowervsp}) the pion rejection power is 20
  corresponding to a pion misidentification rate of 0.05. Combining
  the pion rejection power of the TPC and the two RICH detectors, the
  pion misidentification rate drops from $5\cdot10^{-4}$ (only RICH)
  to $2.5\cdot10^{-5}$ (RICH and TPC) at an electron efficiency of
  0.68 and $1\gevc$ particle momentum. This comparison corroborates
  the great improvements achieved due to the upgrade of the CERES
  spectrometer with the radial drift TPC.

  \section{Summary}
  \label{sec:summary}

  The upgrade of the electron spectrometer CERES/NA45 with a radial
  drift TPC reached its objective to significantly improve the
  momentum resolution and the pion to electron separation
  capability. As a consequence, the study of lepton pair production in
  A-A collisions allowed for the first time to distinguish between
  different theoretical approaches \cite{Yur05,Ada06}. In addition,
  the TPC opened the possibility to access hadronic observables.
  
  The CERES TPC is exceptional in terms of its radial electric field
  and inhomogeneous magnetic field configuration, both posing a great
  challenge to the calibration. The performance reached is close to
  the design specifications. With a spatial resolution of $340\um$ in
  the azimuthal and $640\um$ in radial direction, and with up to 20
  space point measurements per particle track, a momentum resolution
  of $\Delta p/p = \sqrt{\left(1\% \cdot p \right)^2 +
  \left(2\%\right)^2}$ is obtained. This translates into a mass
  resolution of $3.8\perc$\footnote{
The originally envisaged goal of about 2\% \cite{App98} mass resolution 
was not achieved mainly due to multiple scattering in the mirror of RICH2 
and elsewhere in the apparatus.}
for the $\phi$-meson in the \ee-decay channel. 

  In the course of calibration, on the other hand, the
  $\dedx$-resolution of the TPC was brought down to the level of about
  $10\perc$ which allowed the experiment to significantly improve the
  electron/pion identification capability. With the help of the TPC, a
  pion rejection factor of 1:40000 at an electron efficiency of $0.68$
  was achieved for a particle momentum of $1\gevc$.

  \section{Acknowledgment}
  \label{sec:ack}
  
  We would like to thank Ettore Rosso, Wolfgang Klempt and their team
  for their advice during the design of the TPC. The help of Michel
  Bosteels and his team in constructing and taking into operation the
  gas system is gratefully acknowledged. Fabio Formenti was
  instrumental in planning the final readout scheme. We gratefully
  acknowledge the help of Bernd Panzer-Steindel and his team in the IT
  Division and we thank Anton Przybyla for the perpetual on-site
  support.

  This work was supported by GSI Darmstadt, the German BMBF, the
  U.S. DoE,  the Grant Agency and Ministry of Education of the Czech
  Republic, the Israeli Science Foundation, the Minerva Foundation,
  and the Nella and Leon Benoziyo Center for High Energy Physics
  Research.


\end{document}